\PassOptionsToPackage{pdfpagelabels=false}{hyperref} 
\documentclass[useAMS,usenatbib]{mnras}
\usepackage{graphicx}
\usepackage{amsmath}
\usepackage{natbib}
\usepackage{geometry}

\usepackage{txfonts}
\usepackage{times}

\usepackage{xspace}

\newcommand{\msun}{\,M$_{\odot}$\xspace}
\newcommand{\ovi}{OVI\xspace}
\newcommand{\ovii}{OVII\xspace}
\newcommand{\oviii}{OVIII\xspace}
\newcommand{\gr}{\mbox{(g-r)}\xspace}

\title[IllustrisTNG: Highly Ionized Oxygen]{The abundance, distribution, and physical nature of highly ionized oxygen OVI, OVII, and OVIII in IllustrisTNG}
\author[D. Nelson et al.]{Dylan Nelson$^{1}$\thanks{E-mail: dnelson@mpa-garching.mpg.de},
Guinevere Kauffmann$^{1}$,
Annalisa Pillepich$^{2}$, 
Shy Genel$^{3,4}$, \newauthor
Volker Springel$^{5,6,1}$,
R{\"u}diger Pakmor$^{5}$, 
Lars Hernquist$^{7}$, 
Rainer Weinberger$^{5}$, \newauthor
Paul Torrey$^{8}$, 
Mark Vogelsberger$^{8}$, 
Federico Marinacci$^{8}$ \\\\
$^{1}$Max-Planck-Institut f\"{u}r Astrophysik, Karl-Schwarzschild-Str. 1, 85741 Garching, Germany\\
$^{2}$Max-Planck-Institut f\"{u}r Astronomie, K\"{o}nigstuhl 17, 69117 Heidelberg, Germany\\
$^{3}$Center for Computational Astrophysics, Flatiron Institute, 162 Fifth Avenue, New York, NY 10010, USA\\
$^{4}$Columbia Astrophysics Laboratory, Columbia University, 550 West 120th Street, New York, NY 10027, USA\\
$^{5}$Heidelberg Institute for Theoretical Studies, Schloss-Wolfsbrunnenweg 35, 69118 Heidelberg, Germany\\
$^{6}$Zentrum f\"{u}r Astronomie der Universit\"{a}t Heidelberg, ARI, M\"{o}nchhofstr. 12-14, 69120 Heidelberg, Germany\\
$^{7}$Harvard-Smithsonian Center for Astrophysics, 60 Garden Street, Cambridge, MA, 02138, USA\\
$^{8}$Kavli Institute for Astrophysics and Space Research, Department of Physics, MIT, Cambridge, MA, 02139, USA\\
}

\begin{document}

\maketitle

\begin{abstract}
We explore the abundance, spatial distribution, and physical properties of the \ovi, \ovii, and \oviii ions of oxygen in circumgalactic and intergalactic media (the CGM, IGM, and WHIM). We use the TNG100 and TNG300 large volume cosmological magneto-hydrodynamical simulations. Modeling the ionization states of simulated oxygen, we find good agreement with observations of the low-redshift \ovi column density distribution function (CDDF), and present its evolution for all three ions from $z$\,=\,0 to $z$\,=\,4. Producing mock quasar absorption line spectral surveys, we show that the IllustrisTNG simulations are fully consistent with constraints on the \ovi content of the CGM from COS-Halos and other low redshift observations, producing columns as high as observed. We measure the total amount of mass and average column densities of each ion using hundreds of thousands of simulated galaxies spanning \mbox{$10^{11} < \rm{M}_{\rm halo}/$\msun$ < 10^{15}$} corresponding to \mbox{$10^{9} < \rm{M}_{\rm \star}/$\msun$ < 10^{12}$} in stellar mass. Stacked radial profiles of \ovi are computed in 3D number density and 2D projected column density, decomposing into 1-halo and 2-halo terms. Relating halo \ovi to properties of the central galaxy, we find a correlation between the (g-r) color of a galaxy and the total amount of \ovi in its CGM. In comparison to the COS-Halos finding, this leads to a dichotomy of columns around star-forming versus passive galaxies at \textit{fixed} stellar (or halo) mass. We demonstrate that this correlation is a direct result of blackhole feedback associated with quenching and represents a causal consequence of galactic-scale baryonic feedback impacting the physical state of the circumgalactic medium.
\end{abstract}



\section{Introduction}

Oxygen, after hydrogen and helium, is the most abundant element in the universe. It is a fundamental tracer, charting the formation of single stars as well as entire galaxies. As the dominant metal in the interstellar, circumgalactic, and intergalactic media, oxygen is one of the most important elements in astrophysics, and observations of oxygen in its various forms underpin much of our understanding of galaxy formation and evolution. In its lower ionization states, optical emission lines from oxygen ions including OI, OII, and OIII arise in the photoionized gas clouds around young stars, and are therefore common in the spectra of star-forming galaxies. Oxygen emission from the interstellar medium (ISM) provides foundational measurements of the galaxy mass-metallicity relation, and of the evolution of galaxy metallicities across cosmic time.

In its three highest observable ionization states -- \ovi, \ovii, and \oviii, the motifs of this paper -- oxygen traces gas which is either hot, at temperatures above 100,000 kelvin, or gas at low densities, with less than $10^{-4}$ atoms per cubic centimeter. All three ions can therefore arise in the rarefied plasmas which surround galaxies and extend out to large distances: their hot gaseous halos, commonly referred to as the intra-cluster medium (ICM) or circumgalactic medium (CGM), depending on the mass of the dark matter halo. These highly ionized states of oxygen also emerge in the low gas density structures which make up the topology of the cosmic web of large-scale structure: the intergalactic medium (IGM).

Consequently, the ions of oxygen and their observable signatures represent an important regime for theory, particularly in the study of galaxy evolution. As a property of the gas-phase baryonic component, the most direct and most powerful tools for understanding the abundance, distribution, and properties of ionized oxygen across this wide range of scales are hydrodynamical simulations of cosmological volumes \citep{hernquist89,katz91,cen92,navarro93}. Recent large-scale projects including EAGLE \citep{schaye15,crain15}, Horizon-AGN \citep{dubois14}, Illustris, and now IllustrisTNG have reached an impressive level of theoretical and predictive utility. However, much of their focus remains on the stars, which comprise only $\sim$3\% of the total baryonic mass \citep{fukugita04}, already a meager $\sim$5\% component of the total mass budget of the universe \citep{planck2015_xiii}.

Considering the virialized gas contents of dark matter halos, a number of computational works simulating statistically representative volumes have investigated ionized oxygen, focusing in particular on \ovi at low-redshift, $0.1 \la z \la 0.5$. With the `vzw' galactic-winds model and no AGN feedback, \cite{ford16} proposed that \ovi absorption arises mainly from metals ejected from a galaxy in the past, although the occurrence of \ovi was inconsistent with observational constraints, being too low in the simulations in several metrics. With the Illustris model, \cite{suresh17} likewise found that the overall abundances of \ovi were too low with respect to observations. They also explored the effect, as observed, that there is less \ovi around passive galaxies than star-forming galaxies, attributing this dichotomy to AGN feedback. With the EAGLE model, \cite{oppenheimer16} also found that the overall normalization of simulated column densities was below observational constraints. Focused on the differential signatures around red versus blue galaxies, they found an effect qualitatively similar to observations, which was explained as due to halo mass instead of feedback. Using calculations with thermal supernova and no AGN feedback, \cite{cen13} similarly explored the different levels of absorption around red and blue galaxies, finding that the \ovi bearing warm halo gas is transient, requiring constant energy injection.  

\ovi has also been explored with single zoom simulations of just one, or a handful of, individual galaxies. The focus is almost always on the Milky Way mass regime, and always with the caveat of low number statistics. \cite{hummels13} emphasized the possibility to constrain different supernova feedback recipe strengths, finding it difficult with any explored model to reproduce $N_{\rm OVI}$ columns as high as observed. Similarly, \cite{liang16} found simulated Milky Way mass systems unable to reproduce columns or large distance covering fractions as high as observed. Across a larger range in galaxy mass \cite{gutcke17} also concluded a deficit of order half a dex of simulated \ovi columns as compared to observational measurements. 

Overall, numerical studies so far that utilize standard recipes for galaxy physics have failed to produce metals in the temperature and density regime needed to reproduce observations of \ovi. One idea to remedy the apparent deficiency of this ion in simulations is that time-variable radiation fields from AGN coupled with non-equilibrium ionization effects may produce larger ionic abundances than equilibrium calculations alone \citep{vasiliev15}, also demonstrated in hydrodynamical simulations \citep{segers17,oppenheimer18a}. Other possibilities for the physical origin of the \ovi phase include radiative cooling flows \citep{bordoloi17,mcquinn18}, formation in fast shocks \citep{gnat09}, or within thermally conductive interfaces \citep{borkowski90,gnat10,armillotta17}.

A number of recent observational campaigns have explored \ovi in particular \citep[reviewed in][]{tumlinson17}. The line doublet of O$^{5+}$ at 103.19nm, 103.76nm is a resonance transition with permitted absorption transitions out of the electronic ground state in the rest-frame ultraviolet. At low redshift, space observatories are therefore required, and the Cosmic Origins Spectrography (COS) instrument on the Hubble Space Telescope has enabled many new studies beyond the previous abilities of FUSE/STIS. Data on \ovi from quasar spectra is commonly combined with optical surveys to associate gas absorption with close impact parameter galaxy candidates. Notably, \cite{prochaska11} found a covering factor of order unity for \ovi absorption out to the virial radius, or beyond, for their $z < 0.6$ sub-$L^\star$ sample. The targeted COS-Halos survey validated this result over an extended mass range, particularly for close impact parameters $\le$ 200 kpc, and noted the dichotomy of systematically less \ovi around red as opposed to blue galaxies \citep{tumlinson11,werk12,werk13}. The eCGM survey, with increased statistics down to smaller $M_\star$ values and larger impact parameters $\ge$ 200 kpc, also found high covering factors of strong \ovi absorption \citep{johnson15a}. We directly compare to these two datasets in this paper. Connecting to further details of the central galaxy, the Multiphase Galaxy Halos survey \citep{kacprzak15} found an azimuthal dependence of \ovi absorption with respect to the orientation of the galaxy disk, together with an orientation independent kinematic uniformity \citep{nielsen17}.

At larger spatial scales, hydrodynamical simulation volumes also probe the distribution of ionized oxygen in the IGM or `warm/hot intergalactic medium' \citep[WHIM;][]{cen99,dave01}. The random incidence of ionized oxygen absorption, for instance as a function of column density, is quantified in the column density distribution function (CDDF) which provides a benchmark comparison against a robust low-redshift observable \citep{danforth08,thom08,tilton12}. For instance, \cite{rahmati16} provides the $z \simeq 0.2$ \ovi CDDF for the EAGLE model, and \cite{suresh15} likewise for Illustris, both generically under-predicting the incidence of high column density absorption. \cite{oppenheimer12} computes the \ovi CDDF of the `vzw' type models (no AGN feedback), pointing out the sensitivity to the wind parameters as well as to the numerical treatment of metal mixing.

Moving to the higher ionization states, \mbox{He-like} \ovii and \mbox{H-like} \oviii ions both have radiative transitions at x-ray energies. Absorption in both ions from within or near-field to the Milky Way CGM has been measured by Chandra \citep{nicastro02} and XMM-Newton \citep{rasmussen03,bregman07b}. Semi-empirical models often combining \ovii and \oviii tracers in either absorption or emission then determine consistent gas density profiles and other characteristics of our nearby circumgalactic gas reservoir \citep{miller15,li17,faerman17}, emphasizing the possible tension of the `missing baryons' problem.  Simulations have provided predictions for emission from the WHIM as well as from the gaseous halos around larger groups and clusters \citep{kravtsov02,yoshikawa03,fang05,bertone10a} as expected from the theory of collapse \citep{silk77,wr78}. Beyond the Milky Way, resolved measurements of \ovii or \oviii remain difficult due to the spectral resolution and sensitivity of current instrumentation \cite[see review in][]{bregman07}, with several recent efforts towards detection in both absorption \citep[e.g.][]{buote09,williams13} and emission \citep{pinto14}. 

Together, these three highest observable ionization states of oxygen act not only as detectable tracers of the largest baryon reservoirs in the universe, they are also sensitive to the physical details and assumptions of the theoretical modeling, particularly in cosmological hydrodynamical simulations. The purpose of this paper is therefore (i) to provide a comprehensive theory-based census of the abundance and distribution of these ions, (ii) to understand how robust our current computational predictions are, and (iii) to use these models to comment on the physical nature of highly ionized oxygen and its relation to luminous galaxies.

In Section~\ref{sec_methods} we describe the TNG simulations (\ref{sec_sims}) and all our current analysis methodology (\ref{sec_ion_modeling} - \ref{sec_coldens}). Section~\ref{sec_abundance_largescale} then provides a census of the distribution of ionized oxygen on large, intergalactic scales, while Section~\ref{sec_abundance_halos} assesses circumgalactic scales and compares to quasar absorption line observations. Moving into a largely theoretical exploration, Section~\ref{sec_galaxyconnection} connects the halo CGM to properties of the central galaxy. We finish with a discussion in Section~\ref{sec_discussion} and summarize our conclusions in Section~\ref{sec_conclusions}.


\section{Methods} \label{sec_methods}

\subsection{The TNG Simulations} \label{sec_sims}

The IllustrisTNG project\footnote{\url{http://www.tng-project.org}} \citep{pillepich18, nelson18, naiman17, marinacci17, springel18} is the successor of the Illustris simulation \citep{vog14b,vog14a,genel14,sijacki15}. It uses our updated `next generation' galaxy formation model which, in addition to including new physical ingredients, significantly refines the original Illustris model. Full details of the approach used for the TNG cosmological simulations are available in Table 1 of \cite{pillepich17a}, which enumerates the principal differences with respect to the original Illustris model -- in combination with \cite{weinberger17}, which focuses on the high-mass end and the blackhole feedback. Explorations of the model behavior and its sensitivity to variations on smaller test volumes are also presented in those two TNG methods papers. Every aspect of the implementation, including all parameter values and simulation code numerics, are described therein and \textit{entirely unchanged} for our production simulations. For brevity we here identify only the main aspects.

TNG is a series of three large cosmological volumes, simulated with gravo-magnetohydrodynamics (MHD) and incorporating a comprehensive model for galaxy formation physics. It uses the \textsc{Arepo} code \citep{spr10} which solves the coupled equation systems of self-gravity and ideal, continuum MHD \citep{pakmor11,pakmor13}. Gravity is computed using a Tree-PM approach, whereas the fluid dynamics employ a Godunov/finite-volume approach where the spatial discretization is an unstructured, moving, Voronoi tessellation of the domain. The numerical scheme is therefore quasi-Lagrangian; it is also second order in space as well as in time \citep[see][]{pakmor16}, leverages a nested binary hierarchy of individual particle timesteps, and has been designed to efficiently execute massively parallel distributed memory astrophysical simulations.

The TNG simulations include a thorough physical model for the dominant processes which guide the formation and evolution of galaxies. Namely: (i) gas radiative mechanisms at the microphysical scale, including cooling from a primordial/metal-enriched gas plus heating from a time-evolving, spatially uniform background radiation field, (ii) star formation at high gas densities in the ISM, (iii) the evolution of stellar populations and subsequent chemical enrichment following both supernovae Ia, II, as well as AGB stars, while individually tracking the production, return, and gas-phase advection the nine elements H, He, C, N, O, Ne, Mg, Si, and Fe, (iv) galactic-scale outflows driven by stellar feedback energy, (v) the formation, binary merging, and gas accretion of supermassive blackholes, (vi) and multi-mode blackhole feedback which operates either in a thermal `quasar' mode at high accretion rates, or a kinetic `wind' mode at low accretion rates. 

Blackholes are seeded in massive halos and then accrete nearby gas at the Eddington limited Bondi rate. Based on this accretion rate, their feedback mode is determined. When the Eddington ratio exceeds a threshold of $\chi = \rm{min} [ 0.002 (M_{\rm BH}/10^8 \rm{M}_\odot)^2, 0.1]$ thermal energy is injected continuously into the surrounding gas. The rate is $\Delta E_{\rm high} = \epsilon_{\rm f,high} \epsilon_{\rm r} \dot{M}_{\rm BH} c^2$ where $\epsilon_{\rm f,high}$ is the high-state coupling efficiency, $\epsilon_{\rm r}$ is the radiative accretion efficiency, and $\epsilon_{\rm f,high} \epsilon_{\rm r} = 0.02$. Below this threshold, kinetic energy is injected as a time-pulsed, oriented `wind' with a direction which reorients for each event. The rate is $\Delta E_{\rm low} = \epsilon_{\rm f,low} \dot{M}_{\rm BH} c^2$ with $\epsilon_{\rm f,low} \le 0.2$ (its typical value at onset), the efficiency decreasing at low environmental density \citep[see][]{weinberger17}.

Galactic-scale outflows generated by stellar feedback are modeled using a kinetic wind approach, whereby the available energy from SNII is used to stochastically eject star-forming gas cells from galaxies. The injection velocity of such wind particles is $v_{\rm w} \propto \sigma_{\rm DM}$ where $\sigma_{\rm DM}$ is the local dark matter velocity dispersion, subject to a minimum $v_{\rm w,min} = 350$ km/s. The mass loading of the winds is then $\eta = \dot{M_{\rm w}} / \dot{M_{\rm SFR}} = 2 (1 - \tau_{\rm w}) e_{\rm w} / v_{\rm w}^2$ where $\tau_{\rm w} = 0.1$ is the thermal energy fraction and $e_{\rm w}$ is a metallicity dependent modulation of the canonical $10^{51}$ erg available per SNII. Winds particles are hydrodynamically decoupled from surrounding gas until they exit the dense, star-forming environment. The total energy available to drive winds from a gas cell therefore depends on its instantaneous star formation rate as $\Delta E = e_{\rm w} \dot{M}_{\rm SFR}$ which is roughly $\simeq 10^{41} - 10^{42}$ erg/s $\dot{M}_{\rm SFR} / (M_\odot / \rm{yr})$ depending on the local gas metallicity \citep[see][]{pillepich17a}.

The energetic balance between stellar and blackhole feedback as a function of mass and redshift has been explored in \cite{weinberger18}. For example, we find that in galaxies with $M_\star \simeq 10^{10.5}$\msun at $z=0$ the energy injection rate from blackholes dominates that of supernovae at all $z \lesssim 6$, wherein the BH-driven kinetic wind is energetically dominant only at late times with $\sim 3-4 \times 10^{42}$ erg/s versus an energy budget of $\sim 10^{41}$ erg/s from stellar feedback. At this same mass scale galaxies have wind velocities $v_{\rm w} \sim 800$ km/s and mass loadings $\eta \sim 2$ at $z=0$.

The TNG project includes three distinct simulation volumes: TNG50, TNG100, and TNG300. Here we make use of the two currently completed, larger volumes: \textbf{TNG100}, which includes 2$\times$1820$^3$ resolution elements in a $\sim$\,100 Mpc (comoving) box, and \textbf{TNG300}, which includes 2$\times$2500$^3$ resolution elements in a $\sim$\,300 Mpc box. For TNG100 the baryon mass resolution is $1.4 \times 10^6$\msun, the gravitational softening length of the dark matter and stars is 0.7 kpc at $z$\,=\,0, and the gas component has an adaptive softening with a minimum of 185 comoving parsecs. For TNG300 the baryon mass resolution, collisionless softening, and gas minimum softening are $1.1 \times 10^7$\msun, 1.5 kpc, and 370 parsecs, respectively. For the complete numerical details of these two runs and their lower resolution analogs see Table A2 of \cite{nelson18}. 

For TNG we invoke a cosmology consistent with recent observational constraints \citep{planck2015_xiii}, namely $\Omega_{\Lambda,0}=0.6911$, $\Omega_{m,0}=0.3089$, $\Omega_{b,0}=0.0486$, $\sigma_8=0.8159$, $n_s=0.9667$ and $h=0.6774$. Throughout this work we identify gravitationally bound substructures using the \textsc{Subfind} algorithm \citep{spr01} and connect them through time with the \textsc{SubLink} merger tree algorithm \citep{rodriguezgomez15}. Halo masses are always given as $M_{\rm 200,crit}$ and by virial radii we mean the corresponding spherical overdensity $r_{\rm 200,crit}$ values. Stellar masses of the central galaxy are always restricted to a three dimensional aperture of 30 physical kpc. The \gr colors of galaxies are computed according to the fiducial model of \cite{nelson18} accounting for dust attenuation effects.

\subsection{Modeling Metal Ionization States} \label{sec_ion_modeling}

To compute ionization states we use \textsc{Cloudy} \citep[][v13.03]{ferland13} including both collisional and photo-ionization in the presence of a UV + X-ray background \citep[the 2011 update of][]{fg09}.\footnote{This radiation field includes energies from 0.5 to $\sim$4000 Rydberg, and we note that it has a lower intensity than \cite{hm12} at frequencies near the ionization energy of \ovi, by roughly a factor of 3.5. The intensity of the UVB will be important for establishing the amount of photo-ionized \ovi. However, our choice of UVB is the only one self-consistent with the simulations themselves, and we do not explore other options herein.} We follow \cite{bird14} and use \textsc{Cloudy} in single-zone mode and iterate to equilibrium, accounting for a frequency dependent shielding from the background radiation field (UVB) at high densities, using the fitting function of \cite{rahmati13}. As gas cells in the simulation are essentially single-zone (i.e. no internal density structure), it would be inconsistent to assume a multi-zone or more complex geometry in the photoionization calculation. Any physical mechanism producing metal ions in gas structure with very small physical scales ($\ll$kpc) would not be resolved in our current simulations, and not accounted for herein. We do not consider the impact of local sources of radiation beyond the UVB.

We run \textsc{Cloudy} in the `constant temperature' mode, with no induced processes \citep[following][]{wiersma09} and assuming the solar abundances of \cite{grevesse10}. From the simulation output a 4D grid in (n$_{\rm H}$,\,T,\,Z,\,z), hydrogen number density, temperature, metallicity, and redshift is produced with the following configuration: $-7.0 < \log(\rm{n}_{\rm H} [\rm{cm}^{-3}]) < 4.0$ with step size $\Delta$n$_{\rm H}$ = 0.1, $3.0 < \log(\rm{T} [\rm{K}]) < 9.0$ with $\Delta$T = 0.05, $-3.0 < \log(\rm{Z} [\rm{Z}_{\rm sun}]) < 1.0$ with $\Delta$Z = 0.4, and $0 < \rm{z} < 8$ with $\Delta$z = 0.5. Note that the metallicity dependence is minimal, while the redshift dependence captures the changing intensity and spectral shape of the assumed background radiation field. At each grid point we determine and save the ionization fraction $x_{i,j}$ of the j$^{\rm th}$ ion of species $i$. The total mass (or volume number density) of an ion within a gas cell is then given by this ionization fraction times the mass (or number density) of the parent species. We therefore have an estimate of the total mass, spatial distribution of that mass, and kinematics, of any ion such as \ovi throughout the entire simulation volume. For the oxygen ions considered herein, we directly use the oxygen content which is produced, tracked, and output by the simulation, and do not assume solar abundances.

Column densities and number densities, in units of cm$^{-2}$ and cm$^{-3}$ respectively, always indicate the number of oxygen atoms per square or cubic centimeter. 

\subsection{Two-Point Correlation Functions} \label{sec_corrfuncs}

The correlation function of gas metal mass in a given ionization state provides a convenient, statistical way to quantify the distribution of metal ions at a given spatial scale. We measure the three dimensional real-space two-point autocorrelation function - for example, of \ovi mass $\xi_{\rm OVI}(r)$, by taking the positions of all gas cells in the simulation volume, weighting by their total ionic mass. Although typically employed on discrete point sets, the 2-point function is the Fourier transform of the power spectrum and can also be used with a continuous density field as we do herein \citep{peebles73}. In the periodic geometry of the simulation volumes we use the natural estimator 

\begin{equation}
\xi(r_i) = \frac{DD_i}{RR_i} - 1
\end{equation}

\noindent where DD are the number of (optionally weighted) pairwise distances $x_{jk}$ of a given sample between $r_i$ and $r_i+\Delta r_i$

\begin{equation}
DD_i = \sum_{j,k = 1}^{N} w_j w_k \left[ \theta \left(x_{jk} - r_i \right)- \theta \left(x_{jk} - (r_i+\Delta r_i)\right) \right].
\end{equation}

\noindent Here $\theta$ is the Heaviside unit step function, and the $w_j = w_k = 1$ in the case of uniform weighting. We take $N_{\rm bin} = 40$ logarithmically spaced bins from $r_{\rm min} = 2$ kpc\,h$^{-1}$ to $r_{\rm max} = 20$ Mpc\,h$^{-1}$, a dynamic range of 10,000. Around this upper limit box size effects start to compromise spatial clustering measurements for TNG300 \citep{springel18}. For a sample of $N$ gas cells in a periodic cube where each bin is a spherical shell, no actual random samples are needed and the expression for RR$_i$ can be written down directly

\begin{equation}
RR_i = V_{\rm bin} \bar{n} = \frac{4 \pi}{3} \left[ (r_i + \Delta r_i)^3 - r_i^3 \right] (N^2 \bar{w}^2 / L_{\rm box}^3).
\end{equation}

\noindent where $\bar{w}$ is the mean weight. In practice, with possibly tens of billions of individual gas cells, it is both prohibitively expensive and statistically unnecessary to compute the $\xi(r)$ statistic for all pairs. We therefore downsample to $\simeq 3 \times 10^6$ for the $j$ sum, always keeping the $k$ sum complete. To estimate errors, we use the jackknife technique \citep[see][]{norberg09} by successively excluding contiguous particle intervals each of size $N_{\rm gas}/10^5$ for a total of $N_{\rm sub}=100$ resamplings. From the estimated covariance matrix $C$ we plot errors corresponding to one sigma values $\sigma_i = (C_{i,i} / N_{\rm sub})^{1/2}$. 



\subsection{Column Density Distribution Functions} \label{sec_cddf}

We compute the column density distribution function (CDDF) of a given mass tracer $f(N_q)$ as

\begin{equation}
f(N_q) = \frac{F(N_q)}{\Delta N_q \Delta X(z)}
\end{equation}

\noindent where $\Delta X(z) = (H_0/c)(1+z)^2 L_{\rm box}$ is the dimensionless absorption distance for a cubic simulation volume of comoving sidelength $L_{\rm box}$, and $F(N_q)$ is the fraction of gridded columns in some small column density bin $\Delta N_q$ centered around $N_q$. We consider the mass tracers $q \in \{\rm{OVI, OVII, OVIII}\}$ in this work.

In order to calculate $F(N_q)$ we project all the tracer mass in some depth through the simulation volume onto a regular grid \citep[following][]{bird13}, using the standard cubic-spline kernel with support $h = [3V/(4\pi)]^{1/3}$ where $V$ is the volume of each simulation gas cell. Each projection grid cell is therefore an approximation of a single observed line of sight. We use a constant grid cell size of $\simeq 7.4$ ckpc (a grid dimension of 41000x41000 for TNG300), for which we verify the CDDFs are converged by varying this size by a factor of two in each direction. 

Unless stated otherwise, we adopt a depth of 10 cMpc/h, corresponding to $\simeq$ 12.3 pMpc and a velocity range $\Delta v \simeq 750$ km/s at $z=0.2$. This is large enough to cover the typical extent over which observed \ovi absorption components are considered to be single systems, and small enough that multiple absorption systems do not accumulate along individual sight lines due to excessive path lengths \citep[see][and Section \ref{subsec_cddfs} below for the inherent limitations of this approach]{danforth16}.

\subsection{Oxygen Ion Column Densities and Covering Fractions} \label{sec_coldens}

To obtain column densities around a simulated galaxy, \ovi mass is projected onto a square grid with a given transverse size and projection depth. For comparison to COS-Halos, the grid size is 800 pkpc and the projection depth is 3 pMpc, the same depth used for the 2D radial column density profile calculations. For comparison to the eCGM survey, the grid size is 2.6 pMpc and the projection depth is 4 pMpc, which corresponds to a Hubble expansion induced $\Delta v \simeq 250$ km/s at $z\,=\,0.2$. Grid sizes are chosen to cover the range of observed impact parameters, and projection depths in approximate equivalent to the velocity interval over which the observational search for \ovi absorption is undertaken. The pixel scale in all cases is a constant 2 pkpc, and projection is always along the fixed z-axis of the simulation volume (and therefore random with respect to the orientation of the galaxy). As with the CDDF computations, we assume the ion mass in a given gas cell is spatially distributed following the standard cubic-spline kernel with support $h = [3V/(4\pi)]^{1/3}$ with $V$ is the volume of the Voronoi gas cell.

For comparison to column density measurements, a single $N_{\rm OVI}$ value is selected for each simulated realization of each observed galaxy, by randomly choosing among all the pixels which share the same distance from the central galaxy which is closest to the requested impact parameter -- in effect, a random position angle on the plane of the sky. For comparison to covering fraction measurements, the fraction of pixels above a threshold $N_{\rm OVI}^{\rm min}$ in the projected radial distance bin relative to the total number in that bin is the covering fraction $\kappa_{\rm OVI}$, and we show the median and 1$\sigma$ values taken across the simulated ensemble in each radial bin. 


\begin{figure*}
\centering
\includegraphics[angle=0,width=7.1in]{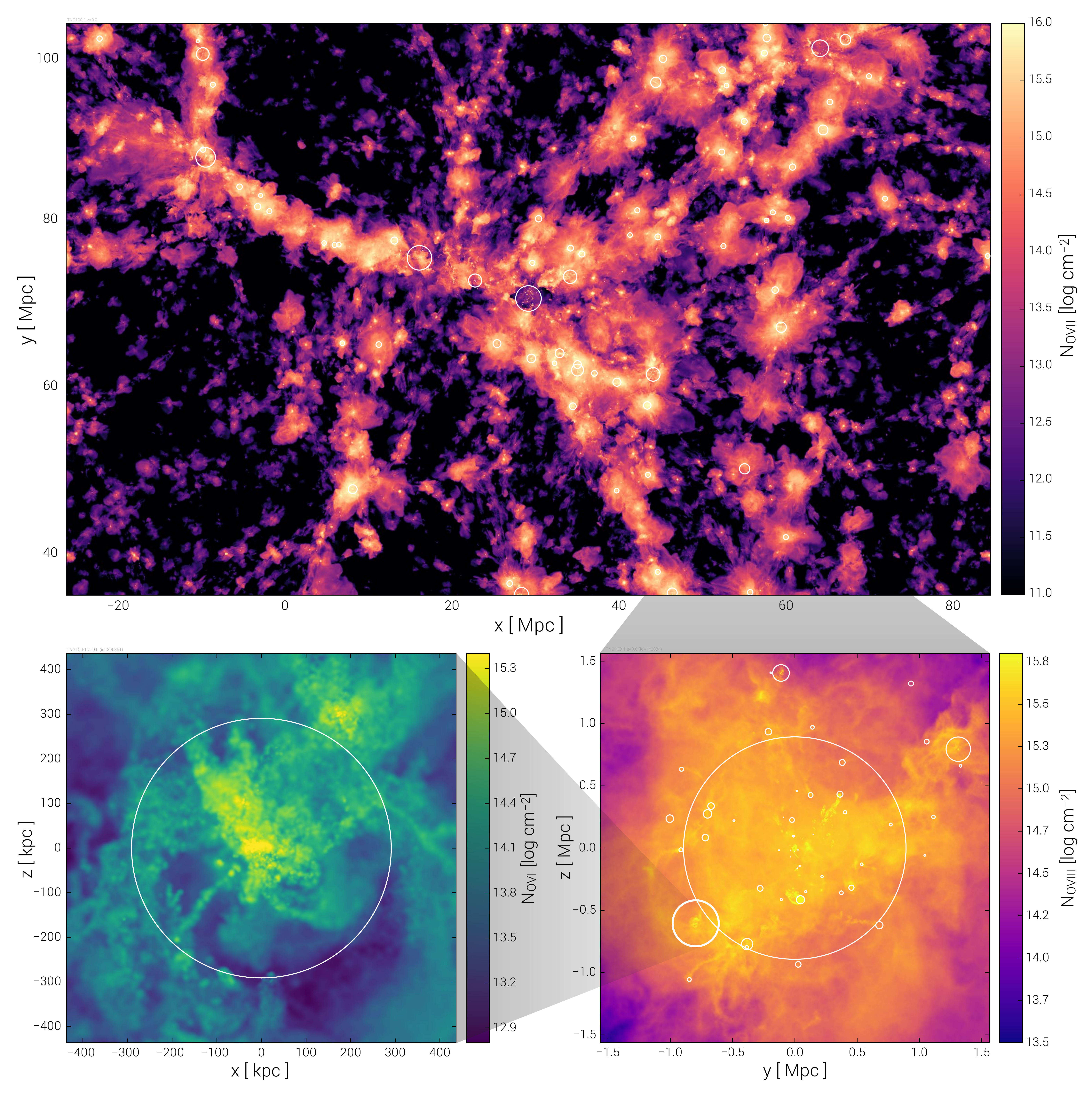}
\caption{ The distribution of highly ionized oxygen in the TNG100 cosmological simulation at $z=0$. In the upper panel, the column density of \ovii tracing the large-scale cosmic web of filaments and collapsed halos. Here we project through a depth of 15 Mpc, a seventh of the entire volume. The 50 most massive visible halos are shown as white circles at their virial radii. In the lower right panel, we zoom onto a small galaxy cluster -- the 23rd most massive -- showing its \oviii column in a box of side length $3.5 r_{\rm vir}$. The 50 most massive satellites of this halo are again shown with smaller circles, while the large circle corresponds to the virial radius ($\sim$ 900 kpc) of the halo itself, which has a total mass of $\simeq 10^{13.8}$\msun. In the lower left panel, we further zoom onto the scale of a small group, whose central galaxy has $M_\star \simeq 10^{10.7}$\msun and total $M_{\rm halo} \simeq 10^{12.5}$\msun, showing the \ovi column density projected in a box of side length $3 r_{\rm vir}$.
 \label{fig_box_composite}}
\end{figure*}

\section{The Abundance and Spatial Distribution of Ionized Oxygen: Intergalactic Space} \label{sec_abundance_largescale}

The production and subsequent evolution of cosmic heavy elements encodes a rich corpus on the interplay between star formation, stellar evolution, gas dynamics, and baryonic feedback processes across a wide range of scales. We therefore begin with a visual impression of the distribution of ionized oxygen from the largest scales of the intergalactic medium down to highly structured ion morphologies within the halos of individual galaxies.

\begin{figure*}
\centering
\includegraphics[angle=0,width=7.1in]{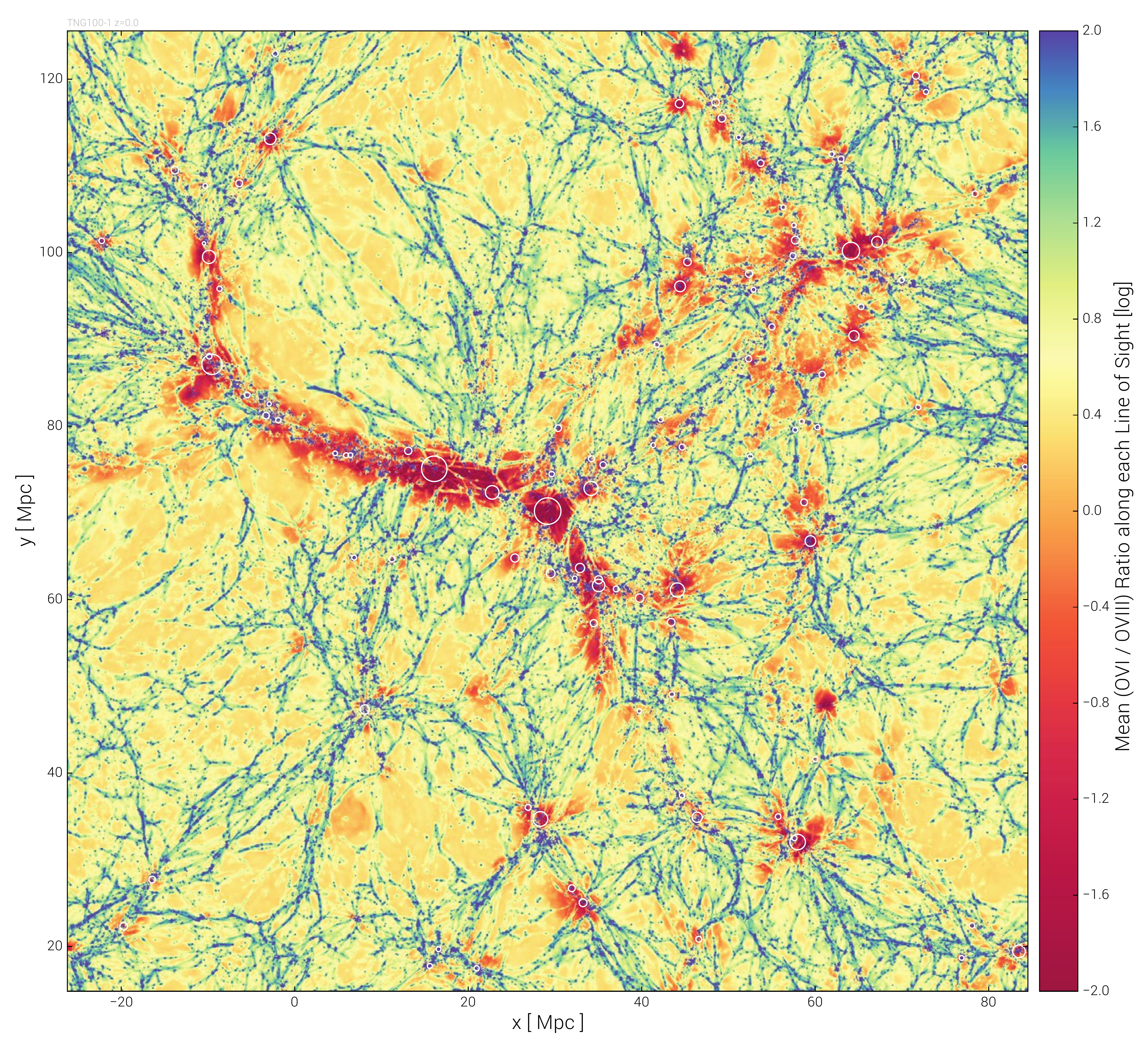}
\caption{ The relationship between highly ionized \ovi and \oviii across the TNG100 cosmological simulation at $z=0$. We show a slice 15 Mpc deep along the line of sight direction, where the color indicates the average ratio of the mass present in the \ovi versus \oviii ions of oxygen. Thin, small-scale cosmic filaments are dominated by \ovi by a factor of 100 or more (blue), whereas the virialized gas in massive collapsed halos, as well as in large cosmic filaments between clusters and the near-field IGM around clusters are dominated by \oviii, also by a factor of 100 or more (red). Regions of the volume near unity (orange) show where these two ions are equally abundant -- in low-density, low-metallicity cosmic voids. The 50 most massive visible halos are shown as white circles at their virial radii.
 \label{fig_box_o6o8}}
\end{figure*}

Figure \ref{fig_box_composite} shows the disposition of several different oxygen ions in the TNG100 cosmological simulation at $z=0$. First, the \ovii column density, tracing the large-scale cosmic web of filaments and collapsed halos (top panel). The most massive cluster in the simulation (frame center; near $x \sim 30$\,Mpc and $y \sim 70$\,Mpc) appears deficient in \ovii as the bulk of its oxygen is at even higher ionization states. As an example, we then zoom onto a small galaxy cluster of total mass $\simeq 10^{13.8}$\msun (and virial radius $\sim$ 900 kpc) to visualize its \oviii column density (lower right panel). As we show later, such a cluster has more \oviii than either \ovii or \ovi, with a total gravitationally bound mass of $M_{\rm OVIII} \sim 10^{8.5}$\msun. On the outskirts of this large overdensity we  further zoom onto the scale of a small group, showing the \ovi column density. The galaxy at the center of this group has a stellar mass of $\simeq 10^{10.7}$\msun (total $M_{\rm halo} \simeq 10^{12.5}$\msun). At this mass scale, \ovi is sub-dominant to \ovii, with a total bound mass of the former of $M_{\rm OVI} \sim 10^{6.5}$\msun. Although centrally concentrated, the distribution of ionized \ovi around a halo like that shown is clearly influenced by the complexities of the nearby environments, including the presence of other, smaller galaxies.

Returning to the largest scales, Figure \ref{fig_box_o6o8} shows the abundance of \ovi relative to \oviii across the cosmological volume of TNG100 at $z=0$. The colormap indicates the mean ratio of $M_{\rm OVI} / M_{\rm OVIII}$ along each line of sight, which varies by more than a factor of 10,000 depending on environment. In under-dense cosmic voids, the two ions are roughly in equipartition (orange), whereas in the thin, dense filaments of the cosmic web \ovi dominates over \oviii by a factor of 100 or more (dark blue). In contrast, the virialized gas halos around massive groups and clusters are dominated by \oviii, again by factors of 100 or more (red), and this is also true in the localized IGM around these high mass halos as well as in the largest cosmic web filaments which bridge between them.

\begin{figure*}
\centering
\includegraphics[angle=0,width=7.0in]{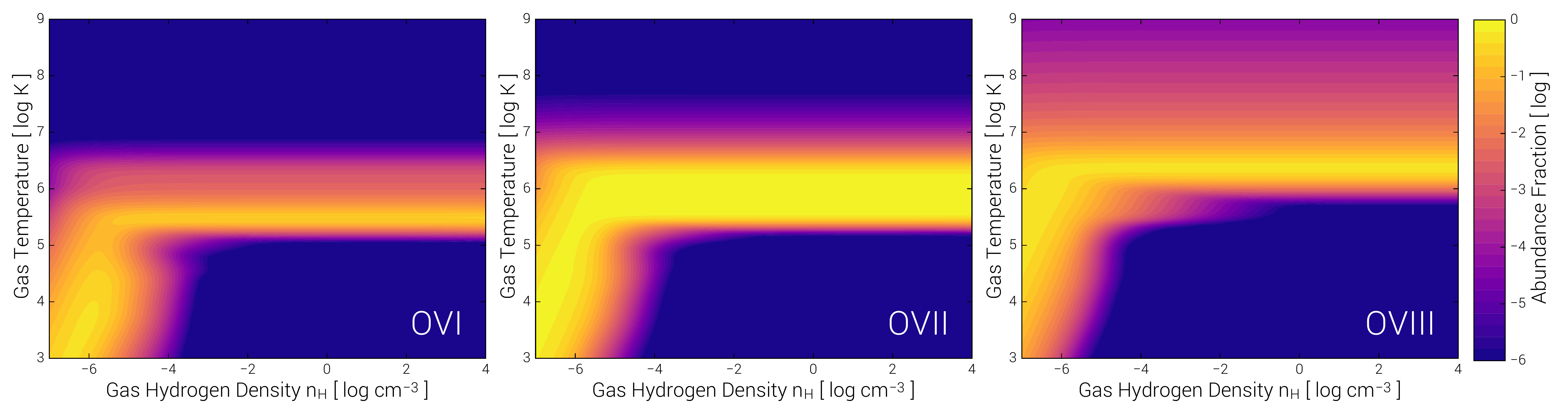}
\includegraphics[angle=0,width=7.0in]{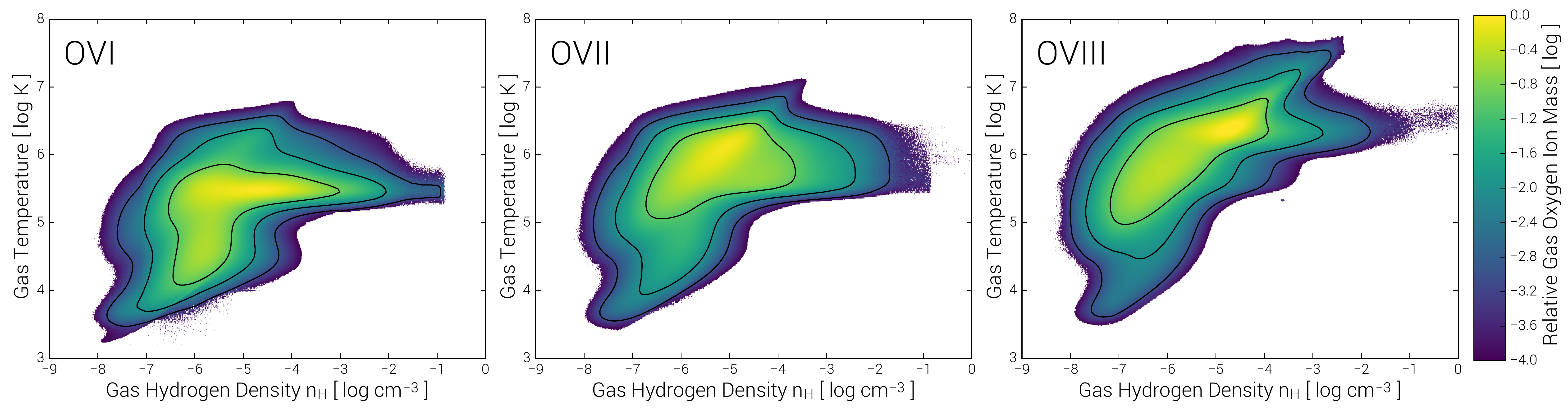}
\caption{ \textbf{Top panels.} Ionization fractions of \ovi, \ovii, and \oviii as a function of density and temperature. The output of the \textsc{Cloudy} modeling given our assumed background radiation field (at $z=0$), and for a fixed metallicity of $Z = 0.1 Z_\odot$. Whereas the horizontal bands represent the regime of predominantly collisional excitation, the largely vertical bands at low hydrogen number densities arise from photoionization; simulation independent. \textbf{Bottom panels.} Phase diagrams in density temperature space for the TNG100 simulation at $z=0$. From left to right, these are weighted by the total per-pixel ion mass of \ovi, \ovii, and \oviii, respectively. The normalized colormap indicates the relative ion mass fraction -- i.e., dark purple values at -4.0 have 10,000 times less ion mass per pixel than peak pixels in yellow. The three black contours enclose relative mass log fractions of $\{-1.0, -2.0, -3.0\}$, from innermost to outermost.
 \label{fig_ion_states}}
\end{figure*}

To better understand the density and temperature regimes within which each oxygen ion arises, Figure \ref{fig_ion_states} shows two diagnostics of ionization fraction for each of \ovi, \ovii, and \oviii. The top panels show the direct output of our \textsc{Cloudy} modeling, as a function of hydrogen number density and temperature, for a fixed metallicity $Z = 0.1 Z_\odot$ and the redshift zero UVB. The largely horizontal band represents the collisional excitation regime, while the low-density vertical branch arises from photoionization. At $z=0$ this process dominates below a hydrogen number density of $n_{\rm H} \la 10^{-5}$ cm$^{-3}$. At redshift two near the peak of the cosmic SFRD this threshold density is approximately two dex higher, before dropping again towards high redshift. Ionization of oxygen in the low redshift CGM is largely controlled by collisional excitation, the abundance fractions therefore peaking at characteristic temperature regimes of $T_{\rm OVI} \sim 10^{5.5}$ K, $T_{\rm OVII} \sim 10^{6.0}$ K, and $T_{\rm OVIII} \sim 10^{6.5}$ K, respectively, which are essentially independent of gas density for sufficiently high densities. There are, however, significant contributions to \ovi, \ovii, and \oviii across widely overlapping ranges of the phase diagram.

\begin{figure*}
\centering
\includegraphics[angle=0,width=5.2in]{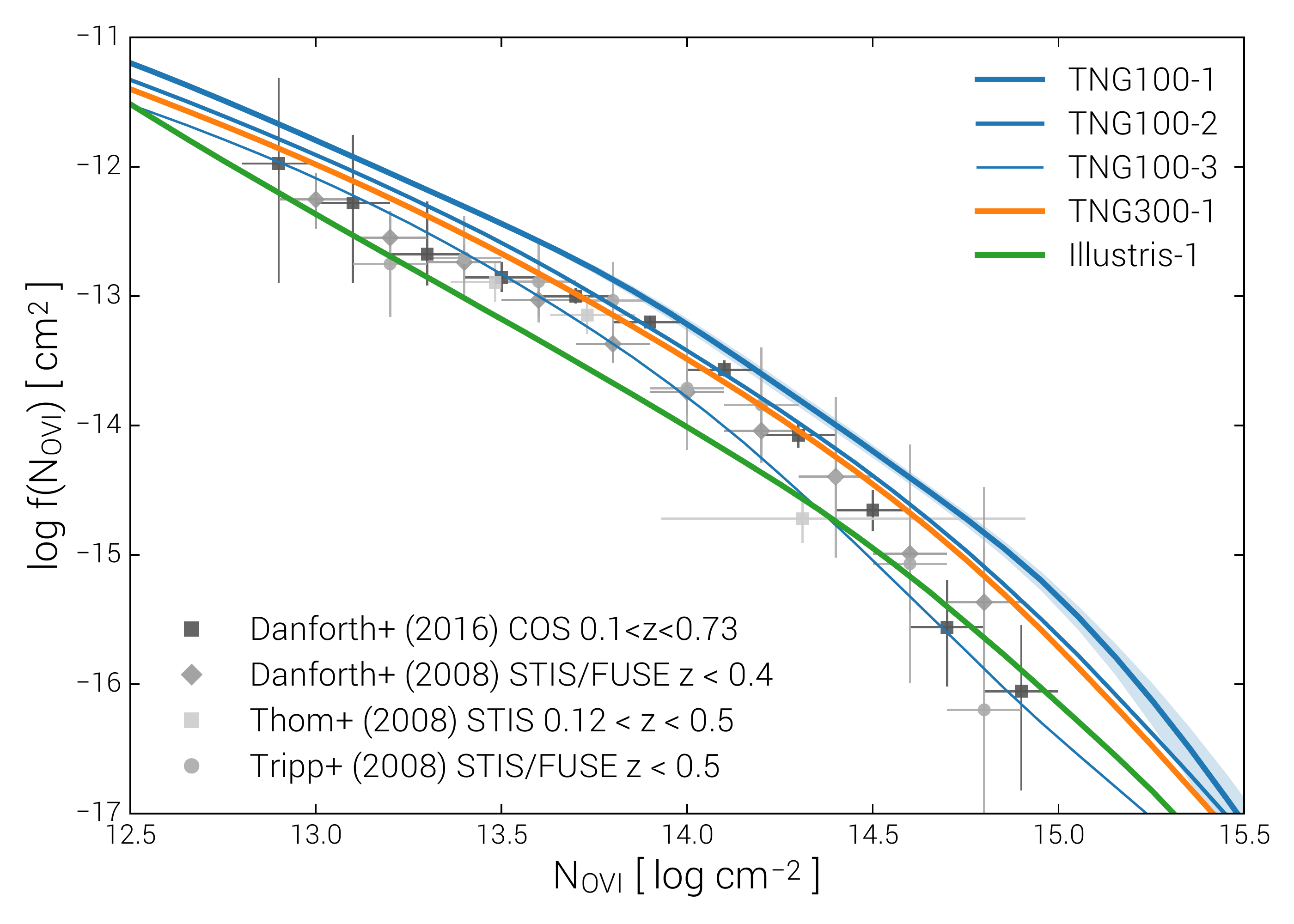}
\caption{ The column density distribution function (CDDF) of \ovi at low redshift, comparing the simulations at a fixed redshift of $z=0.2$ to observed datasets spanning roughly $0.1 < z < 0.7$ \protect\citep{danforth08,thom08,tripp08,danforth16}. There is reasonable agreement with the TNG simulations across the full column density range, possibly with an excess at the highest columns. At the resolution of TNG300 (equivalent to TNG100-2) the TNG model shows the best statistical agreement with the most recent dataset, and marked improvement over the original Illustris simulation. Note that the suffixes -1, -2, and -3 denote decreasing resolution by factors of eight in mass, from the highest (i.e. TNG100-1) to the lowest (i.e. TNG100-3).
 \label{fig_cddf}}
\end{figure*}

\begin{figure*}
\centering
\includegraphics[angle=0,width=2.3in]{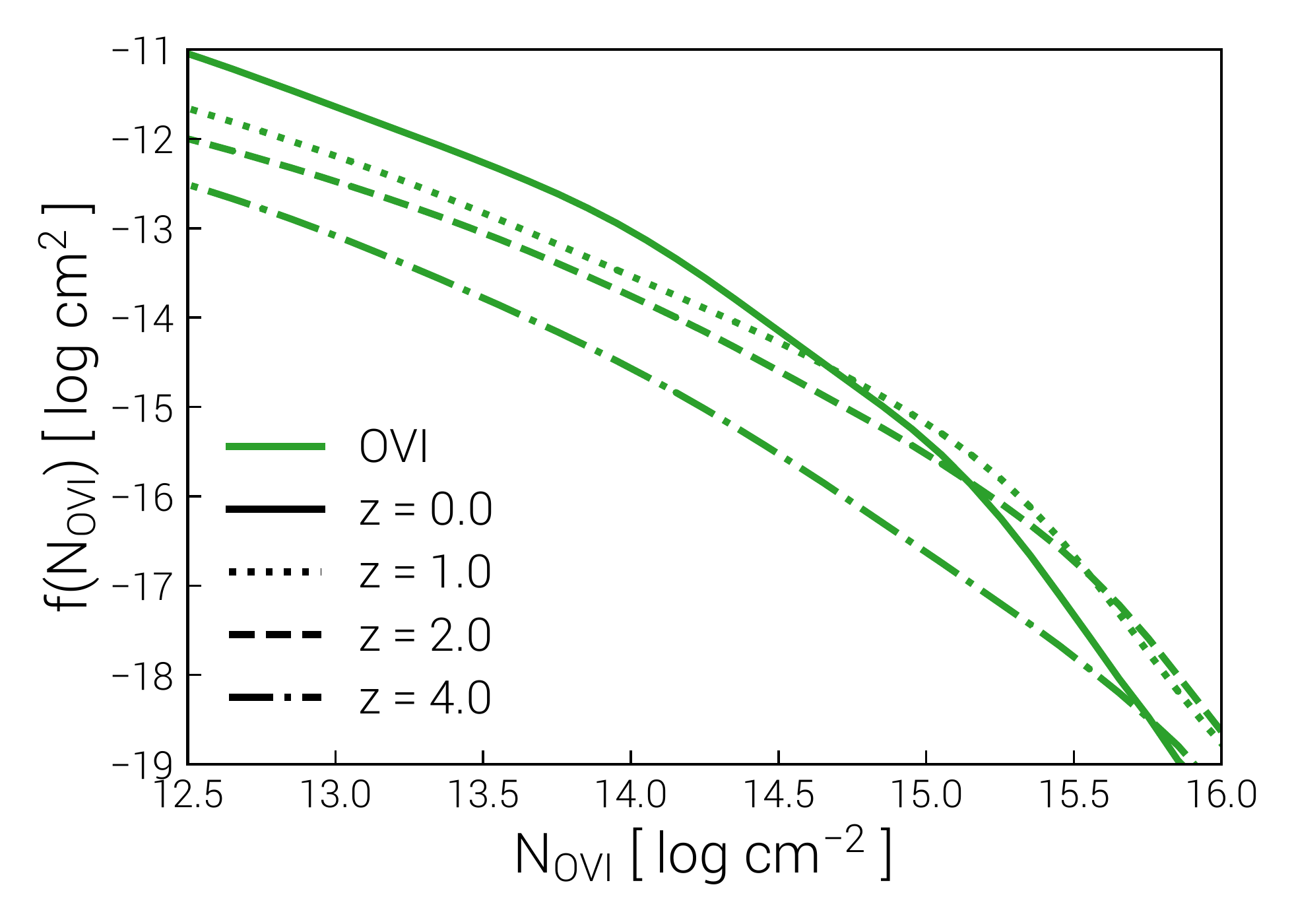}
\includegraphics[angle=0,width=2.3in]{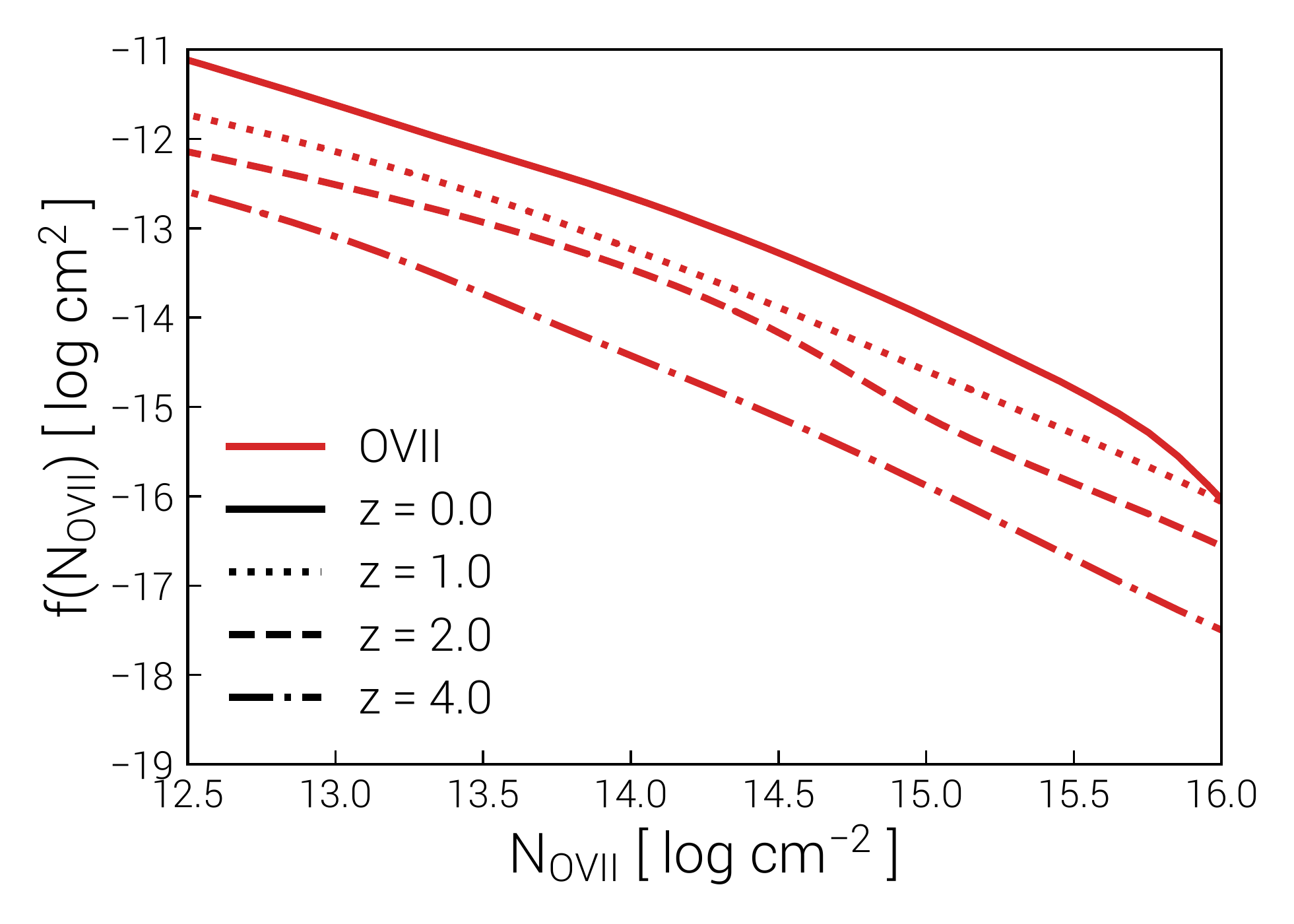}
\includegraphics[angle=0,width=2.3in]{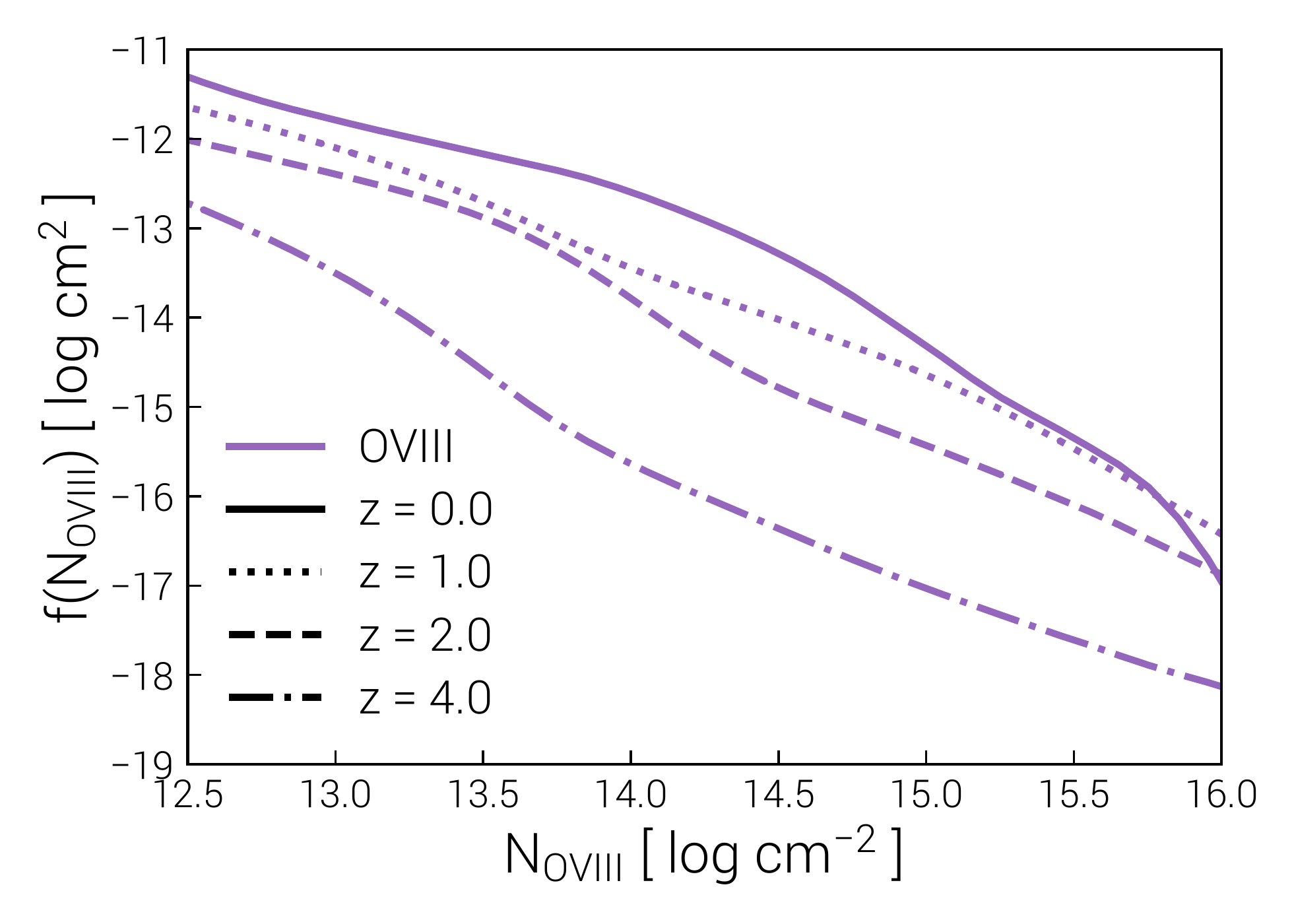}
\caption{ The redshift evolution of the \ovi (left), \ovii (center), and \oviii (right) CDDFs for TNG100 from $z=0$ back to $z=4$. All three ionization states show complex behavior beyond a simple powerlaw in $N_{\rm oxygen}$. In the case of \ovi, for $N_{\rm OVI} < 10^{15}$ cm$^{-2}$ $f(N)$ increases monotonically with time as progressively more oxygen is produced and distributed. The incidence of the highest column densities is fixed prior to redshift zero. The overall redshift evolution of the \ovii and \oviii CDDFs is similar, although there are fewer low columns, and significantly more high columns, in comparison to \ovi. 
 \label{fig_cddf_z}}
\end{figure*}

The occurrence of ionized oxygen in the simulation is a convolution of this ionization fraction with the occupation of gas-phase oxygen in the appropriate density, temperature, and (to a lesser degree) metallicity space. The bottom panels of Figure \ref{fig_ion_states} therefore show the phase-space diagram of the global TNG100 cosmological volume at $z=0$, except weighted in each case by the ion mass in \ovi, \ovii, or \oviii, respectively. The three contour levels on each panel enclose the $(n_{\rm H},T)$ regions where the relative abundances in ion mass are 0.1, 0.01, and 0.001, respectively. The three ionization states of oxygen occupy different, although overlapping, density and temperature regimes. Peak ion mass occurs roughly at $-6.0 < \log(n_{\rm H} \,[\rm{cm}^{-3}]) < -4.0$ and $5.5 < \log(T \,[\rm{K}]) < 6.5$, shifting to lower densities and higher temperatures with increasing ionization. The bulk of \ovi is produced in $10^5 - 10^6$ K gas while \oviii emerges at higher temperatures of $10^6 -10^7$ K. There is little \ovi in the region of T $> 10^6$ K and $n_{\rm H} > 10^{-4}$ cm$^{-3}$. The high density branch of \ovi at $T \simeq 10^{5.3}$ K largely disappears for \ovii and is replaced with the diagonal extension towards the upper right of the panel for \oviii. All three ions extend into the lower-left diagonal branch of the low density WHIM, $n_{\rm H} \le 10^{-7}$ cm$^{-3}$, where the typical gas temperature drops below 10,000 K. However, in comparison to higher densities and temperatures, the relative paucity of highly ionized oxygen in this regime indicates that most of these ions arise in the virialized gas of collapsed structures.

\subsection{Column density distributions of OVI, OVII, and OVIII} \label{subsec_cddfs}

To quantify the relative frequency of different metal column densities on large scales we employ the column density distribution function (CDDF). This 1-point statistic measures the relative occurrence of different columns; observationally, the rate of incidence of different absorbing systems with a given column density, typically in one or more high-resolution quasar sightlines. We compute the CDDFs of \ovi, \ovii, and \oviii from the simulations as described in Section~\ref{sec_cddf}, and Figures \ref{fig_cddf} and \ref{fig_cddf_z} show the result. First, we compare to the observed \ovi CDDF at low-redshift. This is now a well measured quantity, and therefore a first constraint on the accuracy of the \ovi content of the simulations. The CDDF is computed at a fixed simulation redshift of $z=0.2$ to compare to observed datasets spanning roughly $0.1 < z < 0.7$. The corresponding column density values extend from the IGM regime ($N_{\rm OVI} \la 10^{13}$ cm$^{-2}$) to the dense centers of massive halos ($N_{\rm OVI} \ga 10^{15}$ cm$^{-2}$). For instance, the probability of intersecting a low $10^{13}$ cm$^{-2}$ column is more than four orders of magnitude higher than for a $\sim 10^{15}$ cm$^{-2}$ column (for a bin size constant in linear space, or by two orders of magnitude for a $N_{\rm OVI}$ bin size constant in logarithmic space).

In this calculation the \ovi mass from every gas cell in the simulation within a projection depth of $\simeq$ 12.3 pMpc is included. Note that the low column end of the CDDF in particular is sensitive to this depth, and we have here taken a reasonable value for the observational comparison, in order to cover the typical allowed absorber velocity range \citep[see Methods \ref{sec_cddf} and][]{danforth16}. This is only an approximation, for several reasons. Namely, observational columns are determined by absorption profile fitting to individual components or systems or components \citep{tripp08}. By gridding projected ion masses we measure column densities in a different way, such that nearby absorption components are grouped together. A more faithful comparison to these datasets would involve profile fitting in synthetic absorption spectra. However, this would require a fully automated procedure for the detection and fitting of absorption components \citep[e.g.][]{dave97} in both the real and mock spectra. This is not typically done nor readily available, and we do not undertake this additional complexity here.

The uncertainty band around the TNG100-1 line encapsulates the degree of variation of the same calculation repeated with minor methodological changes. Specifically, the size of the griding pixels is varied by a factor of two, in both directions. In addition, the oxygen content tracked in simulated gas cells is ignored and instead solar abundance patterns are assumed. These changes only affect the high column end of the CDDF, and to a negligible degree.

We include comparison to observations from \cite{danforth08}, \cite{thom08}, \cite{tripp08}, and \cite{danforth16} which all use UV quasar absorption spectra from either STIS or COS and provide an essentially random sampling of sightlines through the low redshift IGM. As the most recent dataset with the best statistics, we focus particularly on \cite{danforth16}. In general there is reasonable agreement with the TNG volumes across two dex in column density, with the simulations commonly showing an apparent excess for the highest column absorbers. This is however a difficult observational regime, where line saturation and line blending result in large observational uncertainties at $N_{\rm OVI} \ga 10^{14.5}$ cm$^{-2}$ \citep[see discussion in][]{rahmati16}. Nonetheless, this broad consistency implies that the simulations produce roughly the right amount of \ovi by redshift zero \textit{and} distribute it more or less correctly across different columns.

In Figure \ref{fig_cddf} we include five different simulations: the TNG100 volume at progressively decreasing resolution levels -1, -2, and -3, as well as the large TNG300-1 simulation (equivalent in resolution to TNG100-2), and the original Illustris-1 simulation (equivalent in resolution to TNG100-1). First, we see that the TNG300 line is indistinguishable from its matched resolution TNG100-2 counterpart -- the large volume is clearly not needed for the CDDF statistic. Second, we see that the \ovi CDDF is not invariant to resolution, decreasing at fixed column for lower resolution runs. This is not unexpected, as even the total metal mass or total ion mass in the entire simulation volume will scale with the total stellar mass formed, which itself is not perfectly converged in the TNG model \citep[see discussions in][]{pillepich17a,pillepich18}. Differences in the amount of redistribution, at fixed total mass, due to feedback from BHs, galactic winds, and/or hydrodynamical mixing will also apply.

Although the differences between the TNG100-2 and TNG100-1 curves are of the same order as the observational errors, there is a hint that this statistic is converging towards a result which would be slightly above the uppermost curve. Variation with resolution is sub-dominant to variation of physical model -- the original Illustris line is significantly lower than TNG100, by roughly one order of magnitude at most columns. This leads to a tension of the Illustris simulation when compared against the observed \ovi CDDF \citep[explored in][]{suresh15} which is alleviated in the TNG simulations. Taken at face value, the formal reduced $\chi^2$ values of the \cite{danforth16} \ovi CDDF compared to TNG100-1, TNG300-1, and Illustris from Figure \ref{fig_cddf} are 6.9, 1.5, and 31.1, respectively; the remarkable statistical goodness of fit for TNG300-1/TNG100-2 being a coincidental alignment of the resolution dependent curve with the observational constraints. As neither the total amount nor spatial distribution of oxygen, much less \ovi, was used in the development of TNG, the comparison demonstrates how this particular statistic provides an orthogonal constraint (or check) on our galaxy formation models.

In Figure \ref{fig_cddf_z} we show the redshift evolution of the \ovi, \ovii, and \oviii CDDFs for TNG100 from $z=0$ back to $z=4$. Broadly, all three are increasingly built up over this redshift range with $f(N)$ increasing towards redshift zero. For \ovi (green curves), at $N_{\rm OVI} > 10^{15}$ cm$^{-2}$ the CDDF is established since $z=2$, although not as early as $z=4$. The amplitude of the CDDF at the highest columns even decreases slightly from $z=1$ to the present time, likely due to the late-time redistribution of the densest halo gas by baryonic feedback processes. With respect to \ovi, the higher ionization states of oxygen are more commonly found at high column densities above $\simeq 10^{15}$ cm$^{-2}$. As a result, $f(N)$ follows more closely a powerlaw over a large dynamic range in column -- particularly so for \ovii, and at higher redshifts.

\begin{figure}
\centering
\includegraphics[angle=0,width=3.3in]{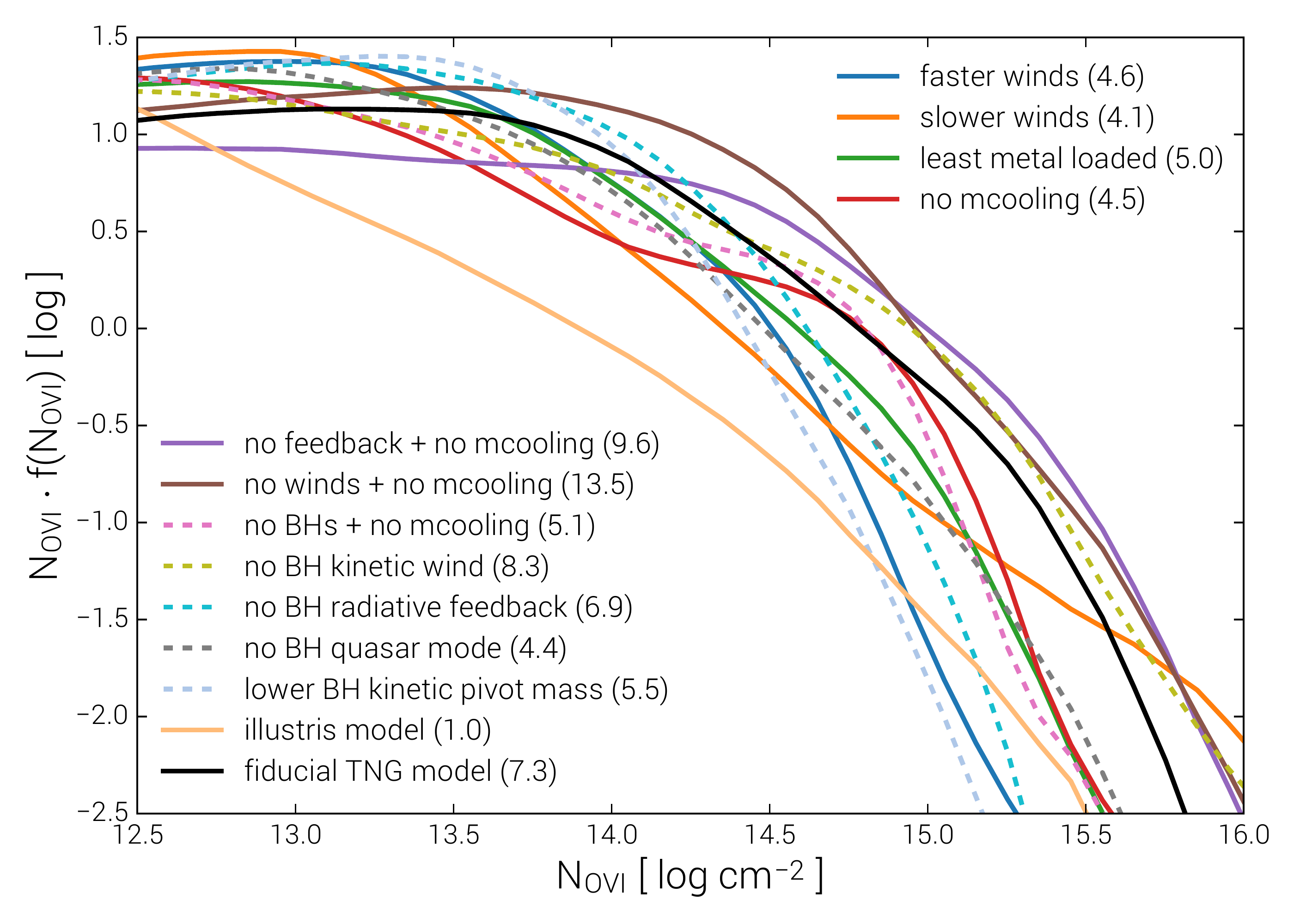}
\caption{ Impact of physical variations in the TNG galaxy formation model on the \ovi CDDF, shown in its first moment $N f(N)$ at $z=0$. Results are given for a series of test simulations in 37 Mpc side-length volumes at the nominal resolution of TNG100-1. The result for the fiducial TNG model is shown in black; colored lines each show a different simulation with one model aspect or parameter changed, except for the old Illustris model result (light orange). In this panel we show the most important variations which lead to significant differences in the CDDF. The number in parentheses for each model gives the $z=0$ global value of $10^7 \times (\Omega_{\rm OVI} = \rho_{\rm OVI} / \rho_{\rm crit,0})$.
 \label{fig_cddf_variants}}
\end{figure}

\subsection{Sensitivity of the CDDF to physical model variations}

The comparison of the Illustris and TNG \ovi CDDFs demonstrated that ingredients of the physical model can have a significant impact. In order to determine the model parameters and/or components which play the largest role, we turn to a large suite test simulations, each of which varies one parameter value or aspect of the model. These variations therefore provide perturbations about our fiducial TNG galaxy formation model. A large number of variants, covering most possible model aspects of interest allows us to unambiguously select those which generate the largest variations in the simulated column density distribution functions.\footnote{Each variant simulation is run at roughly the full resolution of TNG100-1 (6.4 versus 6.2 in log\msun baryon mass), except in a smaller volume with a side-length of $\simeq$ 37 cMpc, requiring therefore $2 \times 512^3$ resolution elements. The exact configuration of these test simulations is the same as, and described fully in, \cite{pillepich17a}. As already demonstrated, the smaller volumes do not affect the CDDF measurement.}

Figure \ref{fig_cddf_variants} shows the result, given in terms of $N f(N)$, the first moment of the $z=0$ \ovi CDDF, together with the global $\Omega_{\rm OVI}$. With respect to the fiducial model realization (in black) even the most extreme changes scatter about the base TNG model, although with significant differences in detail. The most important aspect here of the galactic scale winds is their injection velocity \citep[see Figure B2 of][the impact on late time stellar mass production also being significant]{pillepich17a}. Faster winds (dark blue) which are therefore less mass-loaded steepen the CDDF, strongly suppressing large columns. Conversely, winds slower by a factor of two (orange) flatten the CDDF more than any other variation, producing both the most low and high columns at the expense of intermediate values. In contrast, most changes qualitatively follow that of faster winds, falling below fiducial at $N_{\rm OVI} \ga 10^{14.5}$ cm$^{-2}$; these include less metal rich/loaded winds (green), no metal-line cooling of enriched gas (red), no blackholes nor their energetic feedback, with metal cooling also removed (pink), no BH quasar mode feedback (light gray), no BH radiative feedback (pink), and a lower BH mass for the onset of kinetic-mode FB \citep[light blue; the denominator of Equation 5 in][decreased by a factor of four]{weinberger17}. These changes are all consistent with \ovi mass more broadly distributed in space, thereby increasing the incidence of low columns at the expense of highly localized regions of enrichment.

On the other hand, the three remaining variations act in the other direction, lying above the fiducial model at high columns, indicative of more spatially concentrated \ovi: no galactic winds (i.e. no stellar feedback; brown), no feedback whatsoever -- neither winds nor blackholes (purple), and no kinetic-mode/low-state BH feedback (gold). In these cases, the exact shape of the curves is the result of an overproduction of oxygen and metals in general (with e.g. $\Omega_{\rm OVI} = 9.6 \times 10^{-7}$ in the 'no feedback + no mcooling' model, which is higher than the TNG fiducial case) as well as a more limited spreading of metals throughout the CGM and IGM due to the absence of a physical redistribution mechanism such as winds from either supernovae or black holes. Even in these cases, however, metals can extend far beyond their production sites due to dynamical effects such as tidal stripping, or due to hydrodynamical mixing and/or diffusion.

The Illustris model line (light orange) is an obvious outlier to all the TNG model variations, being too low in normalization and too steep as a function of column. It is qualitatively most similar to the slower winds variant, consistent with the fact that the wind injection velocities are indeed overall much faster in TNG than in Illustris \citep[see Figure 6 of][]{pillepich17a}. Wind velocity has a strong influence on the spatial extent of metal enrichment \citep[as well as on the ISM metallicity;][]{torrey14} and we posit that the `too slow' galactic outflows of Illustris were the primary reason for its tensions with \ovi observations, although the kinetic to thermal balance at injection also strongly modifies the impact of the winds \citep[see also the discussion in Appendix B of][]{suresh15}.

In the regime where the observations are constraining, \mbox{$N \la 10^{14.5}$ cm$^{-2}$}, the variation of the \ovi CDDF between these test simulations is roughly 0.5 dex at most. However, we note that these are not particularly \textit{reasonable} permutations of the base TNG model -- they already represent extreme cases, ruled out by various 1-point statistics related primarily to the stellar content of halos \citep{pillepich17a}. Weaker variations which also fit the $z=0$ galaxy stellar mass function, for instance, show changes in the \ovi CDDF at or below the level of the current observational errorbars. In this regime, systematic differences in analysis methodology and/or the resolution dependence of the simulation results may dominate. Therefore, the current utility of this comparison could be seen largely as a sanity check and not as a model discriminant. However, the \ovi CDDF does probe a physical regime (i) orthogonal to the stellar constraints and (ii) distant from the calibration regimes of the base TNG model. Therefore, agreement gives confidence that other aspects of the model are reasonable, such as the degree to which metals are ejected from halos and enrich the low-density IGM, validating the pursuit of related studies.

Observational constraints on the $z=0$ CDDFs of \ovii or \oviii would require high resolution x-ray absorption spectroscopy, e.g. X-IFU on Athena \citep{barret13,kaastra13} or the grating spectrometer on Lynx \citep{weisskopf15} or Arcus \citep{smith16}. At the same time, a handle on the redshift evolution implies the same capabilities at even lower fluxes. Simultaneous statistical constraints on these three adjacent oxygen ions would provide a powerful and constraining benchmark for hydrodynamical simulations such as TNG which now provide explicit predictions for these statistics.

\subsection{Spatial clustering of oxygen ions}

\begin{figure}
\centering
\includegraphics[angle=0,width=3.3in]{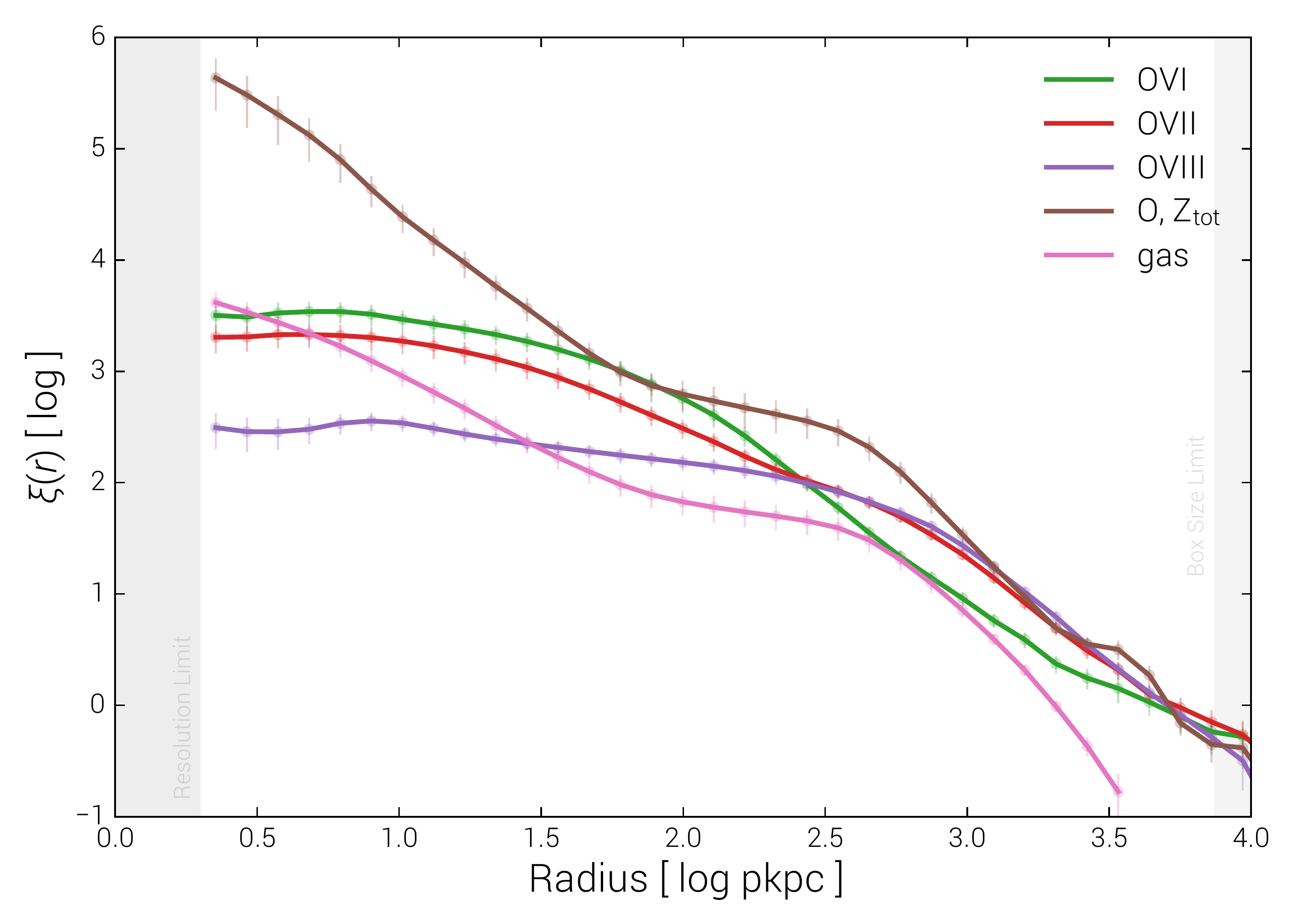}
\caption{ Two point auto-correlation function of several gas mass components in TNG100: \ovi, \ovii, \oviii, total oxygen (or identically total metals), and total gas mass. We give the 3D real-space $\xi(r)$ to characterize the spatial clustering of baryonic mass in these highly ionized states of oxygen, in comparison to the clustering of all heavy elements, and global gas-phase baryonic mass. Note that $\xi_{\rm O}$ and $\xi_{\rm Z}$ are indistinguishable at all separations. In contrast to all gas mass, or all the gas-phase mass in oxygen, the clustering of all three ionized states of oxygen does not significantly increase towards smaller scales below the size scale of a typical host halo.
 \label{fig_tpcf}}
\end{figure}

\begin{figure*}
\centering
\includegraphics[angle=0,width=3.45in]{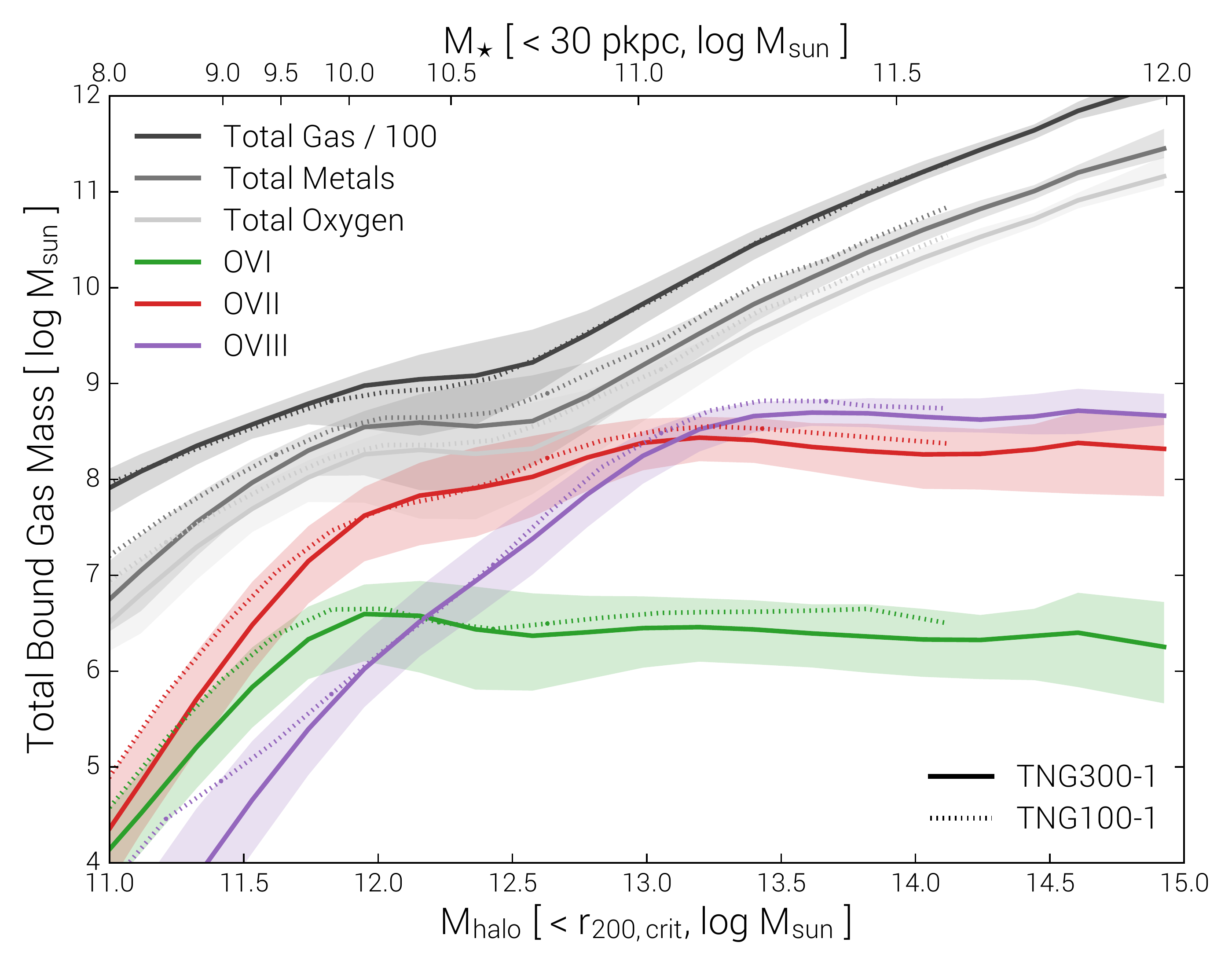}
\includegraphics[angle=0,width=3.45in]{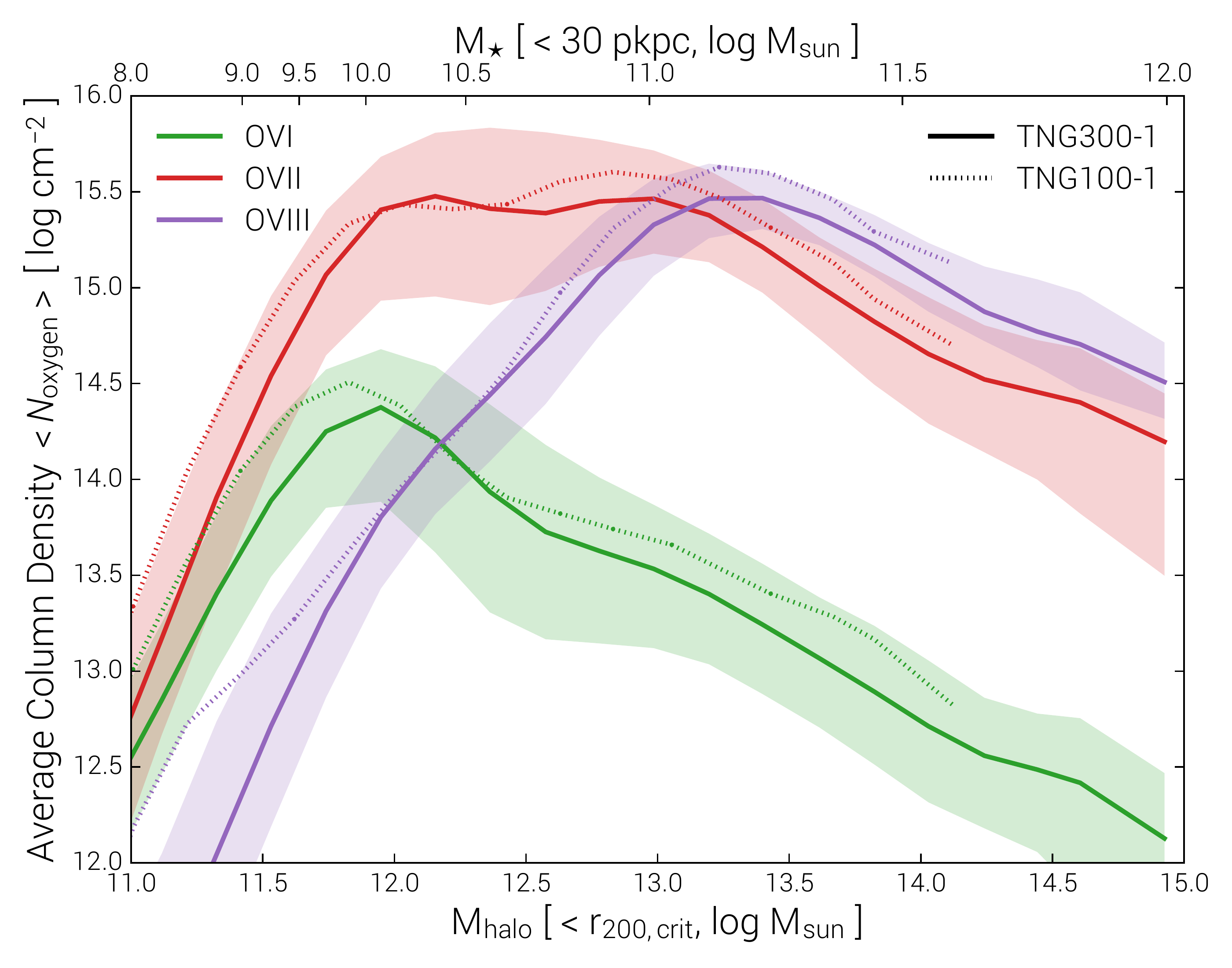}
\caption{ \textbf{Left panel.} Total gravitationally bound mass in different phases: total gas, total metals, total oxygen, and the OVI, OVII, and OVIII ions of oxygen as a function of either halo mass (bottom axis) or galaxy stellar mass (top axis) at $z=0$. We show TNG300 (solid) and TNG100 (dashed). The total gas mass line (dark gray) is multiplied by 0.01 for visual clarity. Below $M_{\rm halo} \la 10^{12}$\msun the total ion masses increase rapidly as a function of halo mass, until an ion-dependent threshold is reached and the mass plateaus to a roughly constant, maximum value. \textbf{Right panel.} The average column density of oxygen ions as a function of halo/stellar mass at $z=0$, computed geometrically as $<$$N_{\rm ion}$$>$ $= M_{\rm ion} / (\pi r_{\rm vir}^2)$. As a result of distributing the mass over a larger projected area, the peak column density for each ion occurs at a characteristic stellar/halo mass. This mass scale for the maximally efficient `production' of highly ionized oxygen by circumgalactic gas halos increases with ionization state from $M_{\rm halo} \simeq 10^{12}$\msun (for \ovi) to $\simeq 10^{12.5}$\msun and $\simeq 10^{13.5}$\msun for \ovii and \oviii, respectively.
 \label{fig_ions_vs_mass}}
\end{figure*}

Beyond simple frequency statistics in terms of observationally oriented column densities, we would like to assess how \ovi, \ovii, and \oviii ions are distributed in space, how they occupy different volumes, and how they cluster together within and exterior to overdensities. To characterize spatial clustering we therefore compute the two point real-space auto-correlation function $\xi(r)$ in 3D, following the procedure described in Section \ref{sec_corrfuncs}. Figure \ref{fig_tpcf} shows this clustering signal for several different components of gas-phase mass: principally, \ovi, \ovii, and \oviii, which are compared against the clustering of total oxygen, as well as total gas mass. The result for total metal mass is omitted as it is indistinguishable from that of total oxygen mass alone. The total gas mass curve (pink) is essentially the redshift zero $\xi_{\rm gas}(r)$ curve from \cite{springel18} -- shown in Figure 1 of that work -- which we decompose here. 

On halos scales ($r \sim 100$ kpc) we see that oxygen is progressively more clustered with \textit{decreasing} ionization state, \oviii having the lowest amplitude signal. Although the $\xi_{\rm O*}(r)$ for all three oxygen ions is maximal at the smallest scales, declining towards larger separations, it is not strongly peaked as $r \rightarrow 0$ and is instead rather constant on scales less than the virial radius of the typical host halo. This behavior is noticeably different from either total metal mass or total gas mass, whose clustering on kpc scales is much stronger (i.e. 1 dex or more) than on halo scales. We tentatively interpret this as a signature of highly ionized oxygen tracing a halo volume filling phase with no substantial smaller scale clumping or density substructure. However, $\xi(r)$ effectively measures a combination of a smooth background density profile and a non-smooth component from substructure, if present, in analogy to the 1-halo term of $\xi_{\rm DM}(r)$ measuring the superposition of gravitationally bound structures on top of a background NFW $\rho(r)$. Therefore, the relatively flat 1-halo $\xi_{\rm O*}(r)$ may also predominantly indicate an ion density profile which is shallower with radius than that of either total oxygen, total metal, or total gas mass.

The volume filling and relatively homogeneous \ovi stands in stark contrast to recent observational hints for small spatial scale origins for the cool gas giving rise to lower ion and HI absorption \citep[e.g.][]{crighton15,battaia15,chen17,stern16}. We postpone a study of cooler ions such as MgII and the `clumpyness' of the TNG CGM for future work. 

\section{The Abundance and Spatial Distribution of Ionized Oxygen: Galactic Halos} \label{sec_abundance_halos}

We have so far focused on the global distribution of \ovi, \ovii, and \oviii, with analyses including all gas-phase baryons throughout the entire cosmological volumes of TNG100 and TNG300, agnostic to any connection with collapsed structures. To understand how highly ionized oxygen arises within the circumgalactic medium of dark matter halos, we now move into a halo-centric frame.

\subsection{Gas phases as a function of halo or stellar mass}

Figure \ref{fig_ions_vs_mass} begins with a measurement of the total mass in each of these three ions (left panel), as a function of halo or stellar mass (bottom and top axes, respectively, the latter derived using the median stellar mass to halo mass (SMHM) relation of TNG300). Instead of summing the different gas phase masses within some arbitrary radial distance, we take the total gravitationally bound mass, as a useful physically motivated definition. For comparison, the total gas mass, metal mass, and gas-phase oxygen mass are also included (gray lines). No distinction is made between gas-phase oxygen in the ISM versus CGM, the former being negligible in relation to the latter ($\sim$10$^{-6}$ by mass if we take the ISM as star-forming gas). The resulting measurements are shown at $z=0$ over four orders of magnitude in halo mass; three orders of magnitude in galaxy stellar mass. The total gas mass is always of order $\sim$ $\Omega_{\rm b}/\Omega_{\rm m} \simeq 16\%$ of the total halo mass, modulated by baryonic effects \citep[see][]{pillepich17a}, of which the total metal mass makes up roughly $\sim$ 0.1\%. The resulting `global halo' gas metallicity ranges from $\sim$0.25 $Z_{\odot}$ to $\sim$0.4\,$Z_{\odot}$, the maximum value occurring near $M_{\rm halo} \sim 10^{12}$\msun. Gas-phase oxygen is a large, nearly constant fraction of these metals; its decomposition into different ions is however more complex.

Below $M_{\rm halo} \la 10^{12}$\msun the total gravitationally bound mass in \ovi, \ovii, and \oviii all steeply increase as a function of halo mass, until a threshold is reached and the mass in a given ion plateaus to a roughly constant, maximum value. Above this point, most halos contain roughly $10^{6.5}$\msun of \ovi mass, with a halo to halo scatter of about 1 dex (indicated by the colored bands). Massive halos can contain significantly more \ovii and \oviii mass, up to $\sim 10^{8.5}$\msun and $\sim 10^{9.0}$\msun, respectively, with higher ions becoming more dominant with increasing halo mass. The constancy of the maximal bound ionic mass with halo mass is striking. We find no significant trend of the scatter with mass, although the scatter of \oviii is about half as large as for its lower ionization counterparts.

These ion masses are also recast in terms of an average column density (right panel), assuming each was uniformly distributed over the area of the projected virial sphere. That is, $<$$N_{\rm ion}$$>$ $= M_{\rm ion} / (\pi r_{\rm vir}^2)$, the resulting values not necessarily reflecting the actual mean column with a given sampling of impact parameter. In contrast to the monotonic behavior of total ion mass, the geometrical average column densities $<$$N_{\rm ion}$$>$ each peak at a characteristic halo/galaxy stellar mass, as a result of distribution over increasingly larger projected areas. For \ovi this occurs just below $M_\star \simeq 10^{10.5}$\msun and $M_{\rm halo} \simeq 10^{12}$\msun. In the case of \ovii this characteristic mass shifts up to $M_{\rm halo} \simeq 10^{12.5}$\msun and for \oviii further to $M_{\rm halo} \simeq 10^{13.5}$\msun. Somewhat in analogy to the peak of the stellar mass to halo mass relation, these halos represent the mass scale at which e.g. the `\ovi production' is maximally efficient. Here, the median halo has an \textit{average} \ovi column density of $\simeq 10^{14.5}$ cm$^{-2}$ across the projected surface of its virial sphere. The average columns of \ovii and \oviii can be an order of magnitude higher.

\subsection{Radial oxygen ion profiles in the CGM and beyond}

\begin{figure*}
\centering
\includegraphics[angle=0,width=3.4in]{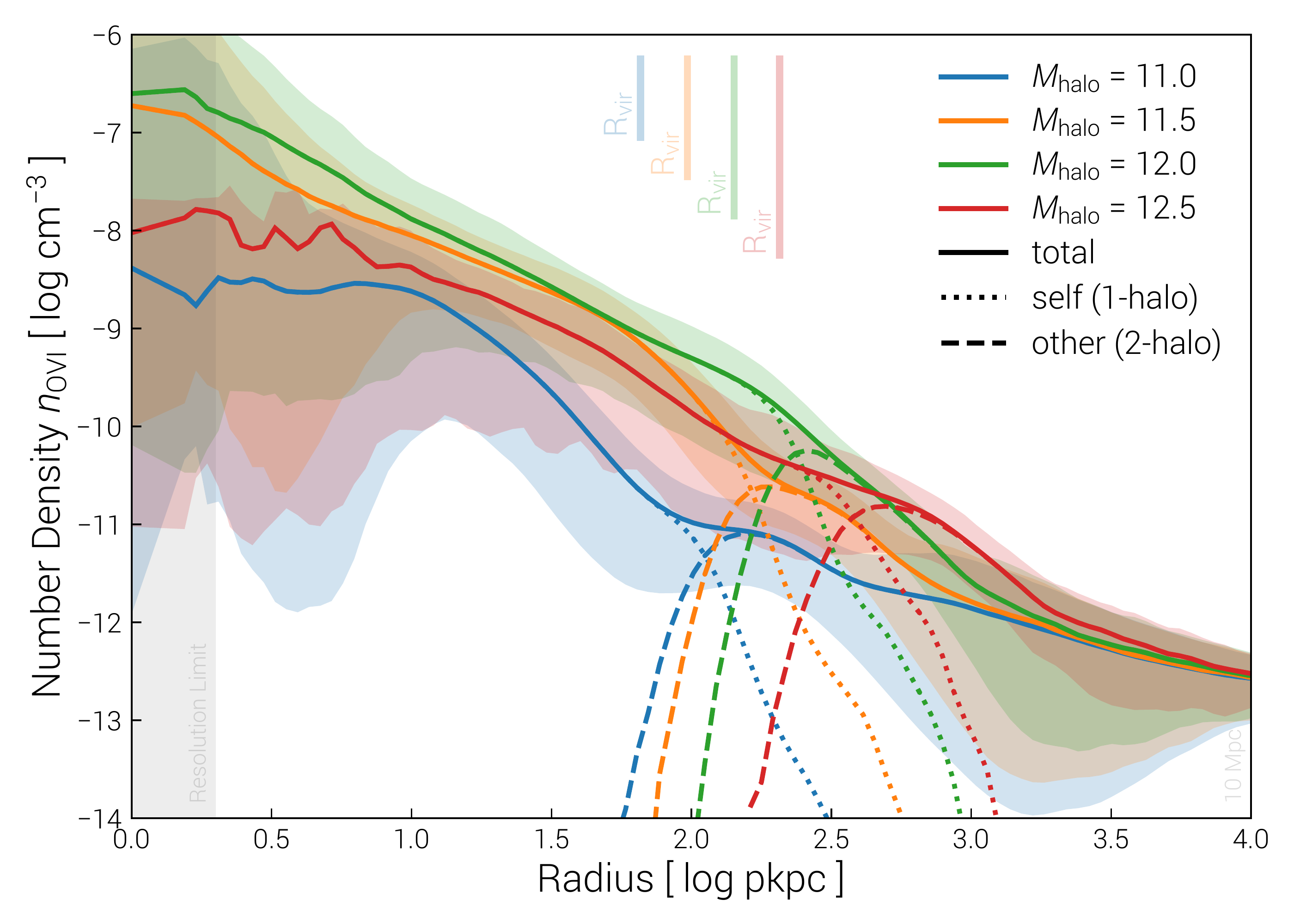}
\includegraphics[angle=0,width=3.4in]{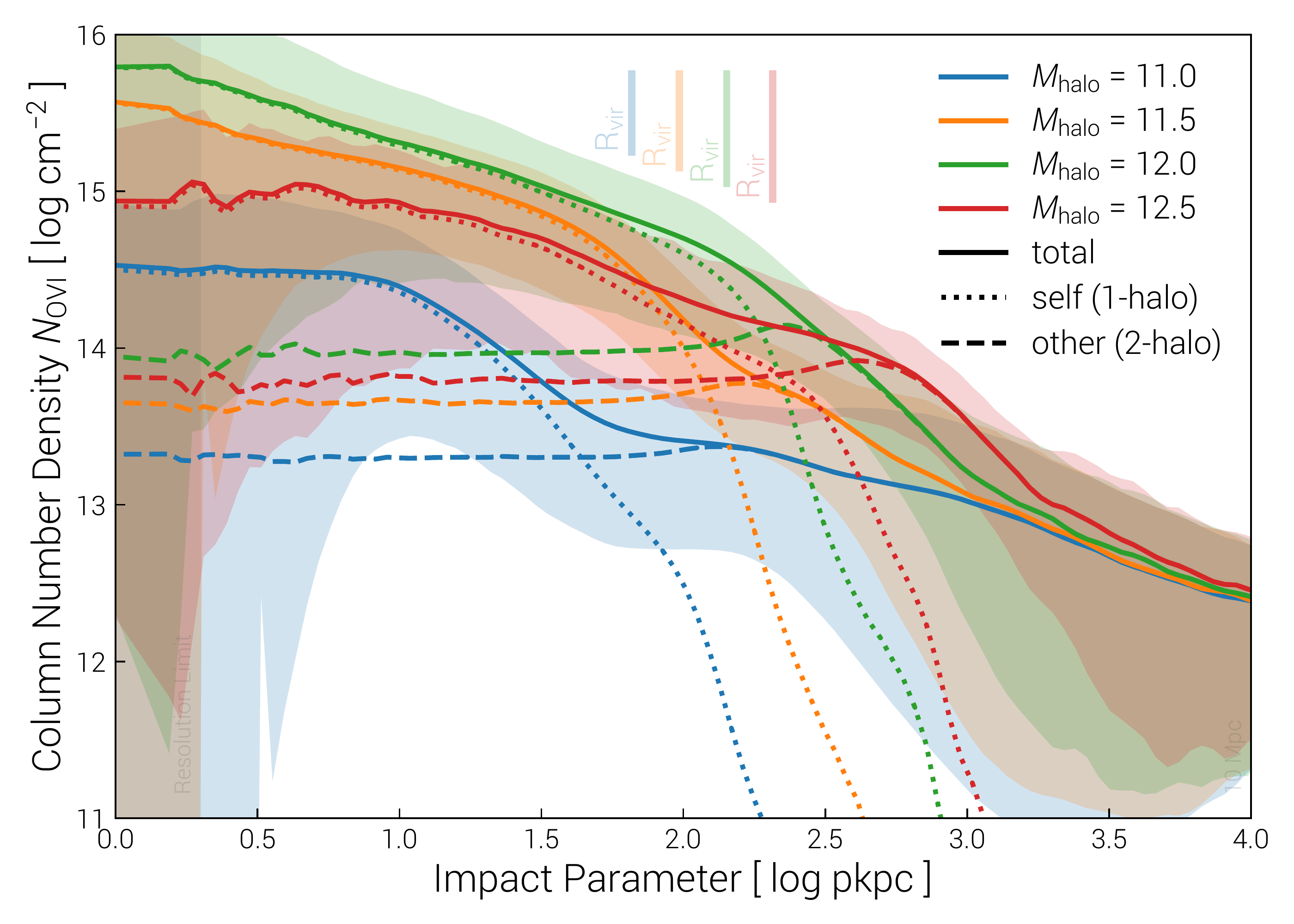}
\caption{ Median 3D radial profiles of the abundance of \ovi around $z=0$ halos. The total \ovi content in the simulation (solid lines) is decomposed into the `self-halo' (e.g. 1-halo) and `other-halo' (e.g. 2-halo) terms. The former includes all gas cells within the stacked halos themselves, while the latter includes the contribution from gas in all other halos as well as from the diffuse IGM. We show profiles stacked in four halo mass bins each with width $\pm 0.1$ dex and centered on $\log(M_{\rm halo}/\rm{M}_\odot) \in \{11.0,11.5,12.0,12.5\}$. In TNG100 these four bins include 5977, 2209, 888, and 245 stacked halos, respectively.
\textbf{Left panel.} Three dimensional number densities, spherically averaged.
\textbf{Right panel.} Two dimensional column densities, projected through a depth of 3 pMpc (corresponding to a total $\Delta v \simeq$ 200 km/s), as in our comparison to COS-Halos below.
 \label{fig_radprofiles}}
\end{figure*}

Figure \ref{fig_radprofiles} presents the radial density structure of \ovi, in particular, for halos of different masses at $z=0$. We first measure median stacked 3D radial profiles in terms of physical number density (left panel). Profiles are stacked in four halo mass bins centered on $\log(M_{\rm halo}/\rm{M}_\odot) \in \{11.0,11.5,12.0,12.5\}$, each with a width of $\pm 0.1$ dex. The measurement in each case is extended from the deep interior of halos, $\sim$ 1 kpc, to large cosmological scales, $\sim$ 10 Mpc, where the signal approaches a constant floor indicative of the mean density (i.e. $\Omega_{\rm OVI}$) of \ovi in the universe. As this is far beyond the virial radii, as well as the turnaround radii, of the considered halos, we decompose the contribution from the `self' (i.e. `1-halo') and `other' (i.e. `2-halo') terms. The former includes all gas cells within the parent friends of friends (FoF) halos of the stacked systems, while the latter includes gas cells in all other resolved FoF halos in the simulation volume, plus the diffuse contribution from gas outside all halos, which we would call the warm/hot intergalactic medium (IGM or WHIM). Note that the exact boundary between these two terms is largely definitional and that our 2-halo term will include contributions from gas at a wide range of distances from the halo itself, starting beyond its FoF boundary.

At a fixed radius within the scale of an individual halo, the abundance of \ovi is a non-monotonic function of halo mass. Moving from halo masses of $\simeq 10^{11}$\msun, to $10^{11.5}$\msun, and again to $10^{12}$\msun, the amount of quintuplely ionized oxygen n$_{\rm OVI}$ increases each time. However, going to still more massive halos with $10^{12.5}$\msun the amount of \ovi then decreases, arriving back to a value which is similar to the lowest halo mass bin considered. This coincidental similarity of the number or column densities of \ovi within sub-L$^\star$ and small group sized halos arises as the former maintains a large fraction of their halo gas at temperatures below the collisional ionization band of Figure \ref{fig_ion_states} (horizontal feature, upper left panel) and at sufficiently low densities that \ovi arises predominantly from photoionization due to the background radiation field. For more massive halos with virial temperatures of order a million degrees or higher the situation is reversed, with most \ovi arising from collisional ionization \citep[see also][]{turner15}.

The halo-centered 3D profiles decline rapidly and smoothly with increasing radius. The $n(r)$ profiles are roughly powerlaws of constant slope from the virial radius inwards, slightly shallower than $r^{-2}$. The lowest and highest halo mass bins, which have overall lower normalizations than the two intermediate mass bins, show a flattening within the inner 10-20\% of $r_{\rm vir}$. In all cases, the self-halo component of the density profile steepens rapidly at the virial radius and drops off as though at a sharp outer boundary \citep[although see][]{nelson16,vog18a}. This contribution equals the 2-halo term (i.e. the dotted and dashed lines cross) just past $r_{\rm vir}$, 50-100 kpc beyond for halos of $10^{12}$\msun. After a local bump in the $n(r)$ profiles due to the maxima of the 2-halo terms, they decline slowly to asymptotic values at cosmological distances.

The right panel of Figure \ref{fig_radprofiles} show the radial abundance of \ovi around the same stacked halo samples, now measured in terms of the median column density in 2D projection. The trends with halo mass and radius are similar, except that the decline of $N_{\rm OVI}(r)$ is much gentler with distance -- for $10^{12}$\msun halos, the column density only decreases by half a dex between 10 kpc and 100 kpc projected. Each profile still declines sharply near the mean virial radius of each mass bin. However, the contribution to $N_{\rm OVI}$ from secondary halos besides the central itself extends all the way to zero projected distance, as a result of the statistical presence of either background or foreground halos along the line of sight, as well as due to the diffuse IGM. For $r \rightarrow 0$ this is a roughly 1\% addition to the column, but in the outer halo [0.5-1.0]$r_{\rm vir}$ this can be a significant or even dominant contribution, depending on halo mass. The diffuse component contributes more to this 2-halo term than other resolved and gravitationally bound halos, by roughly 0.5 dex at large separations (\mbox{b $> 10^{2.5}$ kpc}) and up to 1 dex at small separations (b $< 100$ kpc).

At the rightmost extent of these panels the large, cosmological scales are physically unassociated with the central halos, and reassuringly converge to the same value irrespective of mass bin. The column density value at large spatial extent scales with the projection depth. We note that from \protect\cite{danforth16} the mean absorption expected from random sightlines based on their d$^2\mathcal{N}$/d$\rm{N}$dz analysis is $N_{\rm OVI} \simeq 10^{12.2}$ cm$^{-2}$, which is essentially an observational constraint on the lower allowed limit of the column density profiles at large radius, where we find $N_{\rm OVI} \simeq 10^{12.5}$ cm$^{-2}$ at $\sim$ 10 Mpc.

\subsection{Comparison to Observational Constraints}

Several observational datasets with a statistical sampling of the prevalence of \ovi in and around galactic halos exist. In particular, targeted and untargeted absorption line studies with background quasar sightlines passing through the CGM of intervening galaxies can measure the strength and incidence of \ovi absorption -- at low redshift using the COS instrument on the Hubble Space Telescope. To make a robust comparison to these datasets, we construct a tailored, `mock' survey consisting of a large simulated sample which is statistically consistent with a number of observed galaxy properties we aim to match \citep[similar in spirit to][]{oppenheimer16}. 

\subsubsection{Comparison to the COS-Halos survey}

\begin{figure*}
\centering
\includegraphics[angle=0,width=5.5in]{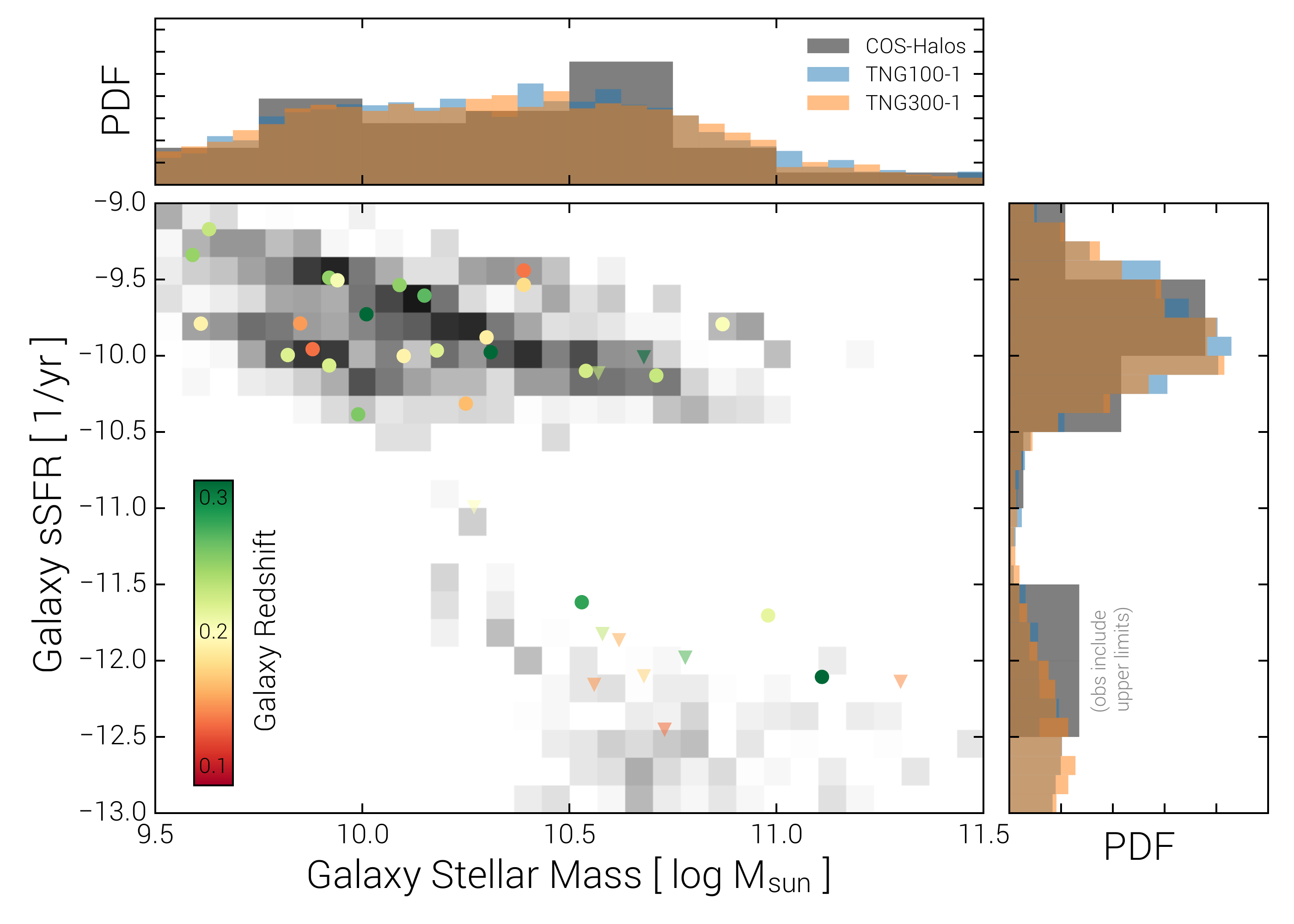}
\caption{ Comparison of the COS-Halos galaxy sample and the simulated galaxy sample constructed to match its characteristics (see text). \textbf{Main panel.} The galaxy sample in the sSFR-$M_\star$ plane. Observational detections are indicated by colored circles, and observational upper limits by downward pointed triangles. Color indicates the redshift of the associated galaxy, while the grayscale histogram in the background shows the distribution of the mock simulated sample from TNG100. \textbf{Top and right panels.} Marginalized histograms of the observed points, in gray, versus the mock samples from TNG100 and TNG300 in blue and orange, respectively. Note that limits are included in the observational histogram as is, and the prevalence of upper limits in sSFR at the high mass end results in the tail of simulated galaxies towards low star formation rates -- consistent with the sample constraints.
 \label{fig_coshalos_sample}}
\end{figure*}

First, we compare to the COS-Halos survey \citep{tumlinson11}, constructing the mock sample as follows. The observed stellar masses and star formation rates are taken from \cite{werk12} and \cite{werk13}, the latter being the Balmer emission line (H$\alpha$) derived values. Both are corrected from a Salpeter IMF, as assumed in that work, to a Chabrier IMF, as assumed herein. A constant uncertainty of 0.2 dex is assumed for all $M_\star$ values, while uncertainties on the SFRs are taken as given. The impact parameters, \ovi column densities, and uncertainties are taken from \cite{tumlinson11}, which also describes the isolation criterion: that COS-Halos galaxies are the most luminous with an impact parameter of $b < 300$ kpc of each quasar sightline at their redshift. We mimic this by selecting only central galaxies which have no more massive companion (measured in terms of $M_\star$ within the 30 pkpc aperture) within a 3D distance of 300 kpc.

To make the mock sample, each observed galaxy is first associated with the simulated snapshot closest to its redshift (these span from $z=0.14$ to $z=0.36$ and therefore coincide with 14 distinct simulated snapshots). For each of $N=100$ realizations we apply a random perturbation to the observed $M_\star$ and sSFR values, normally distributed and consistent with the uncertainties. After applying the isolation criteria, we then consider all galaxies which satisfy any one-sided constraints (i.e. upper and lower limits are enforced). A L1 norm (i.e. Manhattan distance metric) is computed in the space of the remaining bounded parameters (either $M_\star$ alone, or stellar mass and sSFR jointly) between the specific realized observed galaxy and all compatible simulated galaxies, and a match is selected as the system with the smallest `distance'. The COS-Halos sample we adopt has 37 points: the mock sample therefore 3700. 

In Figure \ref{fig_coshalos_sample} we compare the COS-Halos galaxy sample to the mock simulated galaxy sample created to match its characteristics. We consider the systems in the sSFR-$M_\star$ plane, where observational detections are indicated by colored circles, and upper limits by triangles. Behind, with the 2D histogram (grayscale) we show the mock simulated sample from TNG100. In addition, marginalized histograms are given along each axis, where observational limits are included at their values: the prevalence of sSFR upper limits for massive galaxies produces a tail of simulated galaxies towards low star formation rates, consistent by construction with the sample constraints. The finite observational errors similarly produce broadened distributions of both stellar mass and star formation rate in the mock sample realization. Overall, because of the broad mass coverage enabled by our large simulated volumes, as well as the star formation rates of the simulated galaxies across this mass spectrum, we can extract from either TNG100 or TNG300 a mock COS-Halos survey sample with successfully matched characteristics.\footnote{As a sidenote, if we split the mock sample at the usual sSFR value of 10$^{-11}$\,yr$^{-1}$, then the median halo mass of the star-forming sub-sample is 10$^{11.9}$\msun, as compared to 10$^{12.2}$\msun for the non-star-forming sub-sample. This reflects the observational survey selection and the TNG halo mass estimates of the COS-Halos observed galaxies.} This procedure implicitly requires that the simulated galaxies have realistic properties, i.e. that they correctly populate the sSFR-$M_\star$ plane.

\begin{figure*}
\centering
\includegraphics[angle=0,width=7.0in]{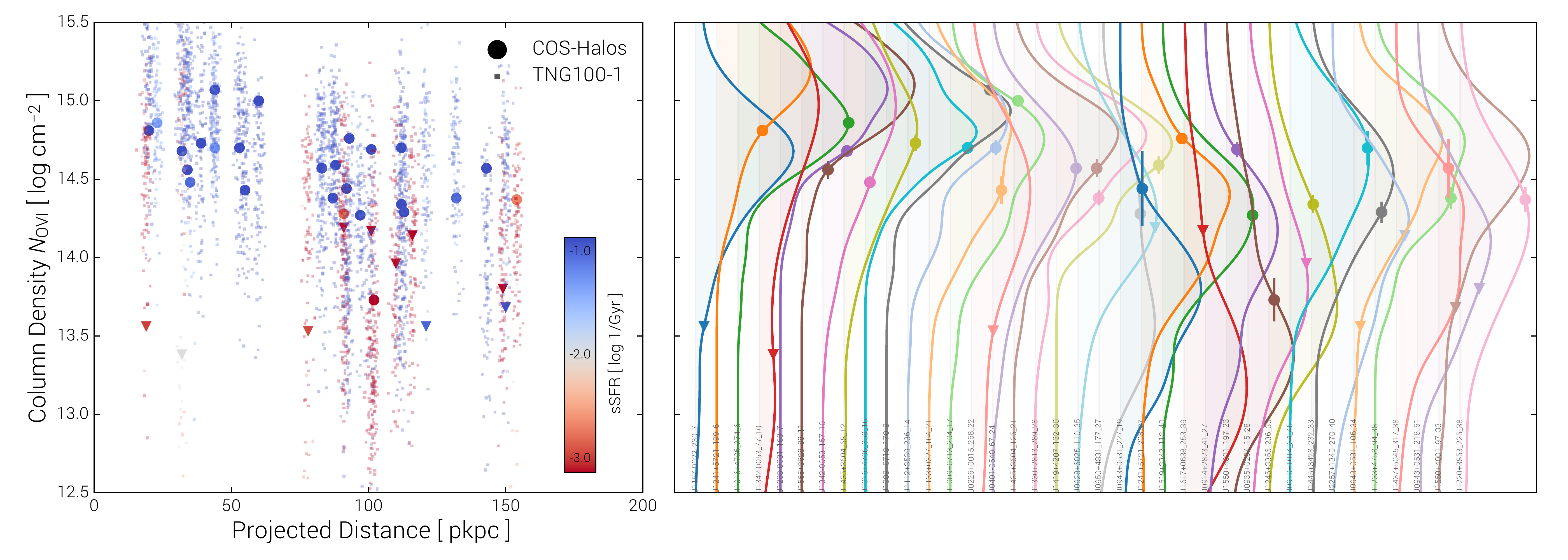}
\includegraphics[angle=0,width=7.0in]{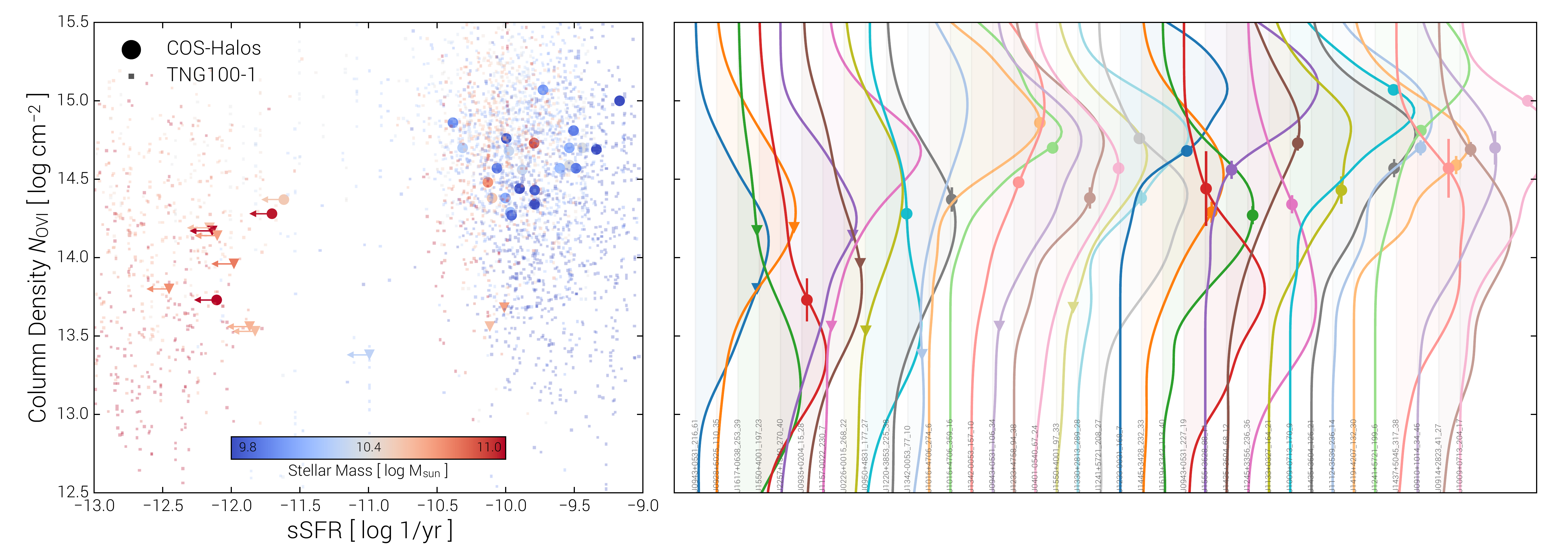}
\caption{ Comparison of the \ovi absorption from the COS-Halos survey versus the simulation results from the mock survey-matched galaxy sample. \textbf{Left panels.} The absorption column density $N_{\rm OVI}$ is shown as a function of impact parameter (top left) and sSFR (bottom left), for the observed sample (large symbols) and the 100 mock realizations from TNG (small squares). In the first case, color indicates sSFR from red/quenched to blue/star-forming, as indicated by the colorbar. In the second case, color indicates galaxy stellar mass, from 10$^9$\msun (blue) to 10$^{11}$\msun (red), as indicated by the colorbar. Observational detections are shown as circles, upper limits by downward-facing triangles. Upper limits on star-formation rates are marked by left-ward facing arrows in the bottom panel, which is the case for the majority of the quiescent COS-Halos sample. Note that agreement along the x-axes is by construction, while agreement along the y-axes represents model validation. \textbf{Right panels.} The area right of each panel shows one-dimensional PDFs of $N_{\rm OVI}$ for the simulated ensemble of each observed system (colored lines). The observed detection or limit is marked on this distribution at its value, and the quasar-galaxy pair name for each system is labeled at the bottom. The horizontal spacing is arbitrary, for visual clarity. 
 \label{fig_coshalos_ovi}}
\end{figure*}

Figure \ref{fig_coshalos_ovi} is the first comparison between the observed constraints from COS-Halos and the TNG100 simulation. In particular, we compare the \ovi absorption column densities $N_{\rm OVI}$ between the COS-Halos survey and the mock simulated survey, as a function of impact parameter (top left panel) or sSFR (bottom left panel). In both cases, large colored symbols represent the observed sample, while each of the 100 mock realizations from TNG are shown as small colored squares. We see that the simulated point clouds broadly overlap with the observational data points -- note that agreement along the x-axes is by construction, while agreement along the y-axes represents model validation. In particular, the lower \ovi columns around low sSFR galaxies, in contrast to the higher \ovi columns around high ($> 10^{-11}$ yr$^{-1}$) sSFR galaxies is similar to the dichotomy observed in the COS-Halos survey.

To enable a quantitative determination of the level of statistical agreement between the amount of \ovi absorption in each observed galaxy and the corresponding realization ensemble from TNG, the right area of each panel shows one-dimensional PDFs of $N_{\rm OVI}$ of each such ensemble (colored lines). The observed detection or limit is marked on this distribution at its value, and the quasar-galaxy pair name for each system is labeled at the bottom. In each panel, the PDFs are ordered along the horizontal direction by ascending order of their value on the x-axis of the corresponding panel (impact parameter, or sSFR). The closer the point to the main density (i.e. peak) of each distribution, the higher the probability that the observed \ovi column is consistent with having been drawn from the mock PDF. An observed point far into the tails of the simulated distribution therefore represents a tension.

To be quantitative, we calculate a statistical metric $\lambda$ which evaluates this likelihood. For an upper limit $x_1$, we take the integral of the PDF in $[-\infty, x_1]$. For a detection $x_2$, we take twice the smaller of two integrals of the PDF over $[-\infty, x_2]$ and $[x_2, +\infty]$, respectively. In both cases, the resulting value is bounded as $\lambda \in [0,1]$. For limits, if the vast majority of the simulated values are consistent with the limit, then $\lambda \sim 1$, whereas as all of the probability density becomes inconsistent with the limit, $\lambda \rightarrow 0$. For a detection at the peak of a symmetric PDF, this value is again unity, whereas if it falls in the extreme tails of the distribution it limits towards zero. For a large ensemble of realizations (with a random mixture of limits and detections) which randomly sample an underlying normal distribution the mean value is $\langle\lambda\rangle$\,$\sim$\,$0.5$. Sufficiently small values therefore represent a high probability that the observed points are inconsistent with the simulations, while values $\lambda \gg 0$ imply that this conclusion cannot be drawn. For example, the leftmost PDF (blue, upper right panel) for J1157-0022\_230\_7 has $\lambda = 0.04$ indicating poor agreement, while the next two PDFs (orange and green) have $\lambda = \{0.38,0.93\}$, indicating much better agreement. 

\begin{figure*}
\centering
\includegraphics[angle=0,width=3.4in]{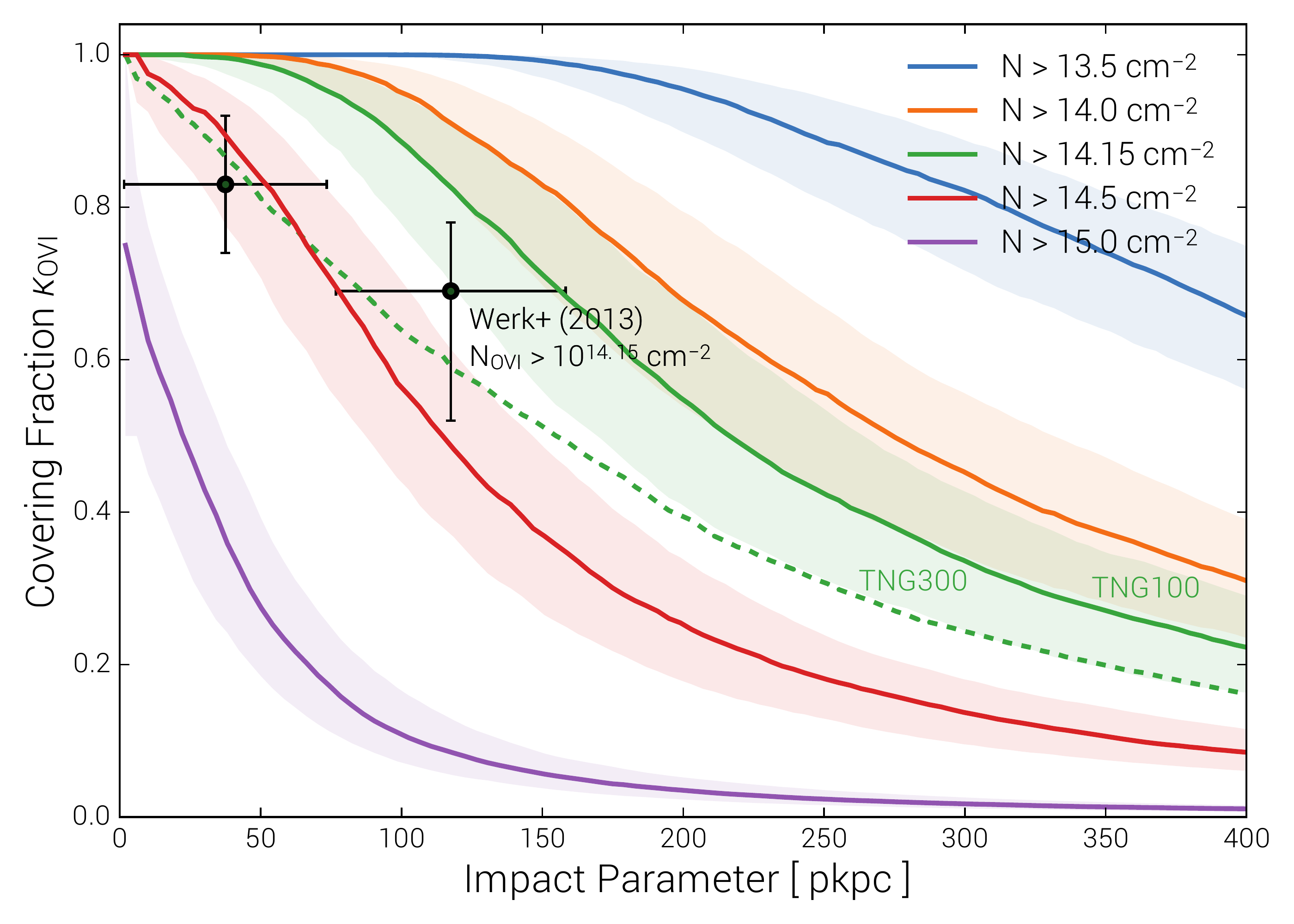}
\includegraphics[angle=0,width=3.4in]{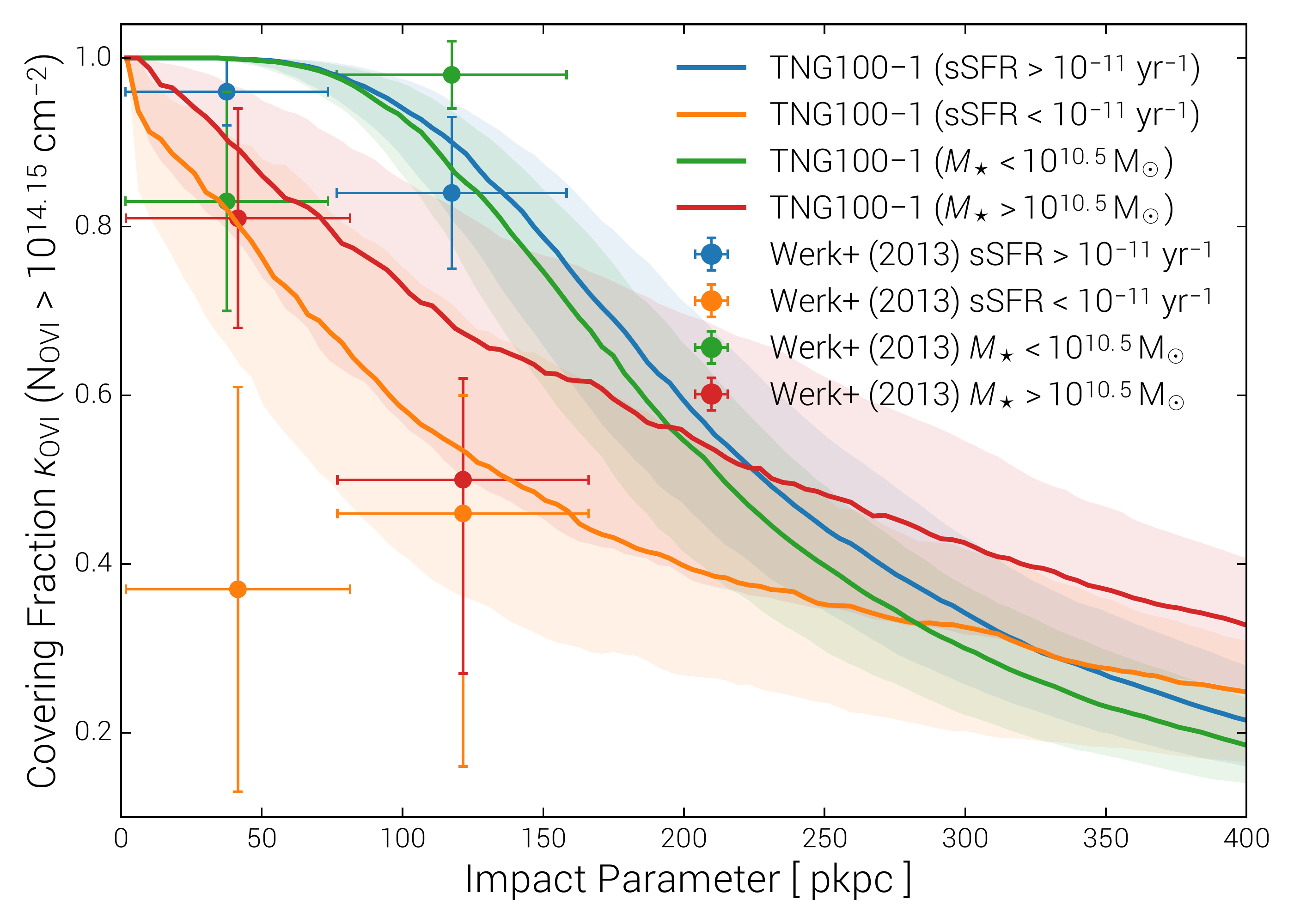}
\caption{ Comparison of the covering fractions of \ovi absorption from the COS-Halos survey versus the simulation results from the mock survey-matched galaxy sample. \textbf{(Left panel).} The $\kappa_{\rm OVI}(r)$ profiles for five different column density thresholds spanning $10^{13.5} \rm{cm}^{-2} < N_{\rm OVI}^{\rm min} < 10^{15} \rm{cm}^{-2}$. The green curve corresponds to the value from \protect\cite{werk13}, shown in black points with errorbars, against which it can be directly compared. We find that the covering fraction of \ovi around the mock COS-Halos sample is at least as high as observed. The analogous measurement from the TNG300 simulation is also given for reference as the dotted green line. \textbf{(Right panel).} Comparison of four specific galaxy sub-samples as explored in \protect\cite{werk13} for a sSFR threshold above (or below) $10^{-11}$ yr$^{-1}$ in blue (orange), and a stellar mass threshold above (or below) $10^{10.5}$\msun in red (green). Using either criterion, we find different predicted $\kappa_{\rm OVI}$ profiles for `red' versus `blue' galaxies. The high sSFR sample has high, steeply declining \ovi columns which extend to large distances, while the low sSFR (high $M_\star$) sample exhibits a shallower decline of high \ovi columns and therefore a higher incidence at all projected distances out to 400 kpc. Note: in all cases, colored bands indicate $\pm$0.5$\sigma$ inter-halo variation, which is significant.
 \label{fig_coshalos_cf}}
\end{figure*}

For the comparison of TNG100 to COS-Halos, the median statistic is $\langle\lambda\rangle = 0.52^{+0.27}_{-0.37}$, the errors giving the 16th to 84th percentiles. Agreement is better for detections than for limits; if we restrict to the former, $\langle\lambda\rangle = 0.62^{+0.18}_{-0.38}$. No observed systems have $\lambda < 0.01$, while two (of 37) have $\lambda < 0.05$. We conclude that the observed data and the simulations are not sufficiently inconsistent to reject the hypothesis that the observed samples could have been drawn from the mock distributions.\footnote{For comparison, we have repeated the entire analysis procedure unchanged on the original Illustris-1 simulation. For detections we find a median $\langle\lambda\rangle = 0.09^{+0.21}_{-0.05}$, indicating that the old Illustris model result is in considerable tension with \ovi observations around a COS-Halos like sample.} Note that these conclusions primarily imply agreement of the distribution medians and not necessarily their widths, to which the $\lambda$ statistic is less sensitive.\footnote{As an alternative comparison, we simply quote the median $N_{\rm OVI}$ values and corresponding 16th/84th percentiles. We caution, however, that this comparison of the sample medians is less robust than the $\lambda$ statistic since the simulated realizations for each observed sightline are only directly comparable to that particular $N_{\rm OVI}$ value. Splitting at sSFR of $10^{-11}$ yr$^{-1}$, the star-forming sample in COS-Halos has $N_{\rm OVI}$ = 14.6$^{+0.2}_{-0.3}$ cm$^{-2}$ and in TNG100 we find 14.6$^{+0.4}_{-0.6}$ cm$^{-2}$, while for the quiescent sample COS-Halos has 14.0$^{+0.2}_{-0.4}$ cm$^{-2}$ compared to 14.0$^{+0.6}_{-0.8}$ cm$^{-2}$ in TNG100.}

A fundamental restriction with background quasar studies of halo absorption is the limit of only one measurement per galactic halo. In order to make a statistical statement about the frequency of absorption at different strengths and across different halo selections, this is typically recast in terms of a covering fraction $\kappa(N > N_{\rm min})$, the proportion of systems exhibiting absorption column densities in a given ion above some threshold value $N_{\rm min}$. Interpretation of this measure can be recast as follows: for a given halo with multiple background sightlines, what fraction have absorption above a threshold column density (that is, a discrete approximation of the corresponding geometrical fraction of projected area). 

The observed systems are always split into sub-samples, and we likewise split the simulated realizations corresponding to each into analogous sub-samples for comparison, calculating $\kappa(r; N > N_{\rm min})$ profiles following the procedure described in Section \ref{sec_coldens}.

In Figure \ref{fig_coshalos_cf} we compare \ovi covering fractions as a function of impact parameter between COS-Halos and TNG. First, for the entire sample, $\kappa(r)$ for five different column density thresholds from $10^{13.5}$ cm$^{-2}$ to $10^{15}$ cm$^{-2}$ is derived out to 400 physical kpc (colored lines, left panel). Each monotonically decreases with increasing distance from the halo center. For all column thresholds $\ge 10^{14.5}$ cm$^{-2}$, the covering fraction at the halo center is unity, and the value of $\kappa$ at large distance plateaus to a value which scales up with decreasing column. For the lowest threshold of $10^{13.5}$ cm$^{-2}$ the covering fraction remains unity even out to $\simeq$ 200 kpc. On the other hand, for the highest threshold of $10^{15}$ cm$^{-2}$, the covering fraction even at zero impact parameter is only $\simeq$ 70\%, dropping to 10\% by 100 kpc and with an asymptotic large distance value of zero. Therefore, for a COS-Halos like sample, these highest columns are found only in the inner halo, and never beyond the virial radius. At the same time, absorption at the low column threshold is ubiquitous within the virial radius, and we would predict it impossible that a sightline falling within such an impact parameter and observed down to this limit would not detect \ovi absorption. Trends both as a function of distance and column density threshold are always continuous.

The two black points with errorbars indicate measurements from \cite{werk13} for a threshold of $N_{\rm OVI} > 10^{14.15}$ cm$^{-2}$. This limit is reproduced with the green curve, against which these observations should be compared. We conclude that TNG100 successfully produces the high observed covering fractions of high column density \ovi absorption. This stands in contrast to the previous Illustris model \citep[see][]{suresh17} as well as the EAGLE model \citep[see][]{oppenheimer16}. Our covering fractions may even be slightly too large, sitting above the observed points at the 1$\sigma$ level, although we show the same analysis repeated on the lower resolution TNG300 run (dashed) which passes through the data points - the same trends of decreasing overall \ovi (and total oxygen) content with decreasing resolution seen in the CDDFs. Note that the colored bands about the median lines indicate, in all cases, halo to halo variation at the $\pm$0.5$\sigma$ level, i.e. decreased from the normal to improve visual clarity, implying that the inter-halo variation of \ovi covering fractions is significant.

\begin{figure*}
\centering
\includegraphics[angle=0,width=3.4in]{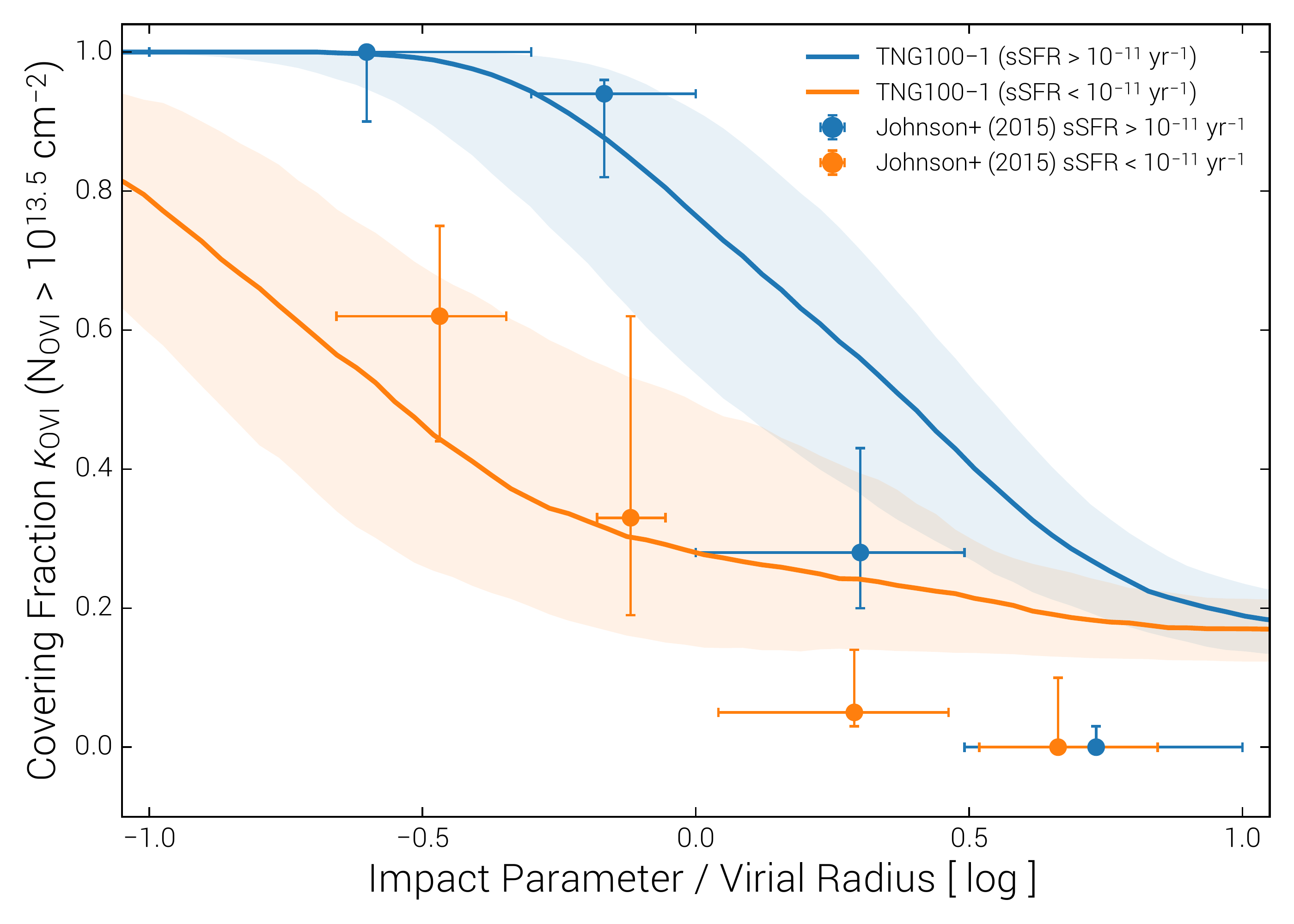}
\includegraphics[angle=0,width=3.4in]{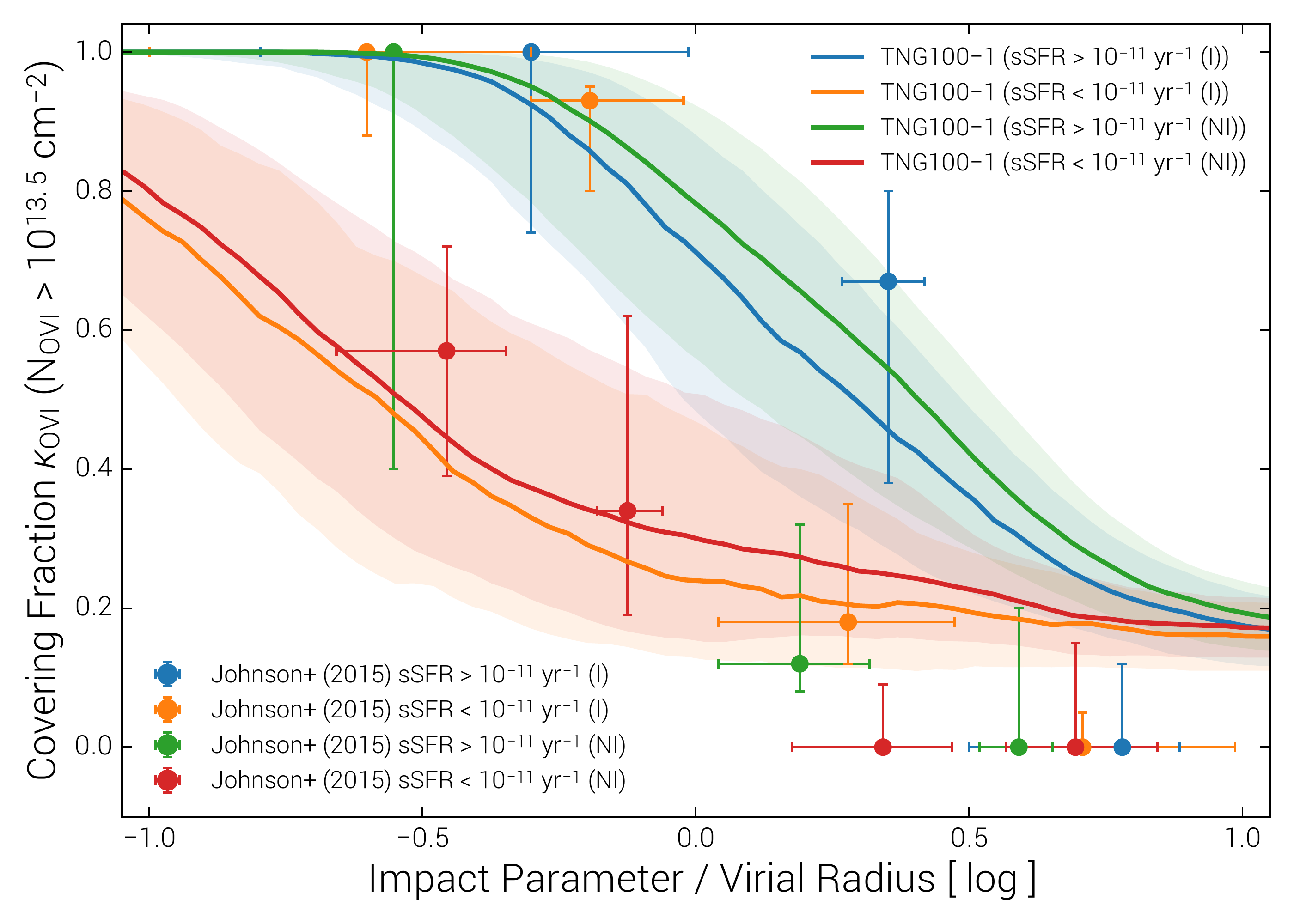}
\caption{ In direct analogy to Figure \ref{fig_coshalos_cf}, the covering fractions of \ovi absorption except for a simulated galaxy sample designed to replicate the `eCGM' survey of \protect\cite{johnson15a}, instead of COS-Halos. With respect to COS-Halos, the galaxy sample has similar distributions in redshift and stellar mass, while the impact parameters probed are mainly at larger separations, extending between 200 and 1000 kpc. Here we show the covering fraction of \ovi absorption above a threshold of $N_{\rm OVI} > 13.5$ cm$^{-2}$ as observed, in symbols with errorbars, compared to median estimates and associated scatter from the simulated sample, shown in lines and colored bands. In the left panel, the entire sample is split into Early (orange) and Late (blue) categories. In the right panel, each of these groups is further split into isolated (I) and non-isolated (NI) sub-samples. For the full sample, measurements from the TNG galaxies split into the Early and Late type classifications are in good agreement with the observations, including the overall radial trends, the separation between these two classes, and a different radial slopes of $\kappa_{\rm OVI}$. Further subdividing based on isolation state, we do not recover a significant signal, in contrast to the observational claim. 
 \label{fig_ecgm}}
\end{figure*}

In the right panel of Figure \ref{fig_coshalos_cf} the COS-Halos sample is sub-divided twice, in one case based on sSFR (blue/orange) and in the other case based on $M_\star$ (green/red). The low-sSFR and high-M$_\star$ selections are largely overlapping if not identical (orange and red points), and likewise for the star forming galaxies. Both criteria result in rather different predicted $\kappa_{\rm OVI}$ profiles: the high sSFR selection having high \ovi covering fractions extending to larger radii, and with a steeper decline, in comparison to the low sSFR selection. Similarly, the high $M_\star$ selection has a shallower decline of high \ovi columns than the low $M_\star$ selection, and therefore a higher incidence at all projected distances out to 400 kpc. The same conclusion may be hinted at by the data in the smallest distance bin (i.e. the red points are above the orange points). However, the apparently flat or even increasing trend of $\kappa_{\rm OVI}$ with distance observed for the low sSFR galaxies (orange points) makes it clear that the statistics are too limited to robustly identify this trend in the observations. We find that the low mass and high sSFR selections result in indistinguishable covering fraction profiles, consistent with the data (i.e. the blue and green lines and points are fall on top of each other). We will return to this connection between halo \ovi and properties of the central galaxy in Section \ref{sec_galaxyconnection}.

\subsubsection{Comparison to the eCGM survey}

Extending our quantitative assessment of the simulated \ovi content of galactic halos in contrast to observational constraints, we move beyond the COS-Halos dataset. In particular, we conduct a similar analysis except with the eCGM survey data of \cite{johnson15a}. This sample has not only different selection functions and probes different regions of parameter space, it also uses independent methods of data reduction and analysis (although still with the COS instrument) and different types of optical followup to locate galaxy counterparts, and different procedures for identifying absorbers with galaxies. It therefore provides a consistency check and additional statistics. This dataset includes 148 galaxies in four quasar fields, 42 of which are taken from COS-Halos directly, which we exclude in the $N_{\rm OVI}$ figures to avoid duplication. The observed and mock samples therefore include 106 and 10,600 galaxies, respectively. With respect to COS-Halos, the sample has similar though slightly broader distributions in redshift and stellar mass, while the impact parameter distribution is mainly between 200 and 1000 kpc and so to significantly larger separations. 

Our sample construction proceeds as before, first matching each observed galaxy in redshift. In eCGM galaxies are flagged as non-isolated based on the presence of a spectroscopic neighbor within a projected distance of 500 kpc, a line of sight velocity difference of less than 300 km\,s$^{-1}$, and stellar mass one third of the target galaxy or greater. Otherwise, the galaxy is considered isolated. We enforce an approximately equivalent isolation criterion by marking isolated galaxies as those which have no companion more massive than a third of their own mass (measured in terms of $M_\star$ within the 30 pkpc aperture) within a 3D distance of 500 kpc. In contrast to the COS-Halos selection, we allow satellite galaxies to be included, as the observed sample includes targets down to $10^{8}$\msun, although since the observations are highly complete only down to $0.1 L^\star$ we require agreement of the isolation state only for $M_\star > 10^{9.5}$\msun and disregard it below ($\sim$20\,\% of the sample). Exact star formation rates are not available, instead the observed sample is split into Early and Late types, based on the presence of emission lines. We convert these two categories into a sSFR threshold of below or above $10^{-11}$ yr$^{-1}$, respectively. We assume a normal (Gaussian) error of 0.25 dex on the reported $M_\star$ values, and 2.0 kpc on the reported impact parameters. 

Following the earlier analysis for COS-Halos, because the \ovi column of each observed point can only be compared to the corresponding distribution of columns from its matching simulated realizations, we measure and quote a few characteristic numbers indicating the level of statistical (dis)agreement. We again compute a $\lambda$ value for each observational point. For the comparison of TNG100 to eCGM, the median statistic is $\langle\lambda\rangle = 0.53^{+0.21}_{-0.19}$, the errors giving the 16th to 84th percentiles. For detections alone $\langle\lambda\rangle = 0.57^{+0.30}_{-0.21}$. No observed systems have $\lambda < 0.01$, while one has $\lambda < 0.05$. We conclude that there is no statistically significant tension between the \ovi absorber statistics of the eCGM dataset and the TNG100 simulation.

To assess the level of agreement in terms of the observed covering fractions, Figure \ref{fig_ecgm} compares the simulated $\kappa(r)$ profiles (solid lines) of \ovi absorption above a threshold of $N_{\rm OVI} > 13.5$ cm$^{-2}$ against those from eCGM (symbols with errorbars) for different galaxy sub-samples. In the left panel, the entire sample is split into Early (blue) and Late (orange) categories. To avoid bias in details of the abundance matching assumptions, we directly apply the assumed $r_{\rm vir}$ for each observed galaxy to each of its TNG realizations for normalizing the simulated curves. For the full sample, measurements from the TNG galaxies split into the Early and Late type classifications are in good agreement with the observations, including the overall radial trends, the separation between these two classes, and a different rapidity of the radial decrease of $\kappa_{\rm OVI}$ between them.

In the right panel, each of these groups is further split into isolated (I) and non-isolated (NI) sub-samples. In all cases radial profiles of $\kappa_{\rm OVI}$ are presented, for consistency with the observational dataset, normalized by the virial radius of their parent dark matter halo. Here we do not find nearly as strong of a signal as observationally claimed. The separation in covering fraction between `I' and `NI' classes, for either the quiescent or star forming samples, is at most $\sim$5\% in the simulations -- on average, it is negligible. At this column and for these samples, the simulations predict an asymptotic covering fraction of $\simeq$15\,\% at large distance, consistent with the binomial errors of the observed points at the $\sim$1$\sigma$ level, which otherwise have too few statistics to verify a nonzero value.

\subsubsection{Sensitivity of the covering fractions to model variations}

\begin{figure}
\centering
\includegraphics[angle=0,width=3.3in]{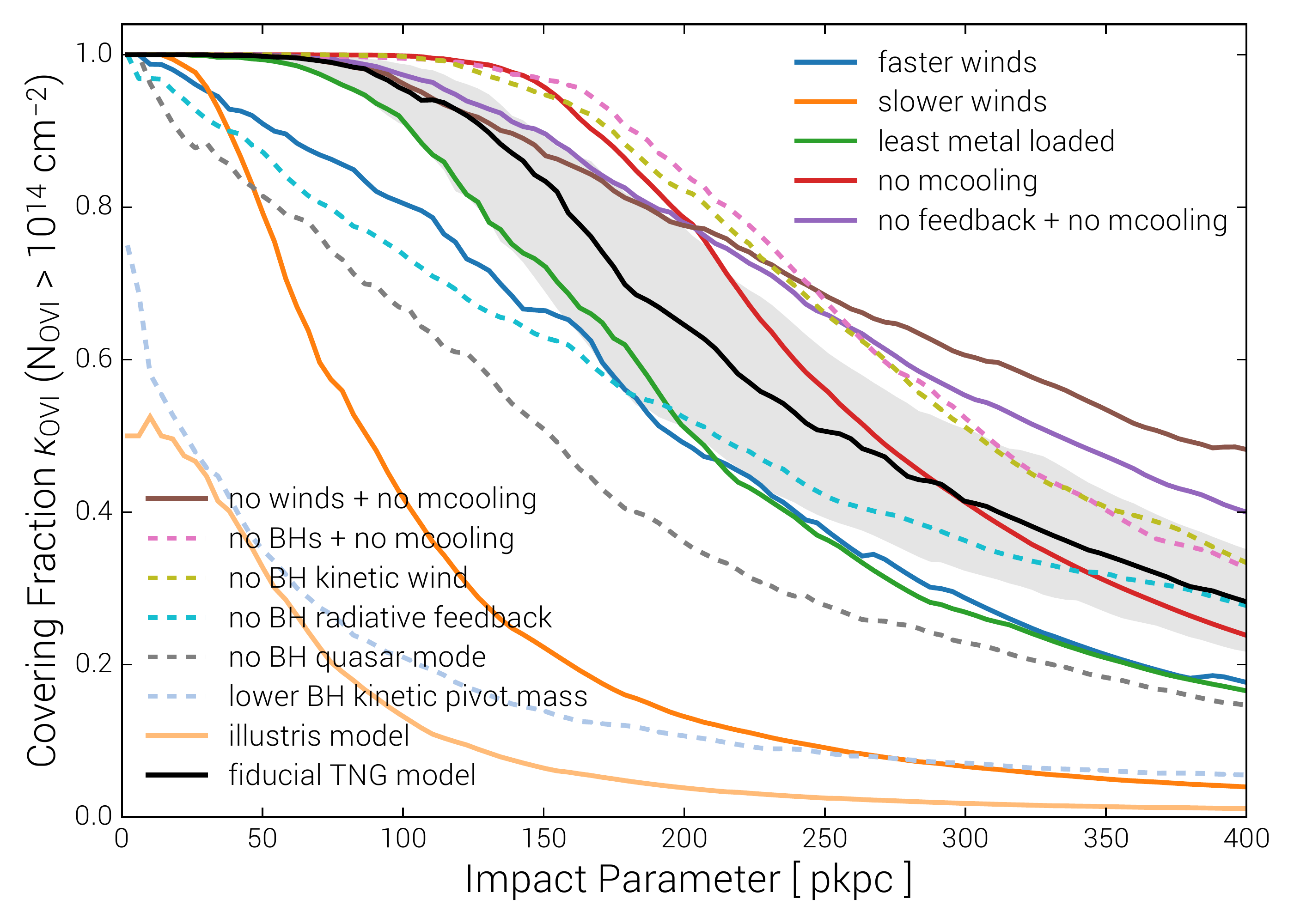}
\caption{ Impact of the physics variations on the \ovi covering fractions. The $\kappa_{\rm OVI}(r)$ profile for a fixed threshold of $N_{\rm OVI} > 10^{14}$ cm$^{-2}$ is shown as a function of impact parameter, stacking all simulated halos with total mass $10^{11.5} < M_{\rm halo}/$\msun$ < 10^{12.5}$, roughly 190 in total for each run. The fiducial TNG model result is given in black, and the same halo to halo variation band is included from this run, in gray, to indicate the relative variability among halos at fixed model. We show the single parameter (or model choice) variations which lead to the most significant differences in $\kappa_{\rm OVI}(r)$, and also include the result of the original Illustris model (light orange).
 \label{fig_cf_variants}}
\end{figure}

Finally, we briefly return to the impact of model variations, extending our previous exploration of the sensitivity of the \ovi CDDF to the halo-centric measurement of covering fractions. Because different simulations can produce rather different galaxy properties including stellar mass and SFR, we do not repeat the mock survey realizations approach on each. Instead, we simply `target' all simulated halos with total mass $10^{11.5} < M_{\rm halo}/$\msun$ < 10^{12.5}$ with one `realization' each, always at $z=0$. This gives a relatively uniform selection across runs, modulo the variable baryonic impact on halo masses. In the fiducial test volume, this selects $\sim$ 190 halos. Figure \ref{fig_cf_variants} shows the result, including the fiducial measurement (in black) with the level of halo to halo variation within this single simulation (gray band). The same eleven model variants are included, and all fall outside this band at some point. 

Four distinct types of behavior are seen, which we discuss from the innermost to outermost. First, substantial suppression of $\kappa$ all the way into the inner halo, as seen for slower winds (orange) and a decrease in the BH mass for the onset of low-state feedback (light blue). Second, flattening of $\kappa(r)$ by distributing most of the halo \ovi to larger distances, as in the cases of no BH quasar mode (gray), no BH radiative feedback (cyan), and faster winds (dark blue). Third, flattening $\kappa(r)$ only by increasing it at large distances, leaving the signal within 100 kpc unchanged: no feedback and no metal-line cooling (purple), and no winds (brown) variants. Fourth, a similar profile of $\kappa(r)$ except with the drop indicative of the halo virial boundary moved outwards, as for no metal cooling (red), no BH kinetic-wind feedback (gold), and no BHs whatsoever (pink); the case of reduced metal loading of the winds (green) is similar, except with the boundary moving inwards.

From the observational point of view, the situation is similar to the \ovi CDDF variations, where the halo-centric absorber statistics such as covering fraction are even more starved for statistics. At present they are therefore only able to discriminate among extreme model variations. From the theoretical viewpoint, we see that similar signatures can result from notably different reasons. For example, the case of the `lower BH kinetic pivot mass' results in substantial suppression of star formation, stellar mass, and so also oxygen production, across the entire volume. Encountering columns of $N_{\rm OVI} \ga 10^{14}$ cm$^{-2}$ then becomes rare, even inside high density circumgalactic gas. On the other hand, the `slower winds' case proceeds differently; it is entirely unable to control physical overcooling, resulting in a significantly too high late time cosmic star formation rate density (SFRD) and overshooting a reasonable stellar to halo mass relation (SMHM) by factors of many. The significant mass of oxygen produced is, however, confined to small galacto-centric distances because the winds cannot appreciably push out metals, resulting in a similar impact on the $\kappa_{\rm OVI}(r)$.

On the other hand, the `no feedback' and `no winds' models show consistent although somewhat counterintuitive behavior. Importantly, these runs produce galaxy populations severely wrong in the majority of measurable quantities, primarily because they fail to regulate star formation in the majority of halos. As a result their redshift zero stellar mass functions, for example, are unrealistic, and excessive stellar mass leads to strong metal overproduction (50\% more gas-phase metals versus fiducial), making any interpretation of the resulting covering fractions difficult. Overall, the resulting \ovi is more concentrated within galactic halos, although at this mass scale and column the covering fractions are actually higher than fiducial for all impact parameters $\le$ 400 kpc. The opposite is true for $\kappa (N_{\rm OVI} > 10^{13} \rm{cm}^{-2})$ at this same mass scale, where the fiducial model has everywhere a higher covering fraction.

Notably, just as we saw before, the Illustris model line (light orange) is the most extreme outlier of the entire ensemble. It is clear that the dearth of \ovi in Illustris relative to the COS-Halos results \citep[explored in][]{suresh15} was primarily driven by physical model deficiencies, and not by any particular issue with analysis methodology. Its qualitative behavior is similar to the `slower winds' variant, and we conclude that modifications to the TNG wind model including their velocities are a dominant reason for the improvements over the previous Illustris results. Nonetheless, the complex coupling between these processes all but guarantees that there is no single, unambiguous change which has led to the global improvements in TNG -- rather, it arises from the interplay of modifications in both stellar and blackhole feedback processes.


\section{Relating Halo Oxygen Content to the Central Galaxy} \label{sec_galaxyconnection}

The connection between the properties of circumgalactic gas and the evolutionary state or ongoing activity of the galaxy at its center directly probes the baryon cycle and the coupling of baryonic feedback processes across a wide range of scales. In both observational surveys we have compared to so far, the differing prevalence of strong \ovi absorption around star-forming versus quiescent galaxies has been noted. Figures \ref{fig_coshalos_cf} and \ref{fig_ecgm}, comparing to COS-Halos and eCGM, respectively, demonstrated that the TNG simulations produce similar signatures. We now endeavor to understand the origin of this particular trend, as well as explore further connections between halo \ovi content and central galaxy properties.

To begin, in Figure \ref{fig_stamps_blue} we show maps of the spatial distribution of \ovi column density around an ensemble of 30 galaxies at $z=0$ from TNG100. Each panel is 800 kpc on a side, oriented such that the central galaxy is face-on. These are selected in the narrow halo mass bin of \mbox{$10^{12}$\,$<$\,$M_{\rm halo}$/M$_\odot$\,$<$\,$10^{12.2}$}, and are the same systems as the first 30 galaxies of Figure 12 in \cite{nelson18}. Namely, all thirty are \textit{blue}, star-forming disk galaxies. We see that halos of extended \ovi gas are always present, typically reaching columns of $\ga 10^{15}$ cm$^{-2}$ in the inner halo, decreasing to $\sim 10^{13}$ cm$^{-2}$ out near the virial radius, and with a variety of morphologies which are nonetheless broadly spherical.

In contrast, Figure \ref{fig_stamps_red} shows maps of thirty different $z=0$ halos from the same halo mass bin, in this case selected to host \textit{red}, quiescent galaxies.\footnote{These are likewise the same systems as the first 30 galaxies in Figure 13 of \cite{nelson18} -- predominantly morphological evolved spheroids.} Visual inspection alone makes it clear that the amount of halo \ovi is suppressed with respect to the star-forming sample. For most halos, the maximal \ovi columns are lower. Average central values are roughly one dex lower at $N_{\rm OVI} \simeq 10^{14}$ cm$^{-2}$. The spatial distribution of the ion is also more frequently disturbed and irregular. Azimuthal profiles at a given radius are less regular, and some halos exhibit elongated (i.e. non-axisymmetric) local depressions of \ovi.

\begin{figure*}
\centerline{\includegraphics[angle=0,width=6.8in]{figures/stamps_TNG100-1_evo-0_red-0_blue-1_rot-face-on_scope_global_sm.pdf}}
\caption{ Projected \ovi column density maps around a sample of $z\,=\,0$ \textit{blue} galaxies, selected as having (g-r)\,$<$\,0.6, and taken from the halo mass bin \mbox{$10^{12}$\,$<$\,$M_{\rm halo}$/M$_\odot$\,$<$\,$10^{12.2}$} of TNG100. These are the same systems as the first 30 of Figure 12 in \protect\cite{nelson18}. Every panel is 800\,kpc on a side, and oriented such that the central galaxy would be face-on. White circles show virial radii. Halos in this mass regime surrounding \textit{blue} galaxies are always surrounded by a circumgalactic reservoir of \ovi, with peak column densities in the central regions commonly reaching $10^{15}$ cm$^{-2}$.
 \label{fig_stamps_blue}} 
\end{figure*}

\begin{figure*}
\centerline{\includegraphics[angle=0,width=6.8in]{figures/stamps_TNG100-1_evo-0_red-1_blue-0_rot-face-on_scope_global_sm.pdf}}
\caption{ Projected \ovi column density maps at $z\,=\,0$ as in Figure \ref{fig_stamps_blue} except here for \textit{red} galaxies, selected as having (g-r)\,$>$\,0.6 and in the same halo mass bin \mbox{$10^{12}$\,$<$\,$M_{\rm halo}$/M$_\odot$\,$<$\,$10^{12.2}$}, corresponding to the low-mass end of the red sequence. These are the same systems as the first 30 of Figure 13 in \protect\cite{nelson18}. The white circle in each panel shows the halo virial radius. The total amount of \ovi around red galaxies is clearly suppressed, as is its average and maximal column densities, coincident with less spherical and more disturbed spatial distributions.
 \label{fig_stamps_red}} 
\end{figure*}

\begin{figure}
\centering
\includegraphics[angle=0,width=3.3in]{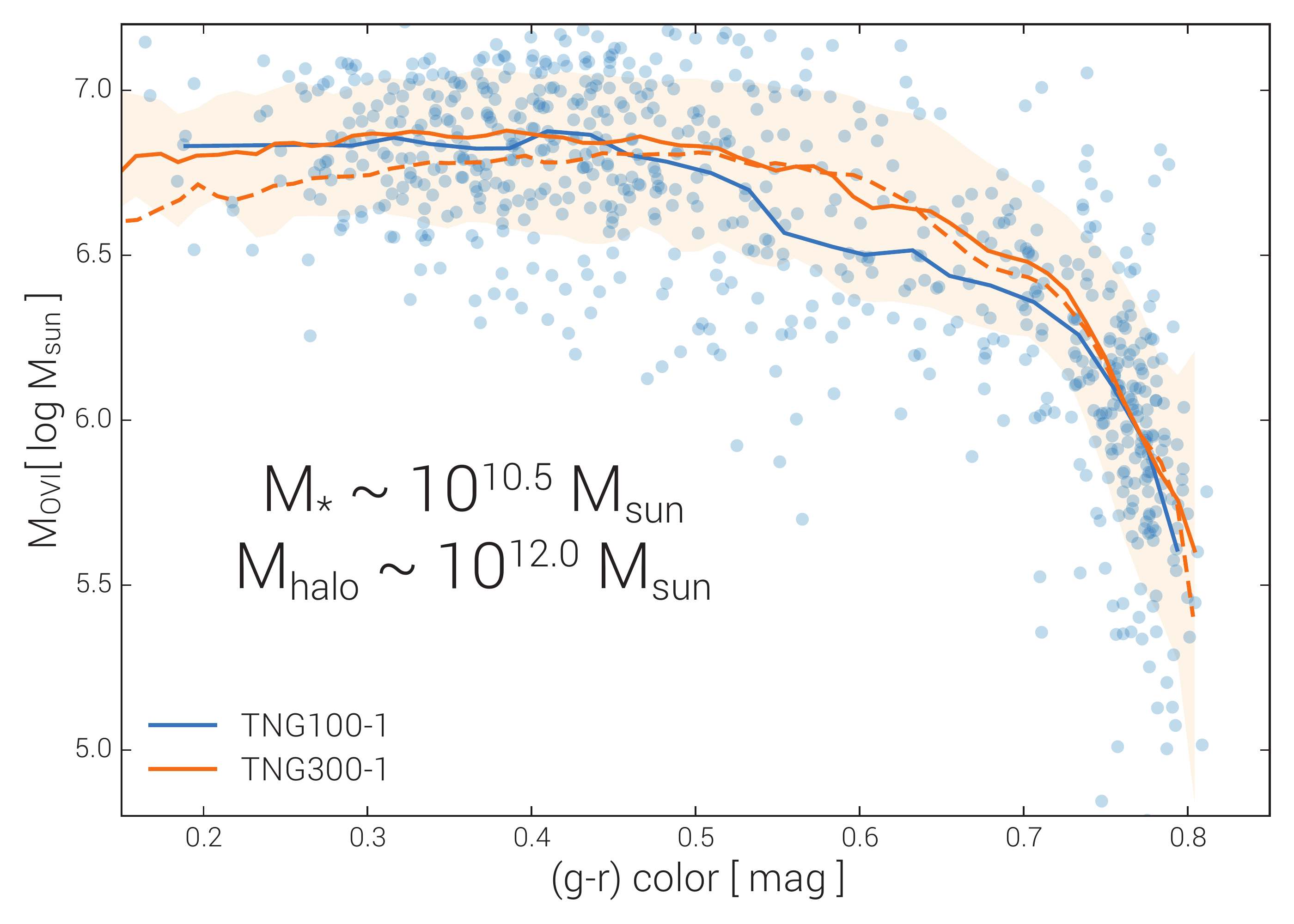}
\caption{ Correlation between the total mass of \ovi in the halo and the \gr color of the central galaxy at fixed mass. Results for both TNG100 and TNG300 are shown. For the former, all the individual systems in these mass bins are shown with individual markers (of order 1000). For TNG300 the 16th to 84th percentiles about the median are given by colored band (including $\sim$ 10,000 objects). We include all galaxies in a narrow stellar mass range of $10^{10.4} < M_\star / \rm{M}_\odot < 10^{10.6}$ (solid lines). To demonstrate that the same trend exists at fixed halo mass, we also show the TNG300 result in the narrow halo mass bin of $10^{12.0} < M_{\rm halo} / \rm{M}_\odot < 10^{12.1}$ (dashed line).
 \label{fig_slice_color}}
\end{figure}

To quantify this dichotomy, Figure \ref{fig_slice_color} presents the correlation between total gravitationally bound \ovi mass in the halo and (g-r) color of the central galaxy, at both fixed stellar mass and fixed halo mass. In both cases, a strong continuous trend is evident: redder galaxies have progressively less oxygen in this particular ion. For blue and intermediate color galaxies with \mbox{(g-r) $\la$ 0.6} we see that $M_{\rm OVI}$ is roughly constant. Galaxies firmly in the red population \mbox{(g-r) $\ga$ 0.7} have, in the median, one order of magnitude less total \ovi mass. At fixed projected halo size, this translates directly into a decrease in the average column density by approximately the same one order of magnitude. This trend is not driven by the well-known underlying correlation of color with halo mass at fixed $M_\star$, i.e. that red galaxies reside in more massive dark matter halos than blue galaxies of the same stellar mass \citep{mandelbaum16}, because the same effect is seen also at fixed halo mass (dashed line).

This is at first glance similar to the key result of \cite{tumlinson11} for COS-Halos, that there is much less \ovi around quiescent galaxies in comparison to their star-forming counterparts. However, this result is based on a population sample spanning more than one dex in stellar mass -- that is, it was never demonstrated, nor claimed, at fixed mass. Indeed, \cite{oppenheimer16} explains the COS-Halos finding as dominated by a simple mass trend, that the quenched galaxies of the sample reside on average in more massive halos, and that these more massive halos have smaller \ovi columns (as also shown here, Figure \ref{fig_ions_vs_mass}). The correlation we predict at fixed mass is therefore fundamentally different, and would require an observational sample with enough statistics (i.e. many tens of systems in a similarly narrow mass bin in $M_\star$ or $M_{\rm halo}$) to either verify or disprove. As we discuss below in Section \ref{sec_discussion}, this correlation also implies the existence of a causal link between energetic feedback from the blackhole of the central galaxy and the \ovi content of its CGM, illuminating an important part of the baryon cycle.

\begin{figure*}
\centerline{\includegraphics[angle=0,width=7.1in]{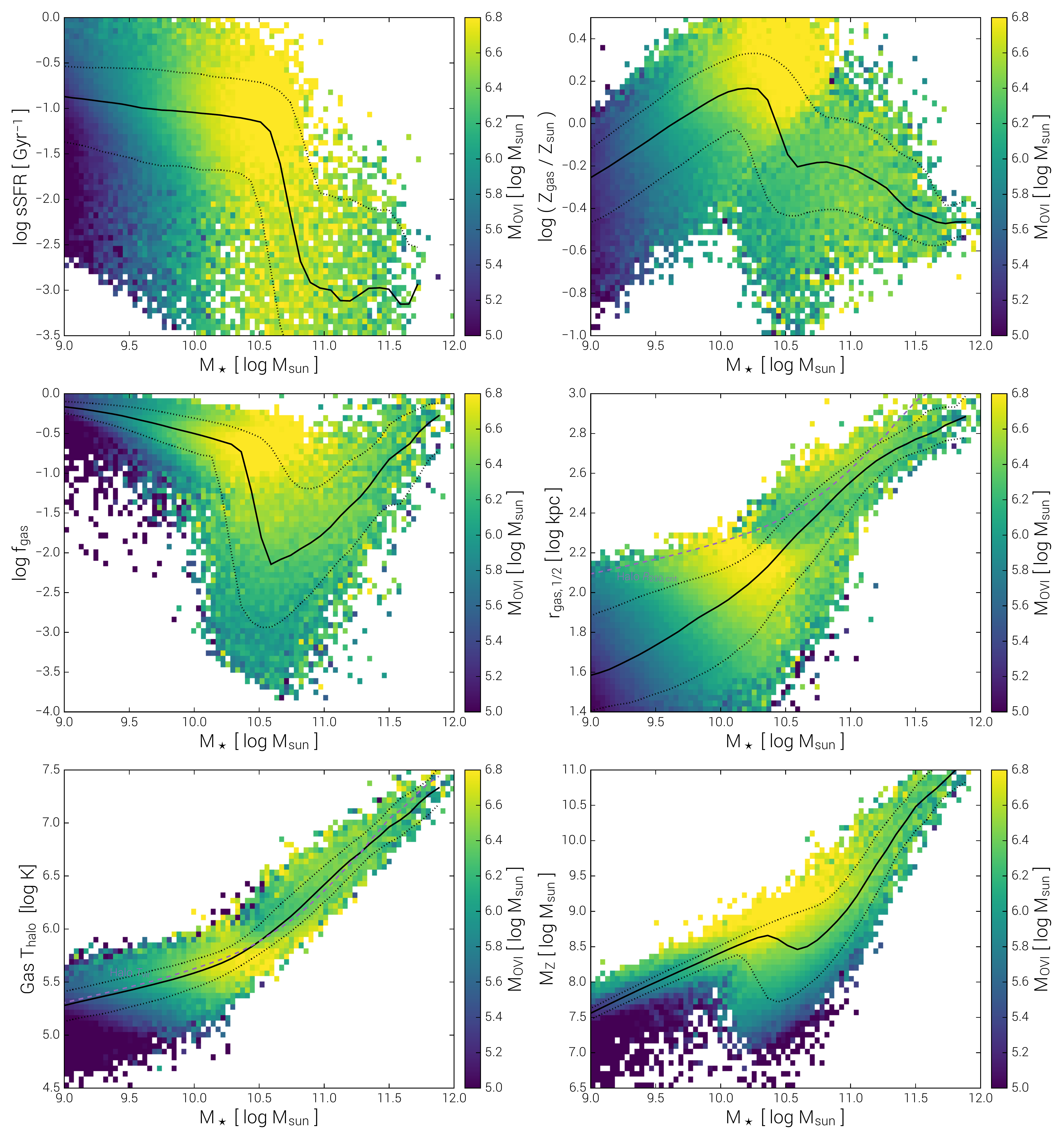}}
\caption{ Six different relations of various galaxy/halo properties as a function of \textbf{stellar} mass for central galaxies at $z=0$. 
In each case, we include the median relation (black solid line) and the 10-90 percentiles (dotted black lines). The background 
color shows the median \ovi mass for all systems in that bin. Here we show
(i) specific star formation rate, 
(ii) galaxy gas metallicity,
(iii) galactic gas fraction,
(iv) size (half mass radii) of the gas,
(v) mean volume-weighted temperature of halo gas,
and (vi) the total gravitationally bound gas-phase metal mass.
 \label{fig_hist2d_mstar_1}} 
\end{figure*}

\subsection{Correlations beyond galaxy color and SFR}

\begin{figure*}
\centerline{\includegraphics[angle=0,width=7.1in]{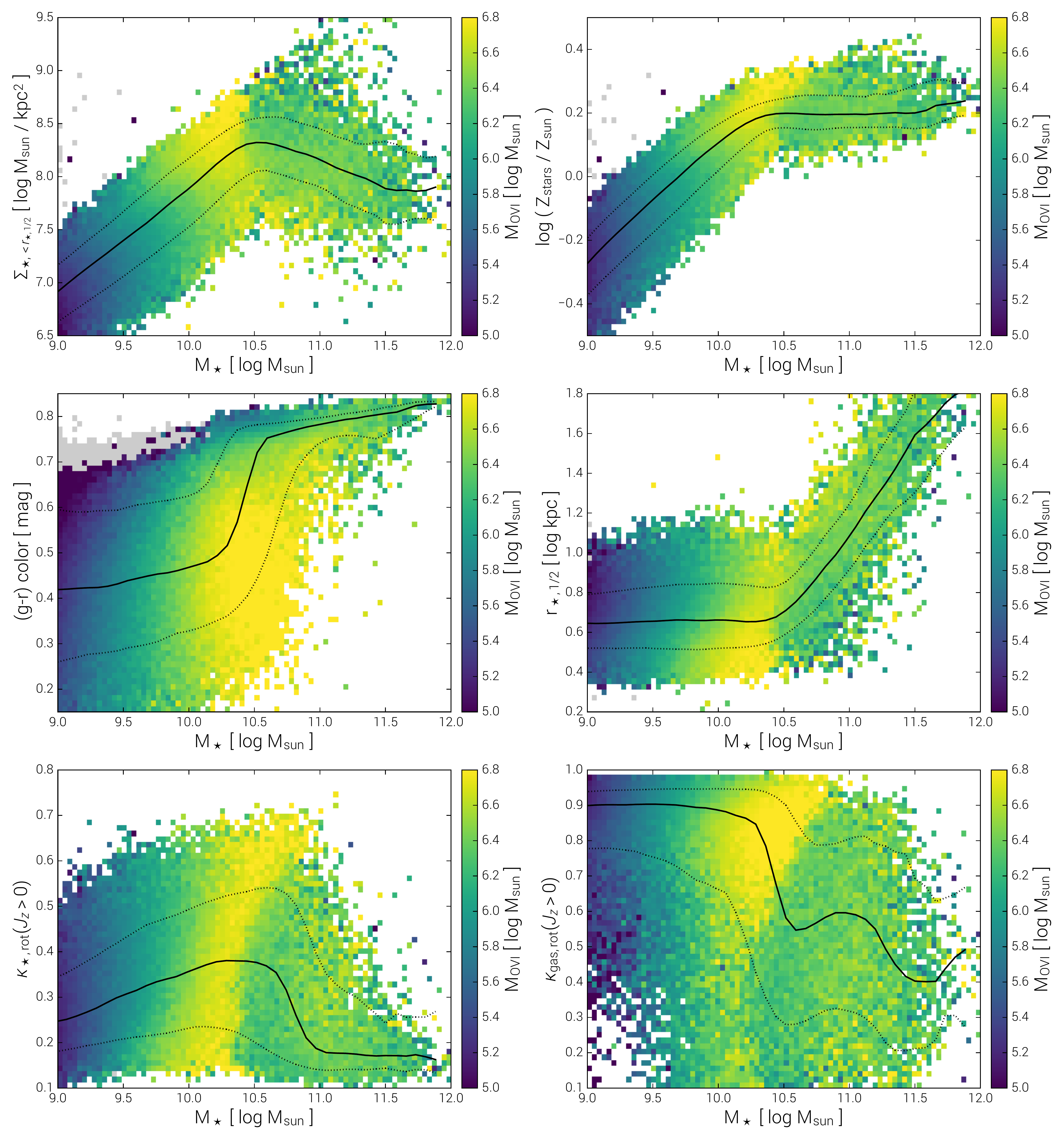}}
\caption{ Six different relations of various galaxy/halo properties as a function of \textbf{stellar} mass for central galaxies at $z=0$. 
In each case, we include the median relation (black solid line) and the 10-90 percentiles (dotted black lines). The background 
color shows the median \ovi mass for all systems in that bin. Here we show
(i) central stellar surface density,
(ii) galaxy stellar metallicity,
(iii) \gr stellar color,
(iv) stellar sizes also in half mass radii,
(v) $\kappa$ rotation measure of the stars,
and (vi) $\kappa$ of the gas.
 \label{fig_hist2d_mstar_2}} 
\end{figure*}

\begin{figure*}
\centerline{\includegraphics[angle=0,width=7.1in]{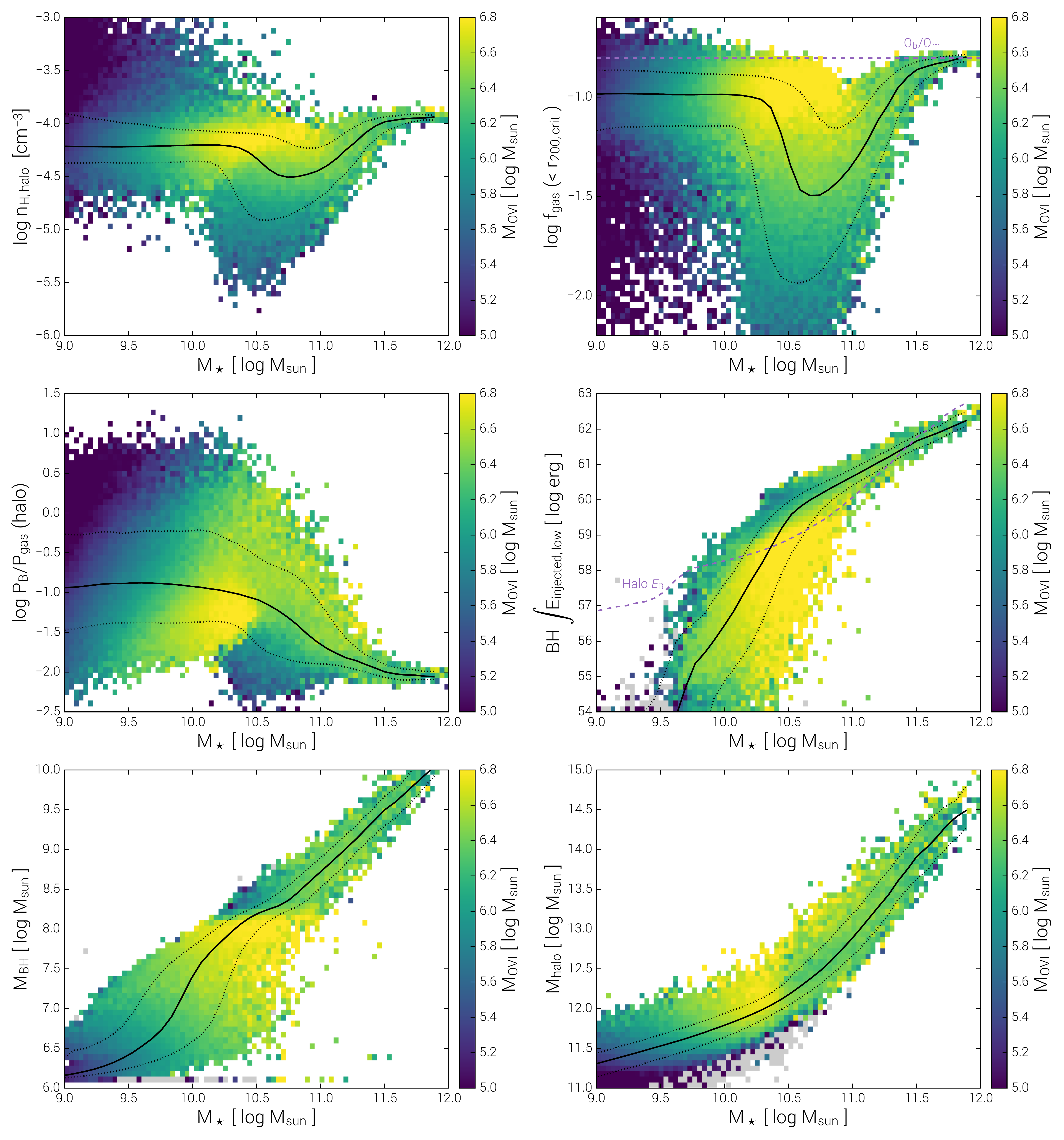}}
\caption{ Six different relations of various galaxy/halo properties as a function of \textbf{stellar} mass for central galaxies at $z=0$. 
In each case, we include the median relation (black solid line) and the 10-90 percentiles (dotted black lines). The background 
color shows the median \ovi mass for all systems in that bin. Here we show
(i) mean volume-weighted hydrogen number density in the halo,
(ii) halo gas fraction within $r_{\rm 200,crit}$,
(iii) volume-weighted $\beta^{-1}$ ratio of halo gas,
(iv) cumulative energy injection in the BH low-state feedback mode,
(v) redshift zero blackhole mass,
and (vi) total halo mass.
 \label{fig_hist2d_mstar_3}} 
\end{figure*}

Notably, optical color is not the only galaxy property which shows clear connections with the total $M_{\rm OVI}$ of the halo. Figures \ref{fig_hist2d_mstar_1}, \ref{fig_hist2d_mstar_2}, and \ref{fig_hist2d_mstar_3} explore several such correlations, with six panels per figure. Each presents a different galaxy or halo property as a function of galaxy stellar mass, for all central galaxies of TNG300 at $z=0$. The solid black line shows the median relation, while the dotted envelopes give the 10th to 90th percentiles. The bulk of the galaxy population therefore falls in this range. In every case, the background color distribution shows the median total gravitationally bound \ovi mass in all the halos which fall into that bin, from $10^5$ to $10^{6.8}$\msun. Since every pixel/bin with at least one galaxy is colored, extreme tails of the distributions are also present. In the \mbox{$9.0 < \log(M_\star / $\msun$) < 12.0$} range, TNG300 contains 162,000 central galaxies at $z=0$ which are included in each panel. Many quantities are computed within some multiple of the stellar half mass radius, denoted by $r_{\star,1/2}$.

Our main focus here is not on the trends of each property with $M_\star$, or the overall trend of $M_{\rm OVI}$ with $M_\star$, but rather the relative abundance of \ovi mass of outliers with respect to `average' galaxies. That is, we are particularly interested in vertical trends -- vertical color gradients -- at fixed stellar mass, indicative of important secondary although non-trivial trends of halo $M_{\rm OVI}$ with a given property. These features may be present only over restricted ranges in stellar mass. Frequently, we observe strong trends at fixed $M_\star \sim 10^{10.5}$\msun, where the bulk of the galaxy population begins to transition from blue, star-forming, main sequence galaxies to the red, quiescent population in TNG \citep{nelson18}. As a result, a heterogeneous mix of both types of galaxies exist over this transitional range of stellar mass, often with very different properties \citep[e.g. sizes;][]{genel18}. In the following discussion we also occasionally comment on trends as a function of halo mass or at fixed halo mass, although these panels are not explicitly shown.

Figure \ref{fig_hist2d_mstar_1} considers galaxy sSFR, with mass and SFR both measured within $2r_{\star,1/2}$ (upper left); galactic gas metallicity (upper right), galaxy gas fraction - gas mass normalized by baryonic mass, both within $2r_{\star,1/2}$ (center left), gas sizes in terms of their half mass radii (center right), mean volume-weighted temperature of the halo gas, defined as $0.15 < r/r_{\rm vir} < 1.0$; (lower left), and the total gravitationally bound metal mass  (lower right). 

The sSFR panel demonstrates how the star formation main sequence of IllustrisTNG is truncated at $\ga 10^{10.5}$\msun, beyond which the majority of galaxies begin to enter the quiescent population \citep{weinberger18}. The overall trend of increasing $M_{\rm OVI}$ with $M_\star$ results in maximal \ovi for star-forming galaxies which are just pre-quiescence. However, at lower masses and at fixed $M_\star$, galaxies above the median relation have more \ovi mass than those below it. This is driven by halo mass -- at fixed halo mass there is no such correlation of $M_{\rm OVI}$ with sSFR (not shown).
The $Z_{\rm gas}$ view shows a correlation between halo \ovi mass and the gas-phase metallicity of the central galaxy at fixed stellar masses around $\sim 10^{10.5}$\msun. Here, unquenched galaxies remain on the median mass-metallicity relation \citep[MZR; explored for TNG in][]{torrey18}, while galaxies whose BHs have transitioned into the low-state have evacuated central, cold, metal-enriched gas mass, allowing relatively less enriched inflows to dilute the remaining $Z_{\rm gas}$. This trend also exists at fixed $M_{\rm halo}$, and outside of this transition regime in mass there is no strong \ovi trend with $Z_{\rm gas}$ at fixed $M_\star$. 

Gas depleted galaxies at all $M_\star \la 10^{11}$\msun also have lower \ovi masses, and the correlation at fixed $M_\star \sim 10^{10.5}$\msun holds also at fixed halo mass, while the correlation at fixed $M_\star \la 10^{10}$\msun is absent at fixed halo mass (i.e. the color gradient is purely horizontal with $M_{\rm halo}$ and has no secondary dependence on $f_{\rm gas}$). This is a common feature in these figures, and represents a localized segregation in halo \ovi content corresponding to whether or not the central blackhole has transitioned into its more effective, low accretion state, feedback mode. This localization is therefore in $M_{\rm BH}$, not stellar mass, as we show below.
As the half mass radii of the gas component monotonically increases with stellar mass we observe the first hint of a horizontal threshold over log($M_\star$/\msun) = $10.5 \pm 0.2$. Galaxies with $r_{\rm gas,1/2} \ga 10^{2.2}$ kpc have suppressed \ovi content relative to their smaller counterparts. Apart from this feature, there is no significant trend of $M_{\rm OVI}$ with galaxy size, in either gas or stars, at fixed halo mass. For comparison, the halo virial radii are shown as a dashed line in this panel. High $r_{\rm gas,1/2}$ outliers have apparently experienced a significant redistribution of gas mass, lowering the central densities (as seen also in $f_{\rm gas}$) and expanding the radius needed to enclose half the gas mass in the halo. 

For the gas temperature in the halo we take the volume-weighted mean of all non star-forming gas cells in the radial range $0.15 < r/r_{\rm vir} < 1.0$. The resulting secondary correlation with \ovi mass is remarkably similar to that just discussed for the gas sizes. There is a broad trend of increasing $M_{\rm OVI}$ with increasing $T_{\rm halo}$ at fixed stellar mass, which arises from scatter in the SMHM as we show below. Around $M_\star \sim 10^{10.5}$\msun, however, we see that halos with gas temperatures above the median are substantially depleted in \ovi content -- in addition to physical redistribution, there is also a thermal effect contributing to the shift in oxygen ionization state occupations (see discussion). For comparison, the canonical virial temperature of the halos are shown as a dashed line.
The final panel shows that over the broad range of $10.0 <$ log($M_\star$/\msun) $< 11.0$ a significant scatter is present in the $M_{\rm Z}-M_\star$ relation of total gravitationally bound gas-phase metal mass to galaxy stellar mass. This scatter is only downward -- i.e., some systems are strongly metal depleted, and as might be expected, that same population is also strongly $M_{\rm OVI}$ deficient. Note that $M_{\rm Z}$ sums over all metals in the central galaxy as well as the CGM. Therefore, this large fraction of missing metals is entirely removed from the halo reservoir of gravitationally bound baryons.

Figure \ref{fig_hist2d_mstar_2} considers six properties primarily of the stars: stellar surface density within $r_{\star,1/2}$ (upper left) and galaxy stellar metallicity within $2r_{\star,1/2}$ (upper right); galaxy \mbox{(g-r)} optical color \citep[center left; using the fiducial model of][]{nelson18}; stellar sizes in terms of the half mass radii (center right); the measurement of oriented rotational support $\kappa_\star$ for the stars (lower left) and the gas \citep[lower right; strong spheroids with $\kappa < 0.3$ and prominent disks having $\kappa > 0.6$, see][]{rodriguezgomez17}.

There is only a weak segregation of \ovi content with $\Sigma_\star$, even in the midst of the transition.
Similarly, the $Z_{\rm stars}$ panel shows no significant correlations, as the metallicities of existing stellar populations cannot be modified by the quenching process.
We have already seen the marginalized relation between $M_{\rm OVI}$ and galaxy \mbox{(g-r)} color at fixed stellar as well as halo mass in Figure \ref{fig_slice_color}. In the transition regime, red galaxies have less \ovi in their halos than blue galaxies. There is little sign of trends at either $M_\star \la 10^{10}$\msun or $M_\star \ga 10^{11}$\msun, \textit{except} that a significant fraction of rejuvenated massive systems recover high \ovi masses. These are the individual isolated pixels in the lower-right region of the panel. Either the oxygen created by the refreshed star formation and expelled in galactic winds has boosted the overall halo oxygen content, or the process which reignited star formation in the first place also positively contributed to the halo \ovi, as might be expected when the circumgalactic media of two halos combine during a galaxy merger \citep[e.g.][]{hani18}.

The half mass radii of the stellar component, as with the gas, monotonically increases with stellar mass \citep[see ][for the size-mass relation in TNG]{genel18}. However, unlike the strong horizontal threshold seen in $r_{\rm gas, 1/2}$, we find only a mild trend of \ovi mass with stellar size at fixed mass. At $M_\star \sim 10^{10}$\msun, more compact systems have systematically more \ovi, while for $M_\star \ga 10^{10.5}$\msun larger galaxies which are also naturally more star forming have larger $M_{\rm OVI}$. We connect this to the blackhole feedback below, which has a direct kinetic impact on the gas in the low accretion state mode. On the other hand, BH feedback has no direct dynamical impact on the stars (other than subsequent orbit evolution due to modification of the total underlying potential), which we see reflected in the lack of a strong correlation between $r_{\rm \star,1/2}$ and $M_{\rm OVI}$ with either stellar or halo mass fixed.

The $\kappa_{\rm \star,rot}$ and $\kappa_{\rm gas,rot}$ measures of rotational support \citep[e.g.][]{sales12} have broadly similar features. Strong gaseous disks with significant angular momentum at low mass are paired with relatively strong stellar disks, both of which disappear with decreasing $\kappa$ for massive galaxies with $M_\star \ga 10^{10.5}$\msun. Each has a diagonal feature in the secondary $M_{\rm OVI}$ correlation indicating that galaxies which have already lost a significant amount of rotational support have likewise lost a significant amount of halo \ovi. That is, disk galaxies have more \ovi than spheroids, at fixed $M_\star$. We saw the same correlation with sSFR previously.

Figure \ref{fig_hist2d_mstar_3} considers key properties of the halo gas and the supermassive blackholes hosted therein. They are: the mean volume-weighted hydrogen number density of the halo gas in $0.15 < r/r_{\rm vir} < 1.0$ (upper left); halo gas fraction within $r_{\rm 200,crit}$ (upper right, i.e. $M_{\rm gas} / M_{\rm tot}$); volume-weighted mean $\beta^{-1}$ ratio of $P_{\rm B} / P_{\rm gas}$ in the halo gas (center left), cumulative energy injected by the central SMBH over its lifetime while in the low accretion `kinetic wind feedback' state (center right), the mass of the SMBH itself (lower left), and the total halo mass M$_{\rm 200,crit}$ (lower right). This final panel shows a view of the stellar mass to halo mass (SMHM) relation.

We have already seen that suppressions of both galactic gas fraction and total metal mass correspond to lower $M_{\rm OVI}$, and the $n_{\rm H,halo}$ panel shows that the average halo gas density traces this same trend. Namely, for $M_\star \ge 10^{10}$\msun there is an outlier population which scatters below the median relation, and these halos with 0.5-1.0 dex lower average gas density also have lower total \ovi mass. The physical process which ejects gas from the central galaxy and metals from the entire halo also clearly lowers the total gas content and so the average density of halo gas.
The amount of this halo gas within the virial radius relative to the cosmic baryon fraction varies as a function of mass. In the most massive clusters, it asymptotes to $\Omega_{\rm b} / \Omega_{\rm m}$, while for the lowest mass galaxies $f_{\rm gas}$ is relatively constant with a median value of roughly half this value. In the transition regime the strong trend of $M_{\rm OVI}$ is similar to that seen in the average halo gas density, such that more baryon depleted halos have progressively less \ovi.

The halo $\beta^{-1} = P_{\rm B} / P_{\rm gas}$ has a more complex behavior: in the transition regime at \mbox{$10.5 \la \log(M_\star [\rm{M}_\odot]) \la 11.0$} galaxies with more \ovi also have higher gas \textit{and} magnetic pressures than the median, and this is true also at fixed halo mass. The increase of $P_{\rm gas}$ is however larger, lowering $\beta$ for these halos. In contrast, at $M_\star \la 10^{10}$\msun we find that $P_{\rm gas}$ increases with \ovi content while $P_{\rm B}$ actually decreases. As a result, halos with $M_{\rm OVI}$ above the median actually have a lower degree of magnetic pressure support, and in net effect $\beta$ increases, causing the crossover of the correlation at $M_\star \simeq 10^{10.5}$\msun. Details of the physics of the magnetic field amplification process and the relative balance between magnetic, thermal, and kinetic energy in halo gas are explored at the Milky Way mass scale in \cite{pakmor17}.

Black hole mass itself, together with the total integrated energy released by BHs in the low feedback state, are the only other properties besides $r_{\rm gas,1/2}$ which display strong horizontal features in the $M_{\rm OVI}$ trend. Namely, at fixed stellar mass over one dex in $M_\star$, galaxies which host a central blackhole more massive than $\simeq 10^{8.1}$\msun show severe depletion of their halo \ovi content. Furthermore, this transition occurs at this same fixed $M_{\rm BH}$ independent of stellar mass, indicating that black hole mass is more strongly correlated with $M_{\rm OVI}$ in the transition regime than $M_\star$ itself.

The physical cause of this association is made clear in the integrated energy injection, which has significant range at fixed stellar mass, from zero up to $\simeq 10^{59}$ erg at $M_\star \simeq 10^{10.5}$\msun. For low values of this injected energy galaxies have their `full complement' of \ovi, as would be expected from the underlying positive trend with stellar mass. Past a well-defined threshold, approximately $\int E_{\rm BH,low} \ga 10^{59}$ erg, the halo loses of order 90\% of its gravitationally bound \ovi mass. This threshold exhibits a diagonal tilt which is nearly linear, increasing by roughly one order of magnitude in total energy from $M_\star = 10^{10}$ to $10^{11}$\msun. It therefore appears that massive galaxies require slightly more massive blackholes before they enter this regime. In this panel we also include an estimate for the redshift zero halo baryon binding energy $E_{\rm B} = (3/5)(\Omega_{\rm b}/\Omega_{\rm m})GM_{\rm 200}^2 / r_{\rm 200}$ with the dashed line. Although this intersects the BH injected energy at the interesting mass scale of $M_\star \sim 10^{10.5}$\msun we caution that this energetics comparison cannot capture the important complexities of how the actual feedback energy has coupled to, and physically affected, baryons in the halo.

Although not shown here, we also observe a scaling of $M_{\rm OVI}$ with the formation time of the dark matter halo, whereby earlier forming halos have less \ovi than later forming halos, at fixed $M_{\rm halo}$. We attribute this to the fact that older halos will have had time to grow more massive BHs and so quench at higher redshift.

The final panel shows the halo mass trend at fixed $M_\star$, which is significant. It is a view of the stellar mass to halo mass (SMHM) relation of TNG at $z=0$, and demonstrates that the scatter has strong residual dependence on $M_{\rm OVI}$. We have already seen the overall trends marginalized over each of the two axes in Figure \ref{fig_ions_vs_mass}. A galaxy has either more or less \ovi in its halo depending sensitivity on its distance away from the median SMHM relation. Of particular importance here is the temperature of the virialized hot halo gas, which increases with $M_{\rm halo}$ such that the presence of \ovi is a strong (although degenerate) indicator of halo mass. Since $T_{\rm vir} \propto M_{\rm halo}^{2/3}$, even a small shift in halo mass can imply a large shift in the temperature-sensitive ionization fraction of \ovi. Although the SMHM relation has a small scatter \citep[see][]{pillepich18} galaxies at a given, fixed stellar mass will exhibit \ovi absorption columns through their CGM which depend on the mass of the parent dark matter halo they inhabit.


\section{Discussion} \label{sec_discussion}

\subsection{The \ovi Content of the Low-Redshift CGM}

The nature of the correlation measured in Figure \ref{fig_slice_color} implies a causal relationship between CGM \ovi and galaxy color (or equivalently, sSFR). This stands in contrast to the result of \cite{oppenheimer16}, which ascribed the dichotomy of $N_{\rm OVI}$ around star-forming versus quenched galaxies as observed in COS-Halos to the underlying correlation of color/sSFR with halo mass. What we find in the TNG model, instead, is a direct effect: the same physical process which strongly modifies the color/sSFR of galaxies as they quench also modifies the physical state of its circumgalactic medium. The clearest culprits from Figures \ref{fig_hist2d_mstar_1}, \ref{fig_hist2d_mstar_2}, and \ref{fig_hist2d_mstar_3} are the two properties, beyond halo mass, which reveal the strongest trends of $M_{\rm OVI}$ at fixed stellar mass. These are the black hole mass, and the total energy injected in the kinetic wind (low accretion state) by the central blackhole. In both cases, the total \ovi mass in the halo drops abruptly above a critical value of $M_{\rm BH} \simeq 10^{8}$\msun, or $\int E_{\rm BH,low} \simeq 10^{59}$ erg, respectively. The panels of Figure \ref{fig_hist2d_mstar_3} for these quantities both show strong horizontal boundaries at these values, indicating that beyond this point (at fixed stellar or halo mass) halos are severely depleted in their \ovi content. Black hole mass as well as total energy injection can only increase with time; such an increase therefore tracks the evolutionary progression of a galaxy.

The physical properties of the halo gas are modified by this energy input, becoming hotter, less dense, and less metal enriched. By inspection of the median radial profiles around red versus blue galaxies at a fixed halo mass of $M_{\rm halo} \simeq 10^{12}$\msun (not shown) we observe that: (i) quenched galaxies are surrounded by a less dense CGM at all radii out to the virial radius; (ii) the metal mass density profiles around quenched galaxies are flatter, being lower within \mbox{$\sim$ 150 kpc} and higher in the near field beyond this radius; (iii) the gas-phase metallicity profiles is likewise shallower, the crossover just outside the galaxy at $\sim$ 20 kpc; and (iv) the CGM around quenched galaxies is less pressurized and hotter at all radii within the virial radius, the latter effect causing an average temperature increase of order 0.1 dex in log K. The cooling properties of the halo gas are also modified. In general, the median cooling time of \ovi halo gas is of order $\sim$ $3-5$ Gyr, increasing gradually with $M_\star$ to as high as $\sim$ 10 Gyr for massive galaxies above $10^{11}$\msun (not shown). Replenishment or maintenance of this gas in its \ovi rich state is therefore required on these timescales. Comparing red versus blue galaxies at $M_{\rm halo} \simeq 10^{12}$\msun we note that the $M_{\rm OVI}$ weighted gas cooling timescales are systematically larger in the halos around quenched systems, by as much as a factor of $2-3$.

These shifts in how halo gas occupies the density-temperature plane lower the ionization fraction of OVI -- the direction of change in Figure \ref{fig_ion_states} (upper left panel) is diagonally upward to the left, moving off the intersection of the photo and collisional excitation branches. At the same time, the mass density of metals including oxygen in the CGM drops within the halo as they are pushed to larger radii. This physical redistribution is the dominant culprit for \ovi suppression, as the total loss of oxygen mass from the halo accounts for the majority of the corresponding lost ionic mass. The low accretion state `kinetic wind' feedback of the central supermassive blackhole therefore impacts the CGM through these two distinct physical effects, and both together result in the dichotomy of \ovi around red versus blue galaxies.

A similar argument related to baryonic feedback processes could also be applied to the galactic winds driven by star formation, rather than blackhole activity. In the context of the TNG model, winds are not the driver of the observed dichotomy around the Milky Way mass regime. However, for less massive galaxies where BHs are sub-dominant, a physical connection between feedback effects and CGM properties may also exist \citep{nelson15a}. Data from the COS-Burst survey offers such a suggestion \citep{heckman17}. Comparing absorption of hydrogen as well as metals in the CGM of starburst galaxies relative to a control sample, they find stronger absorption around the former, consistent with a causal connection between galactic feedback and the properties of its CGM. As their sample spans $\sim 10^{10}$\msun to $\sim 10^{11}$\msun galaxies, precisely the transition regime we have explored in Figures \ref{fig_slice_color} through \ref{fig_hist2d_mstar_3}, we even hypothesize that some of the effects seen are driven by blackhole rather than stellar feedback. Regardless, our findings support the causal nature of the observed differences and highlight the importance of the CGM as a dynamic interface regime between small-scale baryonic effects originating inside galaxies and possibly large-scale consequences in the CGM and even IGM.

\subsection{\ovii and \oviii: Connecting to Milky Way Observations and Models}

We have focused in the second half of Section \ref{sec_abundance_halos} as well as Section \ref{sec_galaxyconnection} on \ovi and the observational connection to UV absorption line studies enabled by COS. Before concluding we return briefly to the two higher ions of \ovii and \oviii. Absorption measurements here require x-ray spectroscopy, and remain difficult due to the spectral resolution and sensitivity of current instrumentation \citep[e.g.][]{buote09,williams13}. However, absorption in both ions in the immediate vicinity of the Milky Way (i.e. from within its CGM) has been detected and characterized.

There have been several subsequent efforts to derive consistent models for the gas distribution in the Milky Way halo. Semi-empirical models often combining \ovii and \oviii tracers in either absorption or emission then determine consistent gas density profiles and other characteristics of our nearby circumgalactic gas reservoir. Particularly, to estimate the total gas mass within the dark matter halo, and therefore assess the apparent `missing baryons' issue relative to the expected cosmic baryon fraction. If the halo hosts a total baryonic mass equal to the universal baryon fraction $f_b = 0.157$ \citep{planck2015_xiii} multiplied by the allowed halo mass range of $10^{11.85} - 10^{12.28}$\msun \citep[$2\sigma$ bounds of][]{mcmillan17}, then the naive expectation is a total $M_{\rm b}$ between $10^{11.0}$\msun and $10^{11.5}$\msun ($\sim$ 2$\sigma$). The observationally constrained total stellar plus cold gas mass is $\sim 6-7 \times 10^{10}$\msun \citep{mcmillan17}, leaving between $4 \times 10^{10}$\msun and $2.5 \times 10^{11}$\msun total baryonic mass expected to reside in the warm/hot circumgalactic medium. 

\cite{miller15} constrain a model with \ovii and \oviii line emission measured with XMM-Newton, with best fit total hot gas masses of $2.9 - 5.3 \times 10^{9}$\msun (within 50 kpc) and $2.7 - 9.1 \times 10^{10}$\msun (within 250 kpc). They note that the best fit radial model would need to extend to several times the virial radius to recover the cosmic baryon fraction. \cite{li17} extend this analysis, also using \ovii and \oviii emission, finding the enclosed baryon mass within 250 kpc to be $3.1^{+0.5}_{-0.3} \times 10^{10}$\msun, which accounts for only 18\% of the missing baryon mass, which is given as $1.7 \times 10^{11}$\msun, after excluding the stars and cold gas. The model of \cite{faerman17} models \ovi absorption from COS sightlines as well as unresolved \ovii and \oviii x-ray absorption to constrain the warm/hot corona of the Milky Way and deduce a total hot gas mass (upper limit) of $3.4 \times 10^{11}$\msun. Their model contains a total gas mass of $1.4 \times 10^{11}$\msun within 250 kpc, which increases to $2 \times 10^{11}$\msun when including the disk baryon mass.

To make a comparison, we construct a simple simulated Milky Way sample by selecting: central galaxies with total $M_{\rm halo}$ in the range given above, a sSFR near the $z=0$ star formation main sequence, specifically $-1.3 < \log{(\rm{sSFR [Gyr^{-1}]})} < -0.3$ to avoid quenched galaxies, and $\kappa_{\rm gas,rot}(J_z > 0) \ge 0.6$ (see Section \ref{sec_galaxyconnection}) to restrict to disk-like morphologies. At redshift zero this sample has 298 galaxies in TNG100. Within 50 kpc, the total gas mass is $3.0^{+1.6}_{-1.2} \times 10^{10}$\msun and the total baryonic mass is $6.0^{+2.3}_{-1.7} \times 10^{10}$\msun, giving the median, 16th, and 84th percentiles. Within 250 kpc, the total gas mass is $1.0^{+0.4}_{-0.2} \times 10^{11}$\msun and the total baryonic mass is $1.3^{+0.5}_{-0.2} \times 10^{11}$\msun. The prediction of TNG is therefore roughly $2/3$ the value of \cite{faerman17}, 4 times the value of \cite{li17}, and fully consistent with the simplest expectation from the cosmic baryon fraction.

As the \ovi, \ovii, and \oviii signatures of our simulated gaseous halos arise self-consistently from the the same baryonic components aggregated in the preceding discussion, this comparison highlights a targeted utility of cosmological hydrodynamical models. In the future, detailed and direct application of the IllustrisTNG simulations to these regimes as well as to detailed empirical modeling of our own Milky Way \citep{miller16} will benefit from a more sophisticated analysis enabled by actual mock absorption spectra \citep[e.g.][]{bird15,hummels17,liang18} as well as explicit modeling of the UV and x-ray emission from extragalactic gas -- \cite{bertone10a}, \cite{corlies16}, \textcolor{blue}{Nelson et al. (in prep)}.


\section{Summary of Conclusions} \label{sec_conclusions}

In this paper we used the TNG100 and TNG300 cosmological magneto-hydrodynamical simulations, part of the IllustrisTNG project, to explore the abundance and distribution of three observable ions of oxygen (\ovi, \ovii, and \oviii) across a wide range of scales, from the circumgalactic medium (CGM) of galaxy halos to the large-scale structure of the intergalactic medium (IGM). We study the prevalence and evolution of the \ovi (O$^{5+}$) distribution in particular, make a quantitative comparison to several observational datasets including COS-Halos\footnote{The data derived herein, including the global scaling relations for different oxygen phases as well as mock UV absorption line surveys, will be released in the future TNG public data release \citep[following][]{nelson15b}.}, and explore the connection between the physical state of the CGM and the properties of the galaxy hosted at its center.

With respect to comparison against current and future observational constraints we find that:

\begin{itemize}
\item The incidence of \ovi absorption $f(N_{\rm OVI})$ is measured in the TNG100, TNG300, and Illustris simulations and contrasted against the observed low-redshift \textbf{column density distribution function (CDDF)}. While Illustris underpredicted the frequency of \ovi absorption at all columns, TNG shows a marked improvement, with TNG300 in excellent statistical agreement, and TNG100 having even slightly too common high column absorption.
\item In addition, we also present the predicted CDDFs of \ovii and \oviii at redshift zero, as well as the \textbf{evolution of all three column density distribution functions} from $z=0$ to $z=4$. High resolution x-ray absorption spectroscopy enabled by future mission concepts, such as the grating spectrometer on Lynx, will provide powerful constraints on these three adjacent oxygen ions for which explicit predictions are now available from hydrodynamical simulations such as TNG.
\item We compare to observed \ovi statistics from the \textbf{COS-Halos} survey by creating a mock survey made up of 100 simulated realizations of each observed galaxy. The resulting sample covers the observed galaxies in the M$_\star$-sSFR plane by construction. Measuring $N_{\rm OVI}$ columns at matching impact parameters for each simulated galaxy, we quantify the statistical (dis)agreement with an average parameter $\langle\lambda\rangle = 0.62^{+0.18}_{-0.38}$, where $\lambda=0$ indicates maximal disagreement and $\lambda=1$ maximal consistency. We conclude that the COS-Halos \ovi absorption data is fully consistent with having been drawn from the simulated probability distributions of the corresponding mock sample. The high observed covering fractions ($\ga$ 60\% within 150 kpc, $\ga$ 80\% within 75 kpc) are successfully produced by TNG100. The observed dichotomy around low vs. high sSFR (or $M_\star$) galaxies is also qualitatively reproduced and quantitatively consistent with the simulations.
\item This same exercise is repeated for the \textbf{eCGM survey}, which samples a wider range of galaxy properties at generally larger impact parameters, and with increased number statistics. For the detections $\langle\lambda\rangle = 0.57^{+0.30}_{-0.21}$, and we conclude that there is no statistically significant tension between the eCGM dataset and TNG100 simulation. We recover with excellent agreement the covering fractions of eCGM split into early-type and late-type galaxy classifications, although we do not find a signal in the differential covering fractions of isolated vs. non-isolated halos as claimed.
\item Our assessment of the TNG model compared to the observed \ovi CDDF and $N_{\rm OVI}$ columns around low-redshift galaxies from COS-Halos and eCGM validates its physical fidelity in a regime the simulation has been neither tested nor calibrated on. 
\end{itemize}

This broad agreement gives us significant confidence that the simulations are a useful, and accurate, tool. We therefore undertake a comprehensive census of these three ions of oxygen and point out several novel, observationally testable predictions of our model:

\begin{itemize}
\item We measure the \textbf{total gravitationally bound mass and average projected column densities} of each ion across the full range of relevant halo (or galaxy) mass: \mbox{$10^{11} < \rm{M}_{\rm halo}/$\msun$ < 10^{15}$} corresponding to \mbox{$10^{9} < \rm{M}_{\rm \star}/$\msun$ < 10^{12}$} in stellar mass. For \ovi, \ovii, and \oviii, the largest projected column densities are reached in halos of $10^{12}$, $10^{12.5}$, and $10^{13.5}$\msun, respectively.
\item We compute the \textbf{stacked radial profiles} of \ovi from the scale of the inner halo to the distant IGM ($\sim$ 1 kpc to $\sim$ 10 Mpc). Combining halos in several mass bins, we provide profiles in 3D number density as well as 2D projected column, both decomposed into `1-halo' and `2-halo' terms. This 2-halo contribution arises from absorption in other secondary halos as well as in the diffuse IGM, and becomes dominant just beyond the virial radius (in 3D number density) or even slightly interior to $r_{\rm vir}$ (in 2D projected column density). For Milky Way mass halos, the median $N_{\rm OVI}(r)$ profile drops slowly from $10^{16}$ cm$^{-2}$ in the halo center to $10^{15}$ cm$^{-2}$ at $r_{\rm vir}$. 
\item Across the global cosmological volume, including both halos and the intergalactic medium, TNG predicts a total matter density of $\Omega_{\rm OVI} = (\rho_{\rm OVI} / \rho_{\rm crit,0}) \simeq 7 \times 10^{-7}$ at $z=0$.
\item Beyond the 1-point statistic of the CDDF, we measure the spatial clustering of \ovi, \ovii, and \oviii mass with the \textbf{two point correlation function} $\xi(r)$. The shape of $\xi$ for all three ions is qualitatively different than that of either total gas mass, or total metal/oxygen mass, flattening at small scales of $\la$ 100 kpc. We interpret this as tracing a relatively homogeneous, shallower mass distribution of ionized oxygen with low intra-halo substructure.
\item Finally, we explore \textbf{how halo \ovi content relates to properties of the central galaxy}. In the transition regime from predominantly blue to red galaxies ($M_\star \sim 10^{10.5}$\msun) we observe that the total mass of \ovi depends strongly on several measurable properties. For instance, galaxies with higher gas fractions, specific star formation rates, gas metallicities, bluer colors, less massive blackholes, or smaller gas sizes all have more \ovi in their CGM than otherwise. Galaxies which are more rotationally supported, in either their stellar or gaseous components, also have more \ovi than their lower $\kappa$ counterparts. These statements all hold at fixed stellar or total halo mass.
\end{itemize}

Finally, investigating the origin of highly ionized oxygen and drawing our main physical inferences:

\begin{itemize}
\item We use a large suite of smaller simulations to explore the \textbf{TNG model sensitivity and dependence on various physical processes and parameters}. In particular, we demonstrate how strongly the \ovi CDDF and covering fractions around Milky Way mass halos change with permutations to the included feedback processes, the details of the galactic winds and blackhole feedback, and more subtle aspects of the magnetic fields, stellar yields, and numerics. While current observations can easily discriminate between extreme model variations, they are most useful at present as an a posteriori consistency check.
\item Most illuminating, we see sharp boundaries in the trends of halo \ovi content with both black hole mass and the total energy injected by supermassive blackholes in their low accretion (kinetic feedback mode) state. At fixed $M_\star$, galaxies with $M_{\rm BH} > 10^{8}$\msun, or total low-state injected energy greater than $10^{59}$ erg have lost $\ga$ 90\% of their \ovi. We conclude that blackhole feedback in its low accretion state directly affects the halo \ovi content, and is the dominant cause of the numerous correlations of $M_{\rm OVI}$ with other galaxy properties. In particular, \textbf{blackhole feedback produces the trend of higher $N_{\rm OVI}$ columns around blue versus red (or star-forming versus quiescent) galaxies}, similar to that first noted in COS-Halos. The dominant reason is direct physical redistribution (i.e. ejection) of metal mass from the CGM; modified ionization state (i.e. hotter gas temperature) is a secondary though non-negligible effect. This dichotomy holds even at fixed stellar mass, and is a causal consequence of galactic-scale baryonic feedback impacting the physical state of the circumgalactic medium.
\end{itemize}


\section*{Acknowledgements}
DN would like to thank Simeon Bird for critical discussions of many related topics over the years, Vicente Rodriguez-Gomez for developing and allowing us to use the \textsc{SubLink} merger tree code, Markus Haider and Nastasha Wijers for spotting mistakes in earlier drafts of this paper, and our referee Dr. Benjamin Oppenheimer for many constructive comments and suggestions.
VS, RP, and RW acknowledge support through the European Research Council under ERC-StG grant EXAGAL-308037 and would like to thank the Klaus Tschira Foundation. 
SG is supported by the Simons Foundation through the Flatiron Institute.
The primary simulations presented herein have been possible due to the Gauss Centre for Supercomputing (GCS) which provided computing time for the GCS Large-Scale Projects GCS-ILLU (2014) and GCS-DWAR (2016) on the GCS share of the supercomputer Hazel Hen at the High Performance Computing Center Stuttgart (HLRS). GCS is the alliance of the three national supercomputing centres HLRS (Universit{\"a}t Stuttgart), JSC (Forschungszentrum J{\"u}lich), and LRZ (Bayerische Akademie der Wissenschaften), funded by the German Federal Ministry of Education and Research (BMBF) and the German State Ministries for Research of Baden-W{\"u}rttemberg (MWK), Bayern (StMWFK) and Nordrhein-Westfalen (MIWF). 
Additional simulations were carried out on the Hydra and Draco supercomputers at the Max Planck Computing and Data Facility (MPCDF). Some additional computations in this paper were run on the Odyssey cluster of the FAS Division of Science, Research Computing Group at Harvard University.

\bibliographystyle{mnras}
\bibliography{refs}

\end{document}